\documentclass[useAMS,usenatbib]{mnras}
\usepackage{graphicx} 
\usepackage{natbib}
\usepackage{multirow}
\usepackage{threeparttable}
\usepackage{rotating}
\usepackage{color}
\usepackage{alphalph}

\title[Molecular Cloud Properties in Simulations and Observations]
  {What is a GMC? Are Observers and Simulators Discussing the Same Star-forming Clouds?}
  
\author[Pan et al.]{Hsi-An Pan$^1$, Yusuke Fujimoto$^1$, Elizabeth J. Tasker$^1$, Erik Rosolowsky$^2$,
  \newauthor 
   Dario Colombo$^2$, Samantha M. Benincasa$^3$, and James Wadsley$^3$ \\
  $^1$ Department of Physics, Faculty of Science, Hokkaido University, Kita 10 Nishi 8 Kita-ku, Sapporo 060-0810, Japan\\
  $^2$ Department of Physics, 4-181 CCIS, University of Alberta, Edmonton, AB T6G 2E1, Canada\\
  $^3$ Department of Physics and Astronomy, McMaster University, Hamilton, ON, L8S 4M1, Canada
}

\begin{document}
\label{firstpage}
\pagerange{\pageref{firstpage}--\pageref{lastpage}}
\maketitle

\begin{abstract}
As both simulations and observations reach the resolution of the star-forming molecular clouds, it becomes important to clarify if these two techniques are discussing the same objects in galaxies. We compare clouds formed in a high resolution galaxy simulation identified as continuous structures within a contour, in the simulator's position-position-position (PPP) co-ordinate space and the observer's position-position-velocity space (PPV). Results indicate that the properties of the cloud populations are similar in both methods and up to 70\% of clouds have a single counterpart in the opposite data structure. Comparing individual clouds in a one-to-one match reveals a scatter in properties mostly within a factor of two. However, the small variations in mass, radius and velocity dispersion produce significant differences in derived quantities such as the virial parameter. This makes it difficult to determine if a structure is truely gravitationally bound. The three cloud types originally found in the simulation in \citet{Fuj14} are identified in both data sets, with around 80\% of the clouds retaining their type between identification methods. We also compared our results when using a peak decomposition method to identify clouds in both PPP and PPV space. The number of clouds increased with this technique, but the overall cloud properties remained similar. However, the more crowded environment lowered the ability to match clouds between techniques to 40\%. The three cloud types also became harder to separate, especially in the PPV data set. The method used for cloud identification therefore plays a critical role in determining cloud properties, but both PPP and PPV can potentially identify the same structures. 
\end{abstract}




\section{Introduction}
\label{sec_intro}

The physical properties of molecular clouds --the birthplace of stars--  are of vital importance, since they reveal clues as to why stars are formed in certain regions, how many stars can form, how star formation proceeds and how it affects the next cycle of stellar production \citep[e.g.,][]{Lar81,Beu02,Lad03,Alv07,And10,Bat14}. Therefore, the method we use to determine these cloud properties is a critical choice.

Traditionally, the best quality data of molecular clouds has come from our own galaxy, where the small distances involved allow galactic molecular clouds with sizes down to $\geqslant$ 10 pc to be resolved even with single dish telescopes. These observations have offered the chance to study molecular clouds in some detail, with previous studies revealing our Galactic molecular cloud population has a mass, size and velocity dispersion of 10$^{4-6}$ M$_{\sun}$, 5 -- 70 pc, and 2 -- 10 km s$^{-1}$, respectively \citep[e.g.,][]{Sol87,Hey09,Rom10}. Despite the quality of this data, it provides information only on one type of galaxy, and the Milky Way is relatively quiescent. Studies of external galaxies are therefore required before we can reveal the whole picture of the relation between molecular clouds and star formation. 

Of course, the drawback of extragalactic studies is the large distances that limit the resolution. However, observations of molecular clouds in these galaxies are now approaching the quality of that in the Milky Way, thanks to instruments and techniques such as the long-baseline interferometry employed with ALMA. ALMA is designed to detect spectral lines such as carbon monoxide (CO) --which traces molecular clouds-- in a normal galaxy like the Milky Way at a redshift $z = 3$ in less than 24 hours. This rate means that the time required to map molecular clouds in a nearby galaxy will shrink to only a few hours. With the highest expected angular resolution of low transition CO lines being about 0.04$\arcsec$, a galaxy $\sim$ 16 Mpc away in the Virgo cluster can be mapped with a physical resolution of $\sim$\,4\,pc. The highest velocity resolution of low transition CO lines at 0.01\,km s$^{-1}$, so assuming the molecular clouds in nearby galaxies have comparable properties to those in the Milky Way, these resolutions should be sufficient to resolve these populations. Note that this assumption seems to be valid based on the current surveys of extragalactic clouds such as PAWS \citep{Sch13,Hug13,Col14} and CANON \citep{Don13}. These observations suggest that extragalactic and Galactic molecular clouds share  similar Larson's scaling relation of cloud properties.

With the spatial and velocity resolution of observations about to resolve a wide population of molecular clouds, comparisons between observations and simulations become a key tool in understanding the results. However, this can be challenging since the two techniques are based on different data structures. Simulations normally have data with three dimensional position ($x$, $y$, $z$) and velocity ($v_x$, $v_y$ ,$v_z$) coordinates, from which they directly calculate the physical properties of the molecular clouds such as mass, radius and volume density. Meanwhile observations have to take their data from the projected image on the sky. This provides only two spatial dimensions (RA and Dec) and typically a single velocity along the line-of-sight. The resultant properties of the clouds are therefore computed in three independent dimensions, rather than six. 

An additional issue is the choice for the molecular cloud boundary. Clouds are typically selected based on their density (in the case of simulations) or intensity (from observations). Yet molecular clouds are not isolated entities and therefore exactly where their edges lie is not obvious. Simulators navigate this issue by selecting a fixed threshold for the cloud edge, while observers consider the ratio of the intensity to the RMS noise\footnote{RMS noise is a thermal, unavoidable noise in observations. Level of RMS noise is controlled by the  weather condition, sky brightness temperature, and receiver temperatures.}. Even within a single technique, there is no consensus for what such thresholds should be. The problem is magnified when the molecular region is extended. Should a continuous contour mark a single cloud or should this be divided into separate peaks? 

A major reason for this degeneracy is the absence of a `correct' answer. Molecular clouds are part of the interstellar medium (ISM); a continuous blend of pressures and densities \citep[e.g.][]{Tas06}. There is therefore no obvious edge to the clouds, which are typically thought to be borderline gravitationally bound at best \citep{Dob11, Hey09}. Likewise, it is not clear which properties should be used to identify the cloud. Simulators typically only use position, taking advantage of their three spatial dimensions. Yet, this excludes the velocity information which could identify two sections of an object moving together. Conversely, observers are plagued by projection effects which can result in well separated bodies being treated as a single object or regions of the same cloud being artifically split due to the internal motions of the cloud. 

The combined result of different data structures and varying choices for cloud identification means that conclusions are being drawn about the star formation conditions in galaxies from results that are not considering the same objects. 

Previous work has tried to assess the extent on this problem on smaller scales, comparing star-forming regions found in synthetic observations of a single simulated molecular cloud with those identified using the full position data. \cite{Ost01} first noticed that the synthetic observed GMCs in their 3D simulations include spatially-unconnected structures  due to the projection effect. When examining dense clumps found in two- and three- dimensional position data, recently \citet{War12} found that projection effects could overestimate the mass of the clumps within the cloud and falsely identify more diffuse regions as clumps. The result was a shift by a factor of three in the clump mass function (CMF), potentially explaining the difference between the observed CMF and the lower stellar Initial Mass Function (IMF) \citep[e.g.,][]{Alv07}. \cite{War12} did note that the addition of the velocity in their two dimensional data improved the match with the three dimensional clump properties compared to when the two dimensional data was used alone. 

\citet{Bea13} further considered the differences in the observed chemistry and gravity of the clump properties by matching clumps in the synthetic observations and three dimensional spatial data if they originate from the same density structure. They concluded that the scatter in the clump properties causes significant uncertainty in the virial parameter (estimating gravitational boundness) of a cloud such that it is difficult to connect this with star formation. 

In this paper, we present a comparison of the physical properties of the molecular clouds identified in three spatial co-ordinates (position-position-position space or PPP) and those in two projected spatial co-ordinates and a single line-of-sight velocity co-ordinate (position-position-velocity space or PPV) in the same simulated galaxy. The goal is to assess whether the same objects can be identified in these two methods and if the properties are consistent.  
This work is on a larger-scale than the previously mentioned studies, comparing the properties of a global cloud populations which extragalactic observations are starting to capture.  

The simulated galaxy was modelled on the barred spiral (SABc) galaxy, M83, using observational data from the 2MASS K-band image \citep{Jar03} to estimate the stellar potential. The simulation was run using the three dimensional adaptive mesh refinement (AMR) hydrodynamics code, \texttt{ENZO} \citep{Bry14} and is presented in \cite{Fuj14}, along with a full description of the run parameters. The gas radiatively cooled down to 300\,K (a limit designed to allow for the lack of small-scale pressure from unresolved turbulence or magnetic fields) but no star formation or feedback was included. 

\begin{figure*}
    \hspace{-4mm}
    \begin{minipage}{0.49\textwidth}
        \centering
		\includegraphics[width=0.9\textwidth]{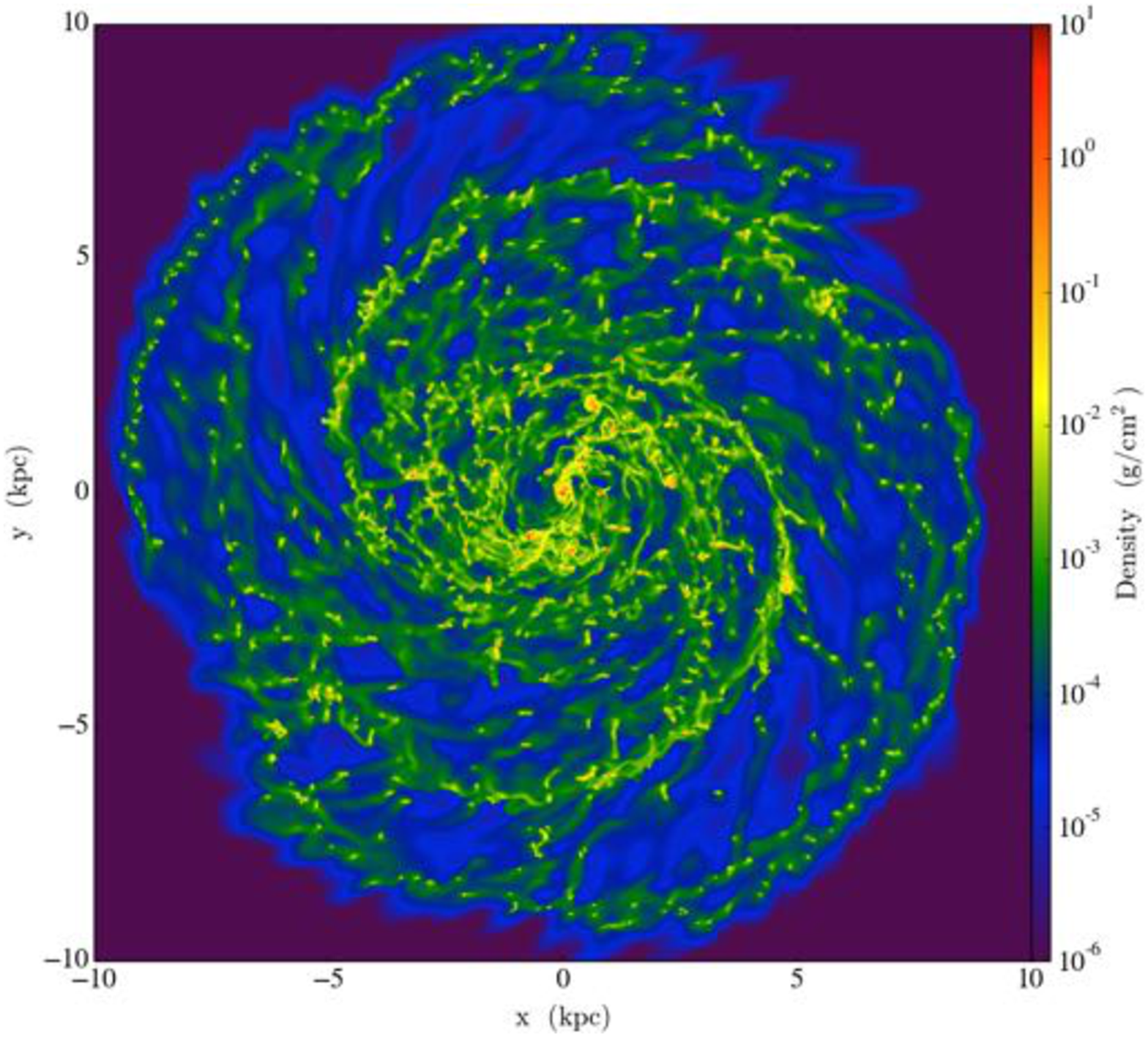}
\begin{center}
 (a)
\end{center}
    \end{minipage}
    \begin{minipage}{0.49\textwidth}
        \centering
		\includegraphics[width=0.9\textwidth]{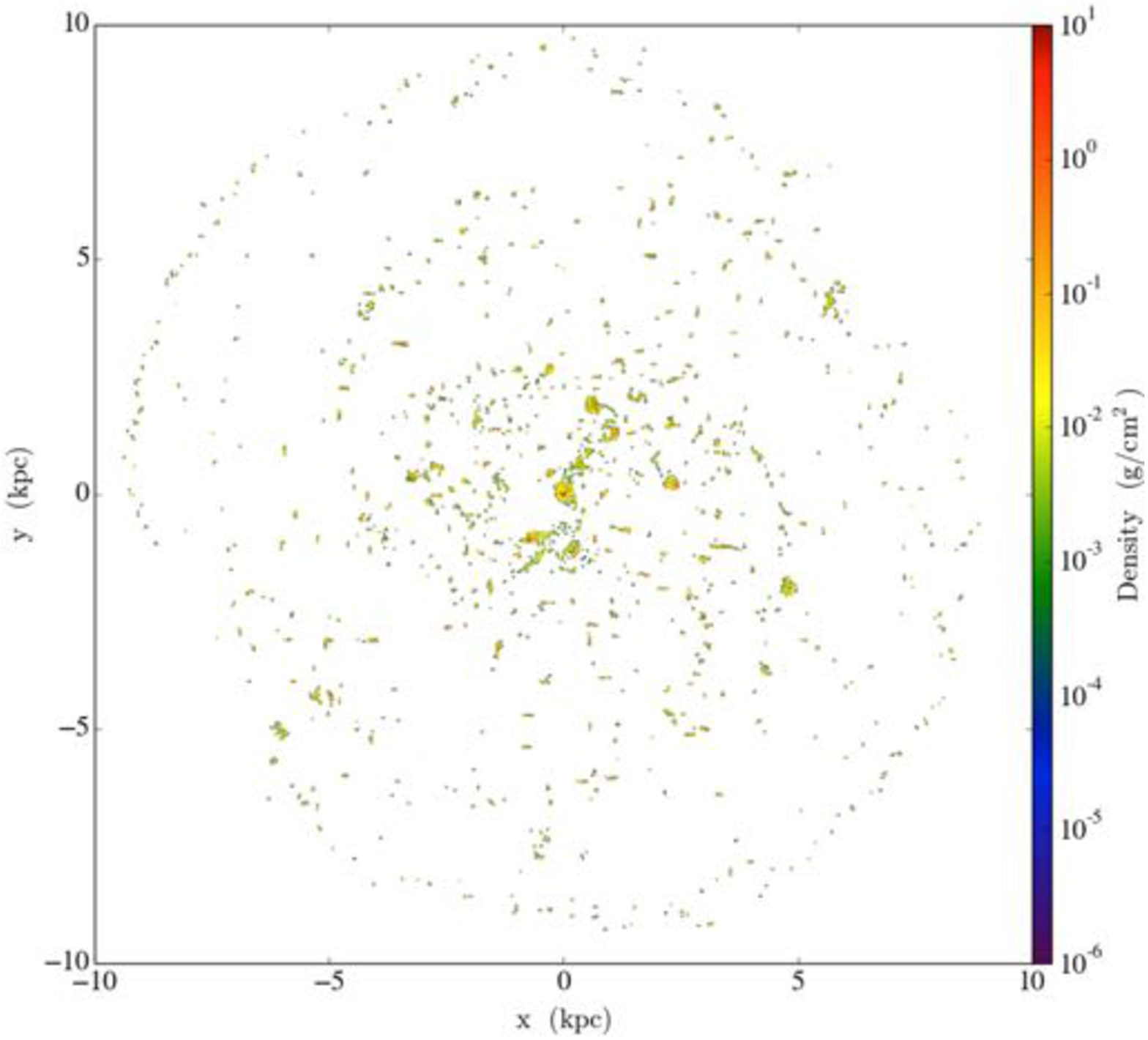}	
\begin{center}
 (b)
\end{center}
    \end{minipage}
    \caption{Projection image of the simulated galaxy used in this work.  (a) Projection image of the simulated galaxy at time of 240 Myr in unit of g cm$^{2}$. All cells are used to make the image. The galaxy was simulated by \citet{Fuj14}.  (b) Projection image of the  galaxy made with cells greater than 100 cm$^{-3}$.}
    \label{FIG_sim_galaxies_full}
\end{figure*}

A projected face-on image of the gas density in the simulated galaxy is shown in Figure~\ref{FIG_sim_galaxies_full}(a). The galaxy's bar and two spiral arms are visible, with the arms extending to a radius of $\sim 10$\,kpc and the bar-end at 2.3\,kpc. M83 is at a distance of 4.5\,Mpc, corresponding to a resolution of 1$\arcsec$ $\approx$ 20\,pc \citep{Thi03}. In the simulation, the smallest cell size is $\sim 1.5$\,pc. 

In \cite{Fuj14}, the molecular clouds are identified in PPP space using two different methods. The main cloud identification method uses a continuous contour at a density of $n_{\rm thresh} = 100$\,cm$^{-3}$. Using this technique, they found GMCs came in three different types:  \emph{Type A} clouds were the most commn GMCs, dominating the cloud populations in all regions and having properties that agreed with typical observations. \emph{Type B} clouds are massive giant molecular associations (GMAs) created through multiple cloud interactions, and \emph{Type C} clouds are  unbound, transient clouds, often found in tidal tails. The second method divided clouds within this contour if there are two peaks separated by at least 20\,pc; a technique used in \citet{Tas09} for cloud identification. We compare these two methods with clouds identified in a PPV synthetic observational data set of the same galaxy (also known as a spectral line data cube). This data set type is commonly used in millimetre to submillimetre observations, which trace the gas in the ISM. The line-of-sight direction which defines the velocity co-ordinate is along the $z-$axis of the galaxy, observing the simulation face-on. The PPV data cube was created with a spatial and velocity spacing of 2\,pc (slightly larger than the simulation cell size) and 1\,km/s and included a Gaussian thermal broadening with a width equal to the velocity dispersion of the gas.

When identifying the clouds in the PPV data set, we assume the galaxy is observed in $^{12}$CO (1--0). $^{12}$CO (1--0) is excited in low density molecular gas, making it is an ideal tracer of the entire molecular cloud. It also has a critical excitation density of $n_{\rm HI} \sim 100$\,cm$^{-3}$, allowing us to use a cloud identification threshold of $n_{\mathrm{HI, thresh}}$ $=$ 100 cm$^{-3}$, in keeping with the PPP schemes. For the PPV cube, only cells with a density greater than the threshold value were included in the data. Such a cut-off produces a projected image shown in Figure~\ref{FIG_sim_galaxies_full}(b). 

\begin{figure}
\includegraphics[width=0.4\textwidth]{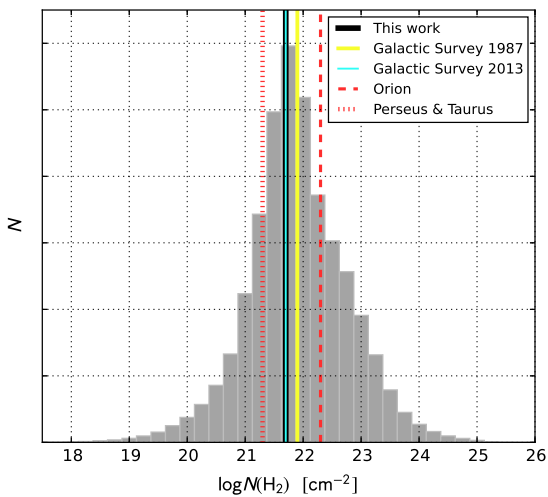}
\caption{Distribution of column density in the PPV spectral line data cube of the simulated galaxy (Figure \ref{FIG_sim_galaxies_full}(b)). Overlaid are the typical column densities of the Galactic clouds. Median column density of our PPV gas is $\log N$(H$_{2}$) $=$ 21.7 cm$^{-2}$ (black line). Typical column densities that \citet{Sol87} and \citet{Bur13} suggest from their Galactic surveys are plotted with yellow and cyan solid lines at $\sim$ 21.9  and $\sim$ 21.7 cm$^{-2}$, respectively. Typical column density of the Orion molecular cloud is $\sim$ 22.3 cm$^{-2}$ \citep[red dashed line,][]{Ber14}. Clouds Perseus and Taurus, have similar column density of $\sim$ 21.3 cm$^{-2}$ \citep[red dotted line,][]{Gol08,Goo09}. All the Galactic values are measured with CO observations.}
\label{FIG_column_density}
\end{figure}

To assess how close our simulated clouds are to those observed in the Galaxy, we plotted the PPV column density distribution and overlaid the typical values found for the Galactic clouds in Figure \ref{FIG_column_density}. The grey histogram shows the PPV distribution, with the typical column density marked by the black line. The Galactic star-forming molecular clouds (color lines include two Galactic surveys \citep[yellow and cyan lines,][]{Sol87,Bur13} and three individual clouds: Orion \citep{Ber14}, Perseus \citep{Goo09} and Taurus \citep{Gol08} (red lines), all estimated by CO observations. The median column density for our PPV data ($\log N\mathrm{(H_{2})}$ = 21.7 cm$^{-2}$) is consistent to the typical values of the Galactic clouds, where $\log N\mathrm{(H_{2})}$ commonly ranges between 21.0 -- 22.5 cm$^{-2}$. This suggests that our results should apply well to genine observations. 

This paper is organized as follows: In Section~\ref{sec_cloud_identification}, we introduce four different cloud identification methods using PPV and PPP, and present two of these as the main methods in this work. Section \ref{sec_results} presents the comparison of the global properties between PPV- and PPP-clouds, as well as the one-to-one match cloud properties between two sets. Classification of molecular clouds based on their physical properties is also shown in Section \ref{sec_results}.  In Section \ref{sec_diff_methods}, we show a brief results of the two minor cloud identification methods. Summary of this work is given in Section \ref{sec_summary}.

\section{Cloud identification in PPP and PPV}
\label{sec_cloud_identification}

For both our data sets in PPP and PPV, we used two cloud identification methods. The first method identifies continuous structures of gas above our chosen density or flux threshold, which we refer to as {\it islands}. The second method can further segregate these bodies into smaller objects such that each contains a single local maxima. We refer to this as the {\it decomposition} method. 

 PPP clouds are identified in the original data from the simulation, while clouds are found in PPV space by first creating a PPV data cube from the simulation data and passing this in FITS format to the CPROPS package \citep{Ros06}.  CPROPS expects the data to be emission intensity, rather than the gas density followed by simulations. To convert between the two, we use a CO-to-H$_2$ conversation factor \citep{Bol13} to change the physical unit (g cm$^{-2}$) to observed $^{12}$CO (1--0) flux (Kelvin). Since we do not consider chemistry (to compute the exact CO abundance) nor radiative transfer and the excitation of molecular lines, this value is for an ideal circumstance. This also means that the conversion factor cancels when we derive the cloud mass, so its precise value does not affect our results.
Since the flux threshold in CPROPS is determined by the ratio to the noise level ($\sigma\mathrm{_{RMS}}$), the so-called signal-to-noise ratio (S/N), a low random noise with level of $\sim$ 1 K is added to the PPV cube. This sensitivity is sufficient to identify the smallest identified Galactic molecular clouds at about 5 $\sigma\mathrm{_{RMS}}$ level with our spatial ($\sim$ 2 pc) and velocity resolution ($\sim$ 1 km s$^{-1}$). These resolutions and sensitivity are higher than current ALMA observations, but technically achievable using the full-operation of ALMA and its long-baseline ability. While there is no doubt that resolution and sensitivity play a strong role in determining GMC properties, they are downstream of this work. Instead, we focus on differences due to the techniques themselves in the absence of external factors.

The cloud boundaries in the island method are defined slightly differently in the two techniques. PPP works from the lowest density, drawing a contour at $n_{\rm HI, thresh} = 100$\,cm$^{-3}$ and defining all cells within a closed section as the cloud. PPV via CPROPS begins by masking the emission with a high S/N, picking out the cloud locations at densities much higher than the background. It then extends this mask to the user defined lowest S/N, which outlines the observed cloud boundary. CPROPS then assumes that the real cloud boundary is larger than the observed cloud boundary, since the cloud outer regions are being obscured by the background noise. It therefore extrapolates from the observed boundary to a sensitivity of 0 K to form the real cloud boundary.

In the decomposition method, the cloud boundaries are identified in a similar manner between PPP and PPV. PPP-clouds are located by searching for local maxima and the adjoining surrounding cells that are above the threshold $n_{\rm HI, thresh} \geqslant 100$\,cm$^{-3}$, assigning these to a single cloud. In the PPP data, maxima closer than 20 pc arere merged into a single object.  PPV via the CPROPS decomposition method initially uses the island method to find the continuous structures and then searches these for separate peaks. For computational speed, the search for island peaks is performed within a cube with a (user-defined) side of 22\,pc $\times$ 22\,pc $\times$ 7 km s$^{-1}$. This size was selected to be comparable to the average GMC  diameter and velocity dispersion in our simulation, ensuring that a single cloud would not be accidentally divided by our numerical choice. If multiple peaks are found within an island, CPROPS finds the lowest contour that surrounds both peaks and then separates them with a contour twice the RMS noise above the shared boundary (the default value suggested by \citealt{Bru03} and \citealt{Col14}).  All pixels within the separated contours are assigned to the peaks. CPROPS can potentially merge peaks within an island if their properties are sufficiently similar, but the high sensitivity of our analysis meant that this was not necessary \citep{Col14}.

The physical properties of the clouds are then derived once the cloud boundaries are set. These derivations are not identical between the two methods, since the assumed raw data (simulation or observation) is measuring different quantities. We do not attempt to correct for this, but adopt the original calculations that are used to describe clouds in simulation and observational studies as part of the comparison between the two methods. 

In both methods and data structures, cloud mass ($\mathrm{M_{c,ppp}}$ and $\mathrm{M_{c,ppv}}$) are calculated by the sum of the cells or pixels in the clouds. We note that a CO-to-H$_{2}$ conversion factor is required for the PPV data when moving between the observed flux and physical gas quantity. A conversion factor of 2 $\times$ 10$^{20}$ cm$^{-2}$ (K km s$^{-1}$)$^{-1}$ is adopted in this work, which is the default value of CPROPS. However, this is used twice in the derivation of the cloud mass and cancels itself out  due to the lack to chemistry,  radiative transfer and excitation of molecular line in this work. We confirmed this by adopting a conversation factor differing by a factor of ten and recalculating the mass to achieve the same result.The mass in both methods is therefore the total mass of gas in cells within our clouds.

For the cloud radii and velocity dispersion properties, the PPP and PPV calculations differ more significantly. The PPP data is calculated from three spatial and three velocity dimensions. It therefore measures the average radius ($R_{\mathrm{ppp}}$) of cloud from its three projected axes, $x-y$, $x-z$, and $y-z$, and computes the mass-weighted one-dimensional velocity dispersion from the average deviations between the gas velocity and the cloud's bulk velocity velocity in $x$, $y$, and $z$ directions. In contrast to this, PPV must measure the properties projected along a single direction. To do this, CPROPS first calculates the geometric mean of the square root of the spatial mass(flux)-weighted second centralized moment\footnote{The second centralized moment is commonly called the variance and is denoted as $\sigma^{2}$. The root-mean-square (RMS) (standard deviation) $\sigma$ is the square root of the variance.} of the intensity distribution along the major and minor axes of the projected cloud boundary. Quantitively, these are defined for each cloud as:  

\begin{eqnarray}
\sigma _{\mathrm{major}} &=& \sqrt{\frac{\sum _{i}T_{i}(x_{i}-\bar{x})^2}{\sum _{i}T_{i}}} \nonumber \\
\sigma _{\mathrm{minor}} &=&\sqrt{\frac{\sum _{i}T_{i}(y_{i}-\bar{y})^2}{\sum _{i}T_{i}}} \nonumber \\
\bar{x} &=&\frac{\sum T_{i}x_{i}}{\sum T_{i}} \nonumber \\
\bar{y} &=&\frac{\sum T_{i}y_{i}}{\sum T_{i}}, 
\end{eqnarray}

\noindent where the $\sigma _{\mathrm{majar}}$ and $\sigma _{\mathrm{minor}}$ are the RMS size of the intensity distribution along the $x$ and $y$ axis, $T_{i}$ is the observed flux of the $i$th pixel within the cloud, $x_{i}$ and $y_{i}$ are the position in the two spatial dimensions of the $i$th pixel and $\bar{x}$Ì and $\bar{y}$ are the flux-weighted mean positions of the cloud. The RMS size of the cloud ($\sigma_{r}$) is the geometric mean of $\sigma _{\mathrm{majar}}$ and $\sigma _{\mathrm{minor}}$, $\sigma_{r}=\sqrt{\sigma _{\mathrm{major}}\sigma _{\mathrm{minor}}}$.

The RMS size is then converted into the effective radius ($R_{\mathrm{ppv}}$) by assuming a mass-centered density profile of a spherical cloud. This is to allow for the fact that the true cloud edge extends beyond the identified boundary of the cloud. The relation is simplified to $R_{\mathrm{ppv}}\,=\,1.91\sigma_{r}$, where the constant is an empirical correction derived from the resolved Galactic GMCs in  $^{12}$CO (1--0) observations \citep{Sol87}. Note that unlike the PPP radii, the $R_{\mathrm{ppv}}$ value is weighted by the flux.  In a similar fashion, the velocity dispersion ($\sigma_{\mathrm{ppv}}$) for the PPV clouds is the flux-weighted RMS velocity dispersion within the defined cloud boundary:
\begin{eqnarray}
\sigma _{\mathrm{ppv}} &=& \sqrt{\frac{\sum _{i}T_{i}(v_{i}-\bar{v})^2}{\sum _{i}T_{i}}} \nonumber \\
\bar{v} &=&\frac{\sum T_{i}v_{i}}{\sum T_{i}},
\end{eqnarray}
where $v_{i}$ is the velocity of the $i$th pixel within the cloud, $\bar{v}$ is the flux-weighted mean velocity of the cloud.

For both the PPP and PPV methods, the mass surface density ($\Sigma_{\mathrm{ppp}}$ and $\Sigma_{\mathrm{ppv}}$), virial parameter ($\alpha_{\mathrm{ppp}}$ and  $\alpha_{\mathrm{ppv}}$) and virial mass ($\mathrm{M_{vir,ppp}}$ and $\mathrm{M_{vir,ppv}}$) are derived from the basic properties (cloud mass, radius and velocity dispersion) in the standard way: $\Sigma$ is defined as the mass per unit area and is therefore simply calculated from the cloud mass and radius. $\alpha$ measures the gravitational binding of a cloud, assume a spherical profile and no magnetic support or pressure confinement. Specifically, $\alpha > 2$ indicates that the cloud is gravitationally unbound while $\alpha < 2$ suggests a bound system. It is defined from the mass, radius and velocity dispersion of the cloud as $\alpha = \frac{1040\times R\times \sigma^2}{M}$. The virial mass comes from this parameter, $\mathrm{M_{vir}} = \alpha \times M$.

\section{Results}
\label{sec_results}
\subsection{General Comparisons}
\label{subsec_general_compare}

\begin{figure*}
    \begin{minipage}{0.45\textwidth}
        \centering
		\includegraphics[width=0.9\textwidth]{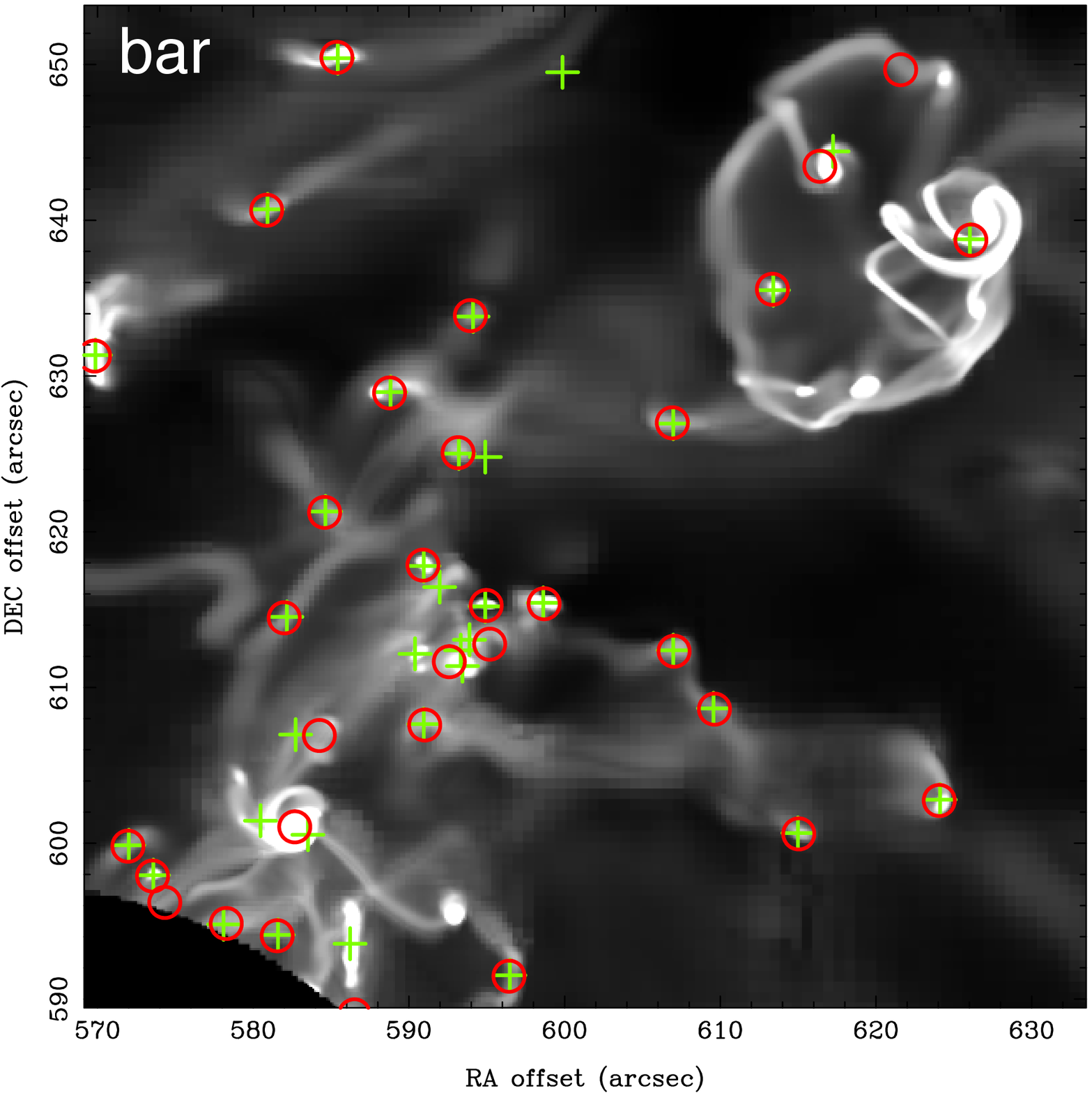}
\begin{center}
 (a)
\end{center}
    \end{minipage}
    \begin{minipage}{0.45\textwidth}
        \centering
		\includegraphics[width=0.9\textwidth]{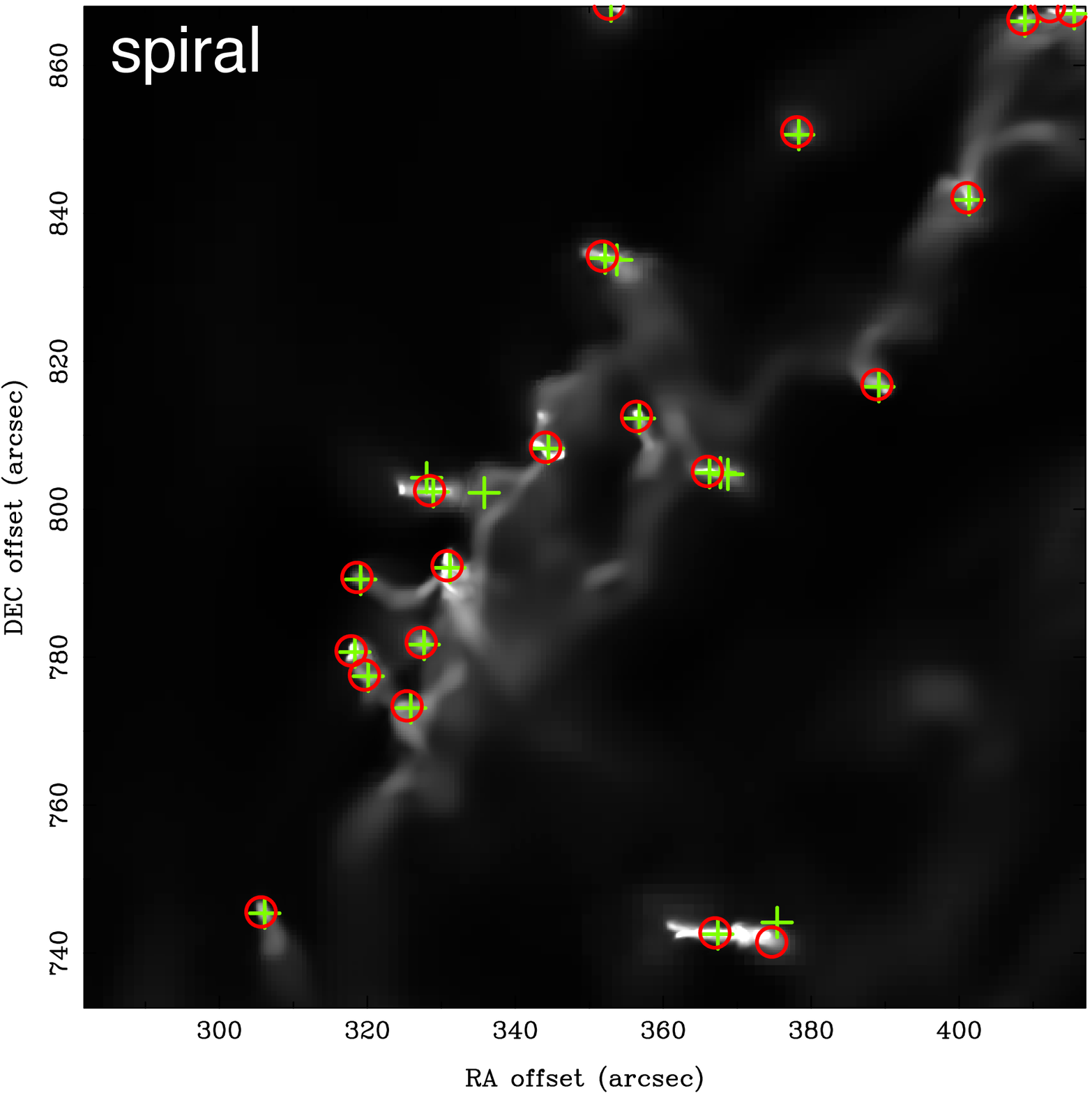}	
\begin{center}
 (b)
\end{center}
    \end{minipage}
    \begin{minipage}{0.45\textwidth}
        \centering
		\includegraphics[width=0.9\textwidth]{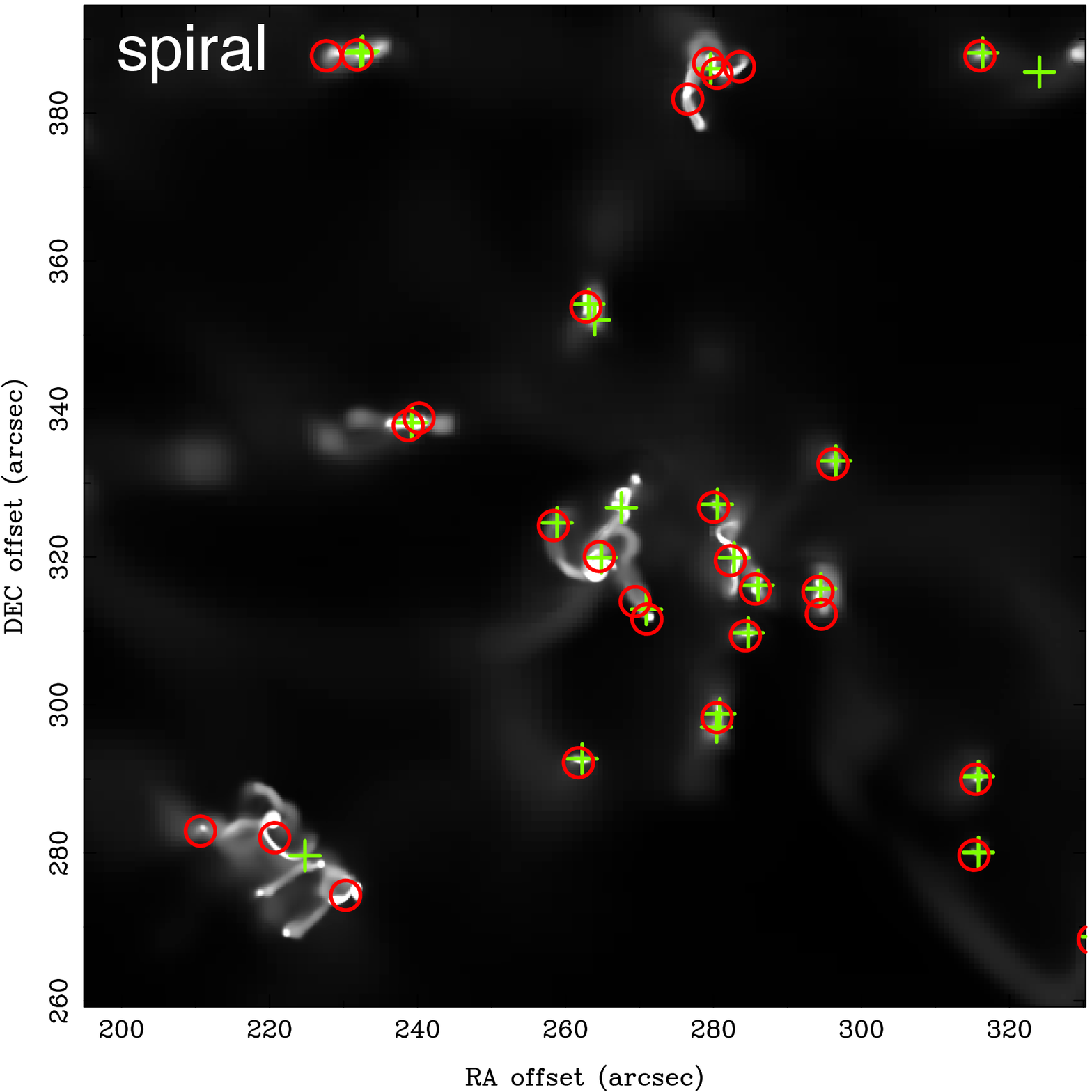}	
\begin{center}
 (c)
\end{center}
    \end{minipage}
    \begin{minipage}{0.45\textwidth}
        \centering
		\includegraphics[width=0.9\textwidth]{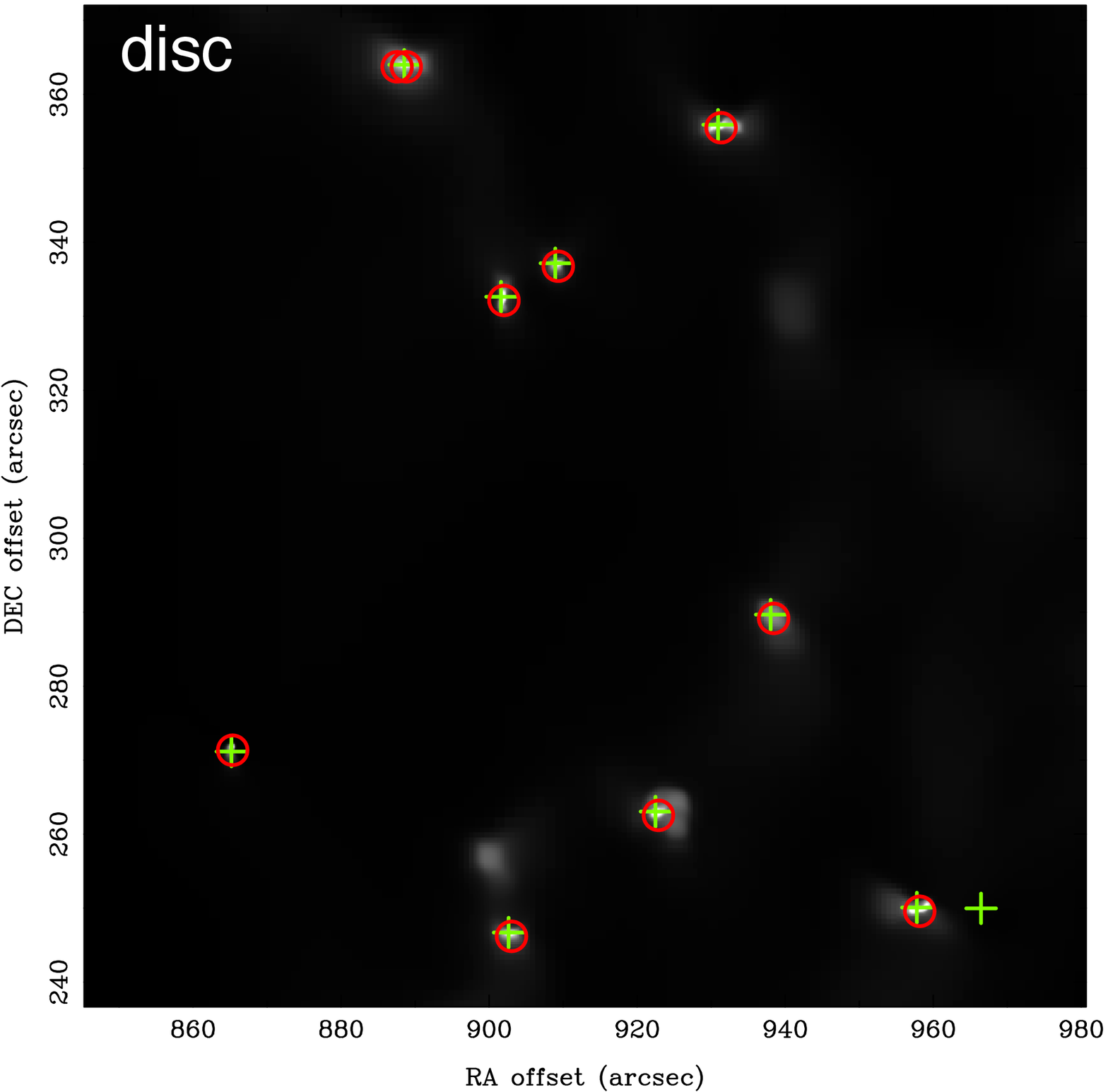}	
\begin{center}
 (d)
\end{center}
    \end{minipage}
    \caption{Examples of cloud distribution in four regions. Each image shows the projected gas density (without adding noise) in gray scale for gas above the cloud definition threshold of 100\,cm$^{-3}$. Overlaid are the cloud positions with red circles for the PPV clouds and green pluses for the PPP clouds. The size and galactic location of the panels are: (a) 1.2 kpc $\times$ 1.2 kpc region $\sim$ 1.2 kpc from the galactic center, (b) 2.4 kpc $\times$ 2.4 kpc region $\sim$ 5.7 kpc from the galactic center, (c) 2.4 kpc $\times$ 2.4 kpc region $\sim$ 6.8 kpc from the center and (d) 2.4 kpc $\times$ 2.4 kpc region $\sim$ 7.7 kpc from the center.}
   \label{FIG_position_examples}
\end{figure*}

A visual comparison of the clouds found in the PPP and PPV data sets for the island method is shown in Figure~\ref{FIG_position_examples}. The panels each show a different region of the galactic disc, with the projected gas density overlaid with red circles for the PPV clouds and green plus signs for the PPP clouds. The location of each symbol marks the cloud centre-of-mass, but the symbol size is not proportional to the cloud properties. Panel (a) shows the inner disc region, around 1.2\,kpc from the galactic centre and within the grand design bar shown in Figure~\ref{FIG_sim_galaxies_full}. Due to the high number density of clouds in this region, this image size is $1 \times 1$\,kpc, whereas panels (b) to (d) show a region of $2.4 \times 2.4$\,kpc. These three panels step outwards from the disc, with panel (b) showing a region at 5.7\,kpc from the galactic centre, panel (c) at 6.8\,kpc from the centre and panel (d) at 7.7\,kpc from the centre and outside the spiral arms. 

Notably, each PPP cloud typically has only one counterpart in PPV, even in the most crowded panel in Figure~\ref{FIG_position_examples}(a). However, this is not universally true. Each panel of Figure~\ref{FIG_position_examples} displays examples of where either the PPP or PPV method has identified multiple clouds where the counter scheme finds a single object.  A unique PPV cloud may have multiple PPP counterparts that are physically close with similar velocities, causing them to merge in the PPV projected data space. Conversely, a PPP cloud may have multiple PPV counterparts if the level of the noise defining the cloud edge in PPV is higher (or within a factor of two) than the gas density between peaks within a single PPP cloud, causing the structure to split. For the gas to be detected in both PPP and PPV, it must be higher than 100\,cm$^{-3}$ and $\sim 2\times 10^{-3}$\,g\,cm$^{-2}$ in projection. Occassionally, this causes a PPP cloud to have no PPV counterpart at all, as shown in the top of panel (a), where the small cloud diameter gave a projected density below the noise limit. These splits may also lead to the centre-of-mass of the clouds in a region not overlapping, producing an off-set as seen in the lower left corner of panel (c).

This agreement in cloud locations also extends to the cloud properties. Table~\ref{TAB_phy_prop_gal_str} shows a comparison between the median properties of the clouds found by the PPV and PPP island methods. The top row shows the properties for the entire cloud population, while the next three rows takes an separate look at the bar, spiral and disc populations (see next section). In all areas, PPP finds slightly more clouds than in the PPV case, although the numbers differ by less than 10\,\%.  This indicates that the effect of merging PPP clouds due to projection effects is slightly more pronounced than splitting PPP clouds due to noise. Both PPP and PPV clouds have masses around  $4 \times 10^5$\,$M_\odot$ and radii around 15\,pc. The cloud surface density peaks at 525\,M$_\odot$pc$^{-2}$, with velocity dispersion around 4 - 5\,km s$^{-1}$ and the clouds are found to be borderline gravitationally bound.

\begin{table}
\caption{Median physical properties of GMCs.}
\label{TAB_phy_prop_gal_str}
\begin{tabular}{llllll}
\hline
                        &        & PPV                   & PPP              &     \\
\hline
\multirow{7}{*}{Total}                                      & number  & 971 & 1029 & \\
                      &  M$\mathrm{_{cl}}$    & 3.7 $\times$ 10$^{5}$ & 3.6 $\times$ 10$^{5}$ & [M$_{\sun}$]\\
                      & radius & 14.2                  & 15.9             &  [pc]     \\
                      & $\mathrm{\Sigma_{cl}}$     & 525                   & 525             &  [M$_{\sun}$ pc$^{-2}$]  \\                    
                      & $\sigma$    & 4.2                   & 5.0                & [km s$^{-1}$]   \\
                      &  M$\mathrm{_{vir}}$  & 2.4 $\times$ 10$^{5}$ & 3.9 $\times$ 10$^{5}$  & [M$_{\sun}$]\\                   
                     & $\alpha$  & 1.1                   & 1.1            &       \\

\hline
\multirow{6}{*}{bar}   & number   & 69 & 77 \\
                       & M$\mathrm{_{cl}}$    & 1.8 $\times$ 10$^{5}$ & 2.7 $\times$ 10$^{5}$ & [M$_{\sun}$] \\
                        & radius & 13.9                  & 14.2               &  [pc]   \\
                        & $\mathrm{\Sigma_{cl}}$     & 329                   & 615            &   [M$_{\sun}$ pc$^{-2}$]       \\
                        & $\sigma$   & 3.6                   & 4.6            & [km s$^{-1}$]       \\
                        & M$\mathrm{_{vir}}$   & 2.0 $\times$ 10$^{5}$ & 3.3 $\times$ 10$^{5}$ & [M$_{\sun}$] \\
                        & $\alpha$  & 1.4                   & 1.5            &       \\
\hline
\multirow{6}{*}{spiral}  & number & 471 & 515 \\
                        & M$\mathrm{_{cl}}$    & 5.5 $\times$ 10$^{5}$ & 4.5 $\times$ 10$^{5}$ & [M$_{\sun}$]\\
                        & radius & 14.7                  & 16.3               &  [pc]   \\                        
                        & $\mathrm{\Sigma_{cl}}$     & 799                   & 666         &   [M$_{\sun}$ pc$^{-2}$]          \\
                        & $\sigma$   & 5.5                   & 5.7          & [km s$^{-1}$]         \\                        
                        & M$\mathrm{_{vir}}$   & 3.9 $\times$ 10$^{5}$ & 5.0 $\times$ 10$^{5}$ & [M$_{\sun}$]\\
                        & $\alpha$  & 1.2                   & 1.2         &          \\
\hline
\multirow{6}{*}{disc}   & number & 98 & 102 \\
                        & M$\mathrm{_{cl}}$    & 5.3 $\times$ 10$^{5}$ & 5.4 $\times$ 10$^{5}$ & [M$_{\sun}$]\\
                        & radius & 13.6                  & 15.9             &  [pc]     \\
                        & $\mathrm{\Sigma_{cl}}$     & 708                   & 586             &   [M$_{\sun}$ pc$^{-2}$]      \\                        
                        & $\sigma$   & 5.7                   & 5.4           & [km s$^{-1}$]        \\
                        & M$\mathrm{_{vir}}$   & 4.5 $\times$ 10$^{5}$ & 4.7 $\times$ 10$^{5}$ & [M$_{\sun}$]\\                       
                        & $\alpha$  & 0.9                   & 1.0         &          \\
                        
\hline
\end{tabular}
\end{table}

\subsection{Environmental dependence of cloud comparisons}

We classified clouds based on whether their local environment was in the bar, spiral or outer disc structure of the galaxy. Which region a cloud belongs to is dictated by its physical location, as described in \cite{Fuj14}. {\it Spiral clouds} are defined as those located within the galactocentric radii of $2.5 < r < 7.0$\,kpc. The {\it bar cloud} population live in a central rectangular section of size 5.0\,kpc $\times$ 1.2\,kpc at an angle of 135$^\circ$\,degrees. The outer {\it disc clouds} are those beyond 7.0\,kpc.

The median cloud properties for each galactic environment are shown in bottom three rows of Table~\ref{TAB_phy_prop_gal_str}. In both PPV and PPP, the median mass of clouds in the bar ($\sim$ 2 $\times$ 10$^{5}$ M$_{\sun}$) is about two times lower than those in the spiral and disc. This is due to a population of low density clouds that form in the tails of tidal interactions between clouds in this close-packed region (see \citet{Fuj14} for the full discussion). These small clouds are also less gravitationally bound, raising the average value of $\alpha$ for this area, and have lower velocity dispersions. In the spiral and disc regions, the median cloud radii and velocity dispersion sit around $13 - 16$\,pc and $5 - 6$\,kms$^{-1}$ for both sets of clouds. However, the PPV clouds tend to be smaller than the PPP clouds by around $0.5 - 2.5$\,pc ($\sim$ 3 -- 20 \%). There is also a marked difference in the surface density, with PPV clouds taking higher values, except in the bar region, where the median PPV surface density is significantly lower than the PPP population. 

\begin{figure*}
    \begin{minipage}{0.32\textwidth}
        \centering
		\includegraphics[width=0.9\textwidth]{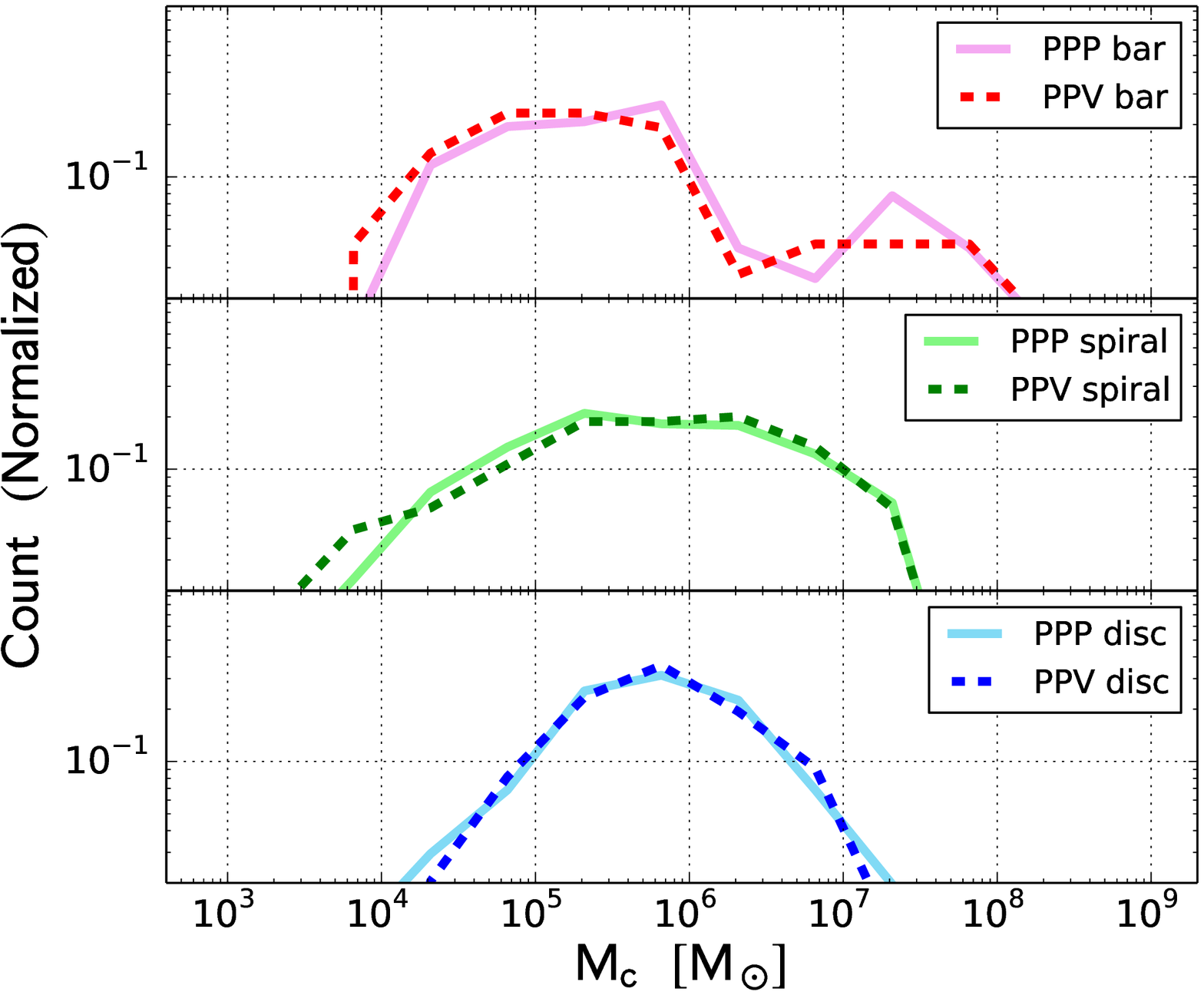}
\begin{center}
 (a)
\end{center}
    \end{minipage}
    \begin{minipage}{0.32\textwidth}
        \centering
		\includegraphics[width=0.9\textwidth]{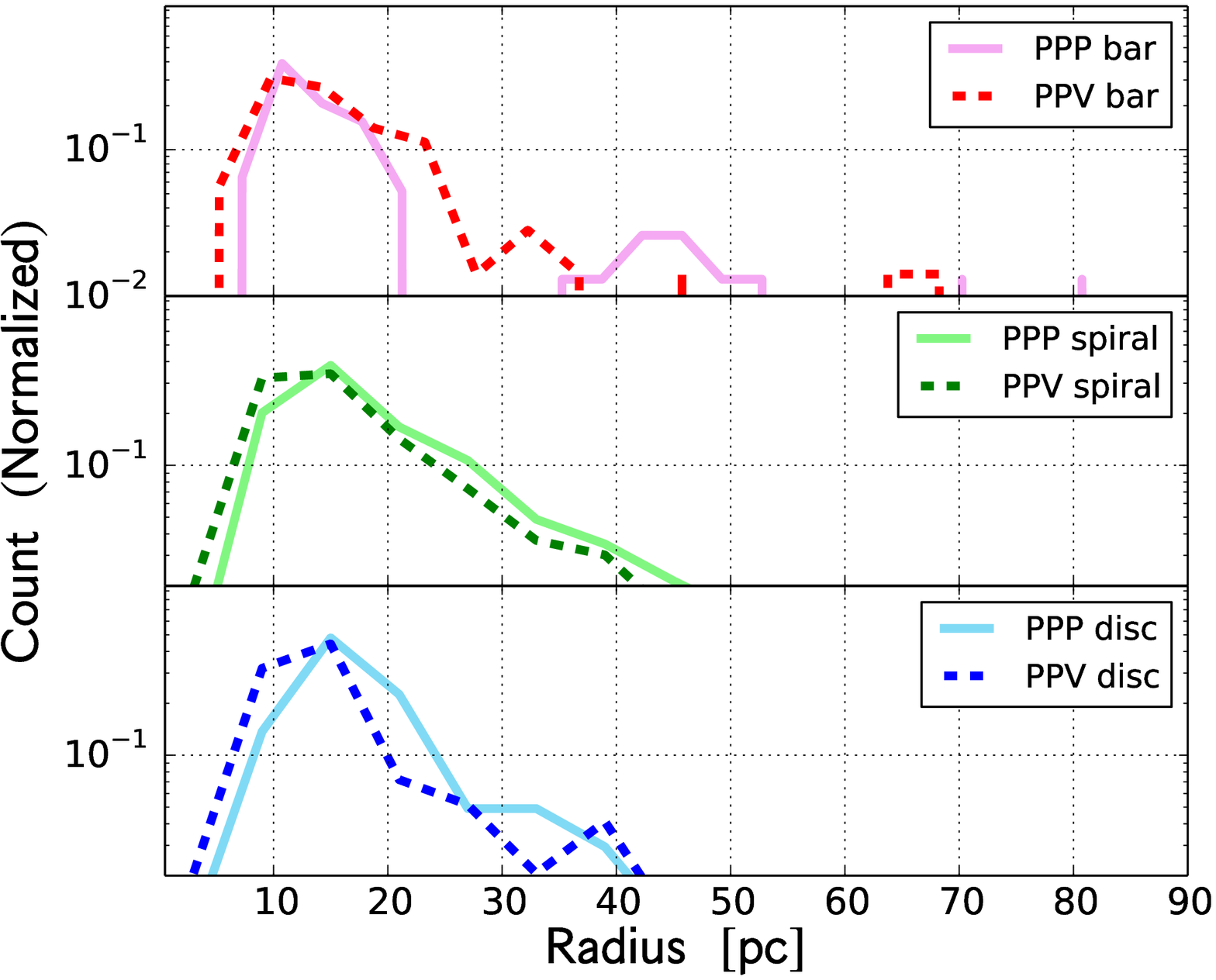}	
\begin{center}
 (b)
\end{center}
    \end{minipage}
    \begin{minipage}{0.32\textwidth}
        \centering
		\includegraphics[width=0.9\textwidth]{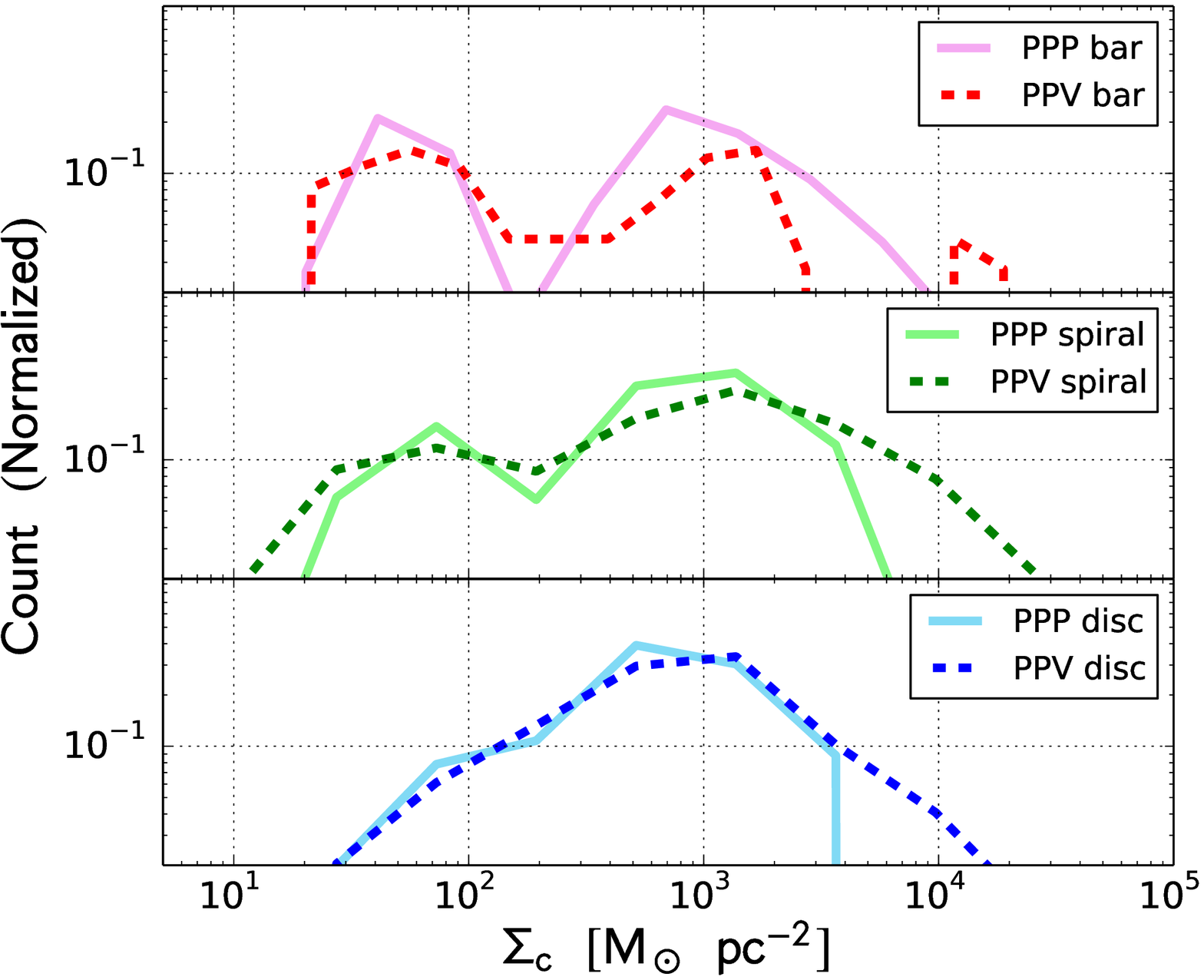}	
\begin{center}
 (c)
\end{center}
    \end{minipage}
    \vspace{15pt}
    \begin{minipage}{0.32\textwidth}
        \centering
		\includegraphics[width=0.9\textwidth]{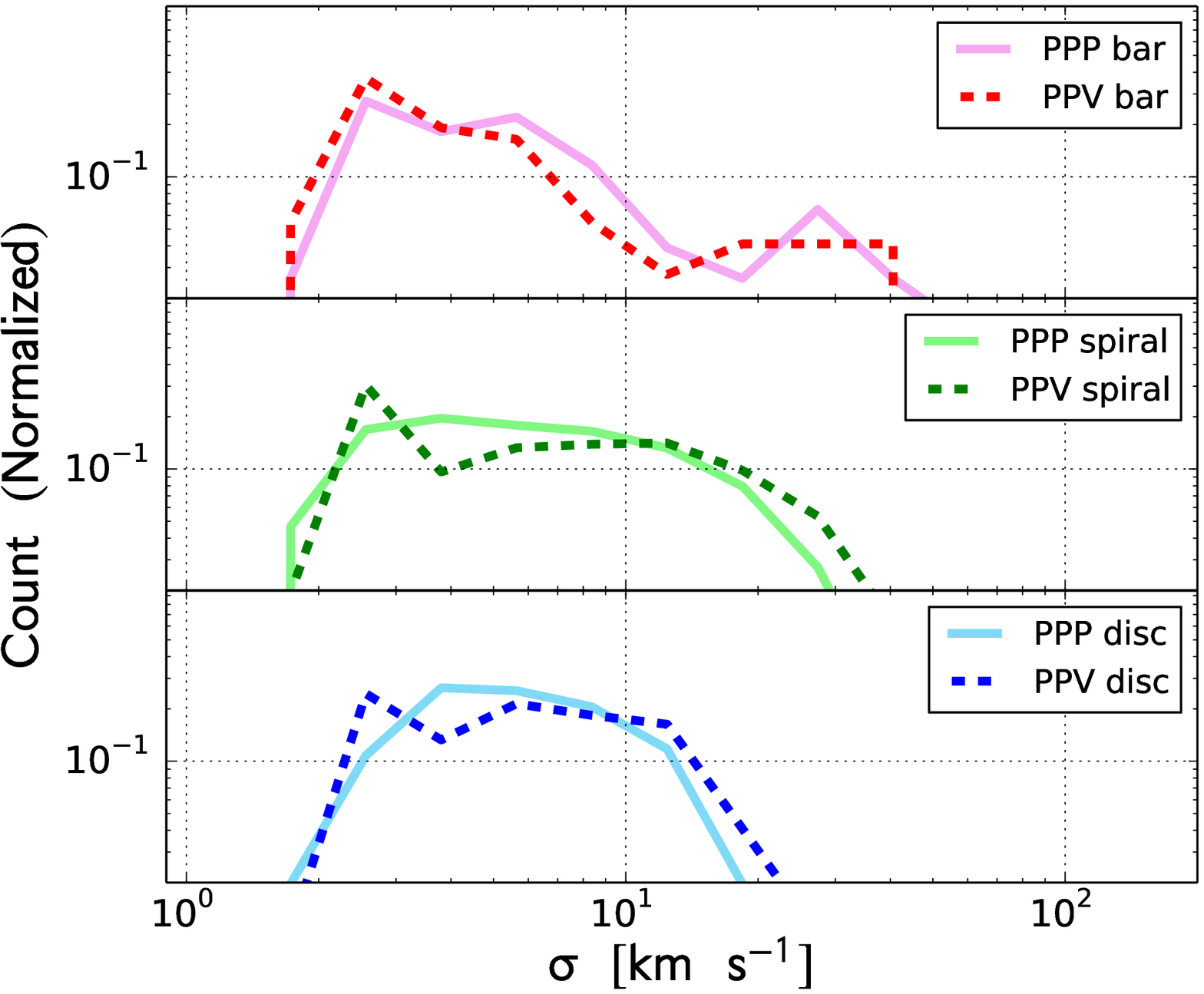}	
\begin{center}
 (d)
\end{center}
    \end{minipage}
        \begin{minipage}{0.32\textwidth}
        \centering
		\includegraphics[width=0.9\textwidth]{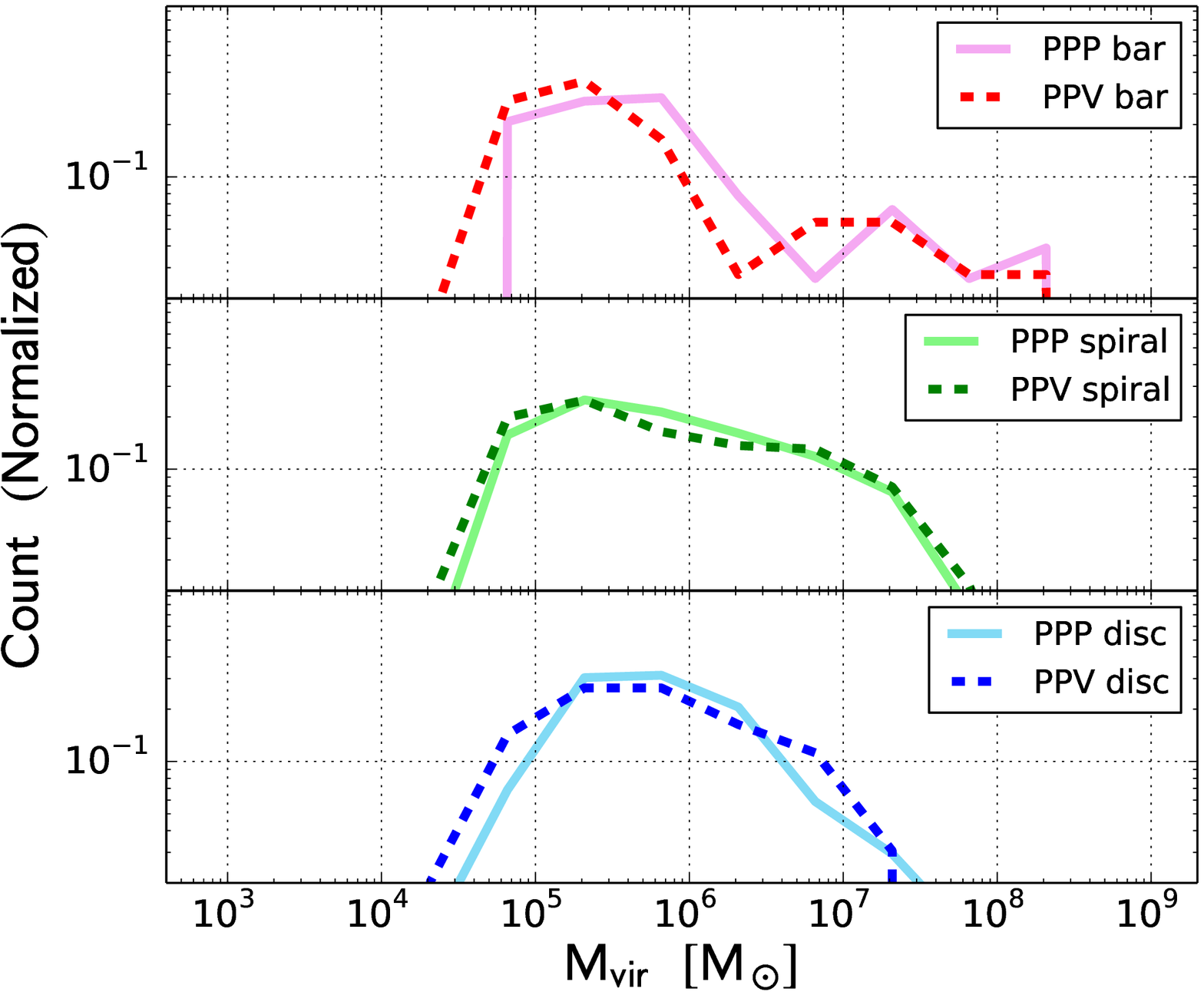}	
\begin{center}
 (e)
\end{center}
    \end{minipage}
    \begin{minipage}{0.32\textwidth}
        \centering
		\includegraphics[width=0.9\textwidth]{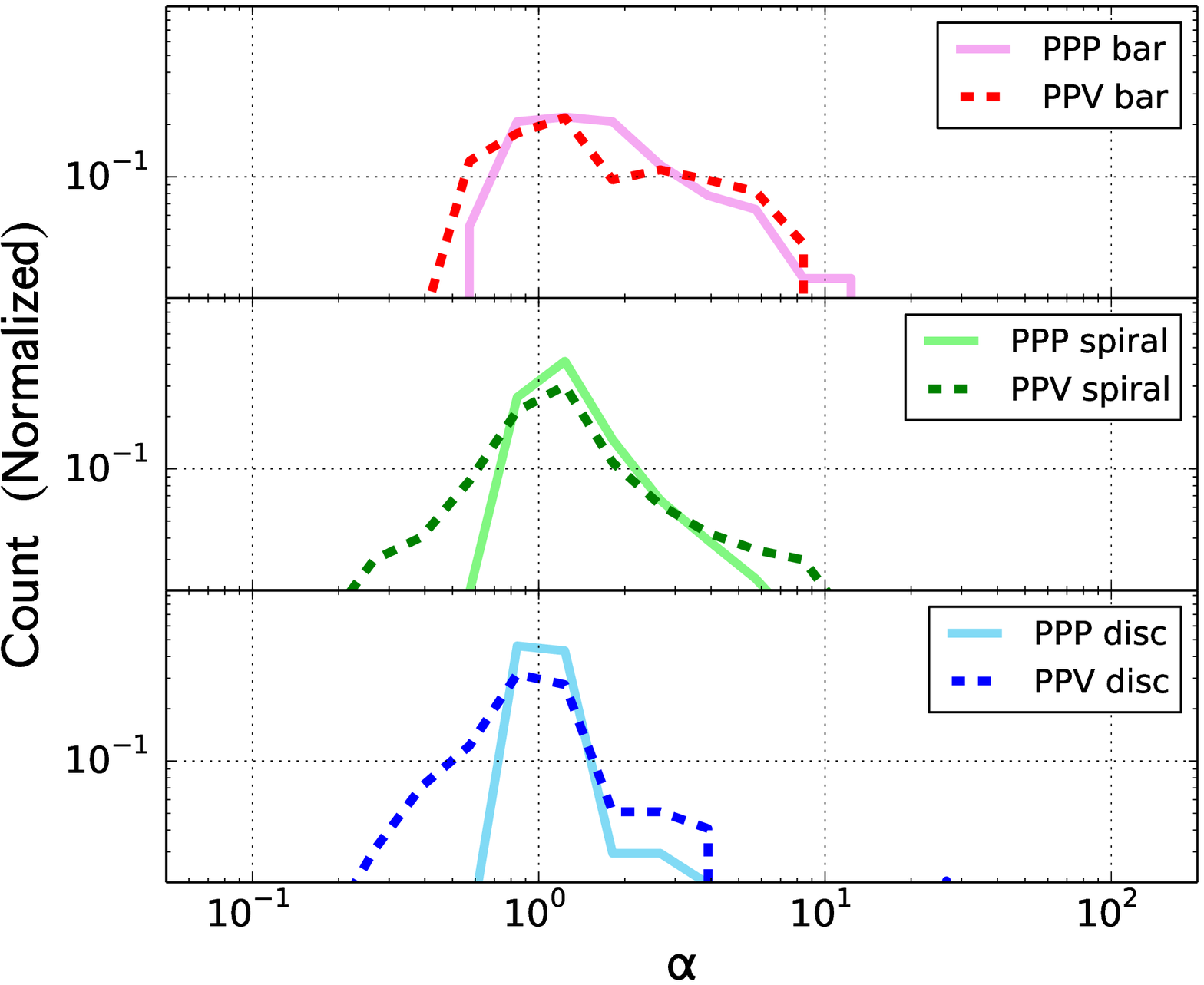}	
\begin{center}
 (f)
\end{center}
    \end{minipage}
	\caption{Normalized distribution of cloud properties in bar region (red), spiral region (green), and disc region (blue). PPV and PPP clouds are shown with dashed and solid lines, respectively. Derivation of cloud properties are summarized in Table \ref{TAB_phy_prop_gal_str} for PPV-clouds and Table \ref{TAB_phy_prop_gal_str} for PPP-clouds. Panels present: (a) molecular cloud mass,  (b) radius, (c) surface density of mass, (d) velocity dispersion, (e) virial mass, and (f) virial parameter.}
	\label{FIG_stucture_prop}
	\end{figure*}

These differences and similarities in the cloud properties can be seen in more detail in the distribution plots in Figure~\ref{FIG_stucture_prop}. Each of the distributions for mass, radius, surface density, velocity dispersion, virial parameter virial mass are divided into bar cloud populations (red), spiral (green) and outer disc (blue), with the dashed line marking the PPV data while the solid line shows the PPP. 
  
The cloud mass distribution shown in Figure~\ref{FIG_stucture_prop}(a) show a high similarity between the two data sets for all three environments. Indications of a bimodal population is seen by a dip in the bar cloud profile, due to a population of very massive merger remnants above $10^7$\,M$_\odot$. Since mergers are far more common in the high density bar area, the number of these massive clouds is much less in the spiral arm and non-existant in the disc, producing a continuous distribution in these environments (more details regarding the formation of the cloud populations can be found in \citealt{Fuj14}). In the lower interaction environment of the outer disc, the cloud mass range has a narrower spread about the median value of $5 \times 10^{5}$\,M$_{\sun}$. These features are produced equally well in both the PPV and PPP data sets, showing such results are independent of the identification technique.

The radii of the clouds also appears similar in the two data sets, as shown in Figure~\ref{FIG_stucture_prop}(b). The largest difference is seen in the bar region, where the PPV profile finds more clouds at both small ($\leq$ 10 pc) and large ($>$ 60 pc) radii. The former of these is due to the use of mass-weighting in the PPV calculation for the cloud radius, compared to the non-weighted measure in the PPP calculation. It is an effect most common in the bar, where the high fraction of long-lived massive clouds are more centrally concentrated than their smaller counterparts. 
At the other end of the scale, the crowded cloud population in the bar combines PPP clouds in the projected PPV data set, creating more extended structures. These differences are less marked in the spiral and disc regions as the clouds are less packed with fewer mergers.

\begin{figure*}
    \begin{minipage}{0.43\textwidth}
        \centering
		\includegraphics[width=0.9\textwidth]{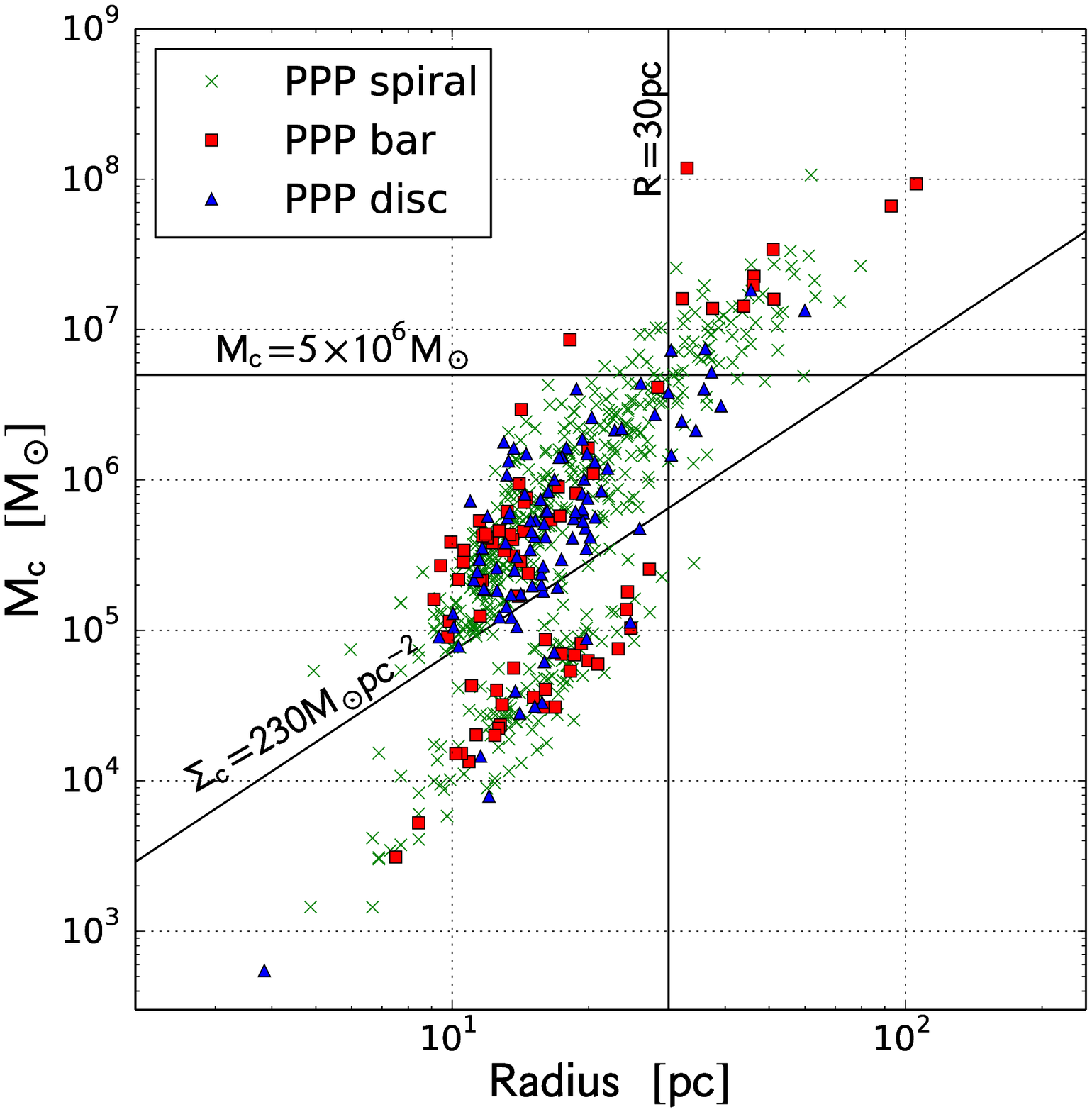}
\begin{center}
 (a)
\end{center}
    \end{minipage}
    \begin{minipage}{0.43\textwidth}
        \centering
		\includegraphics[width=0.9\textwidth]{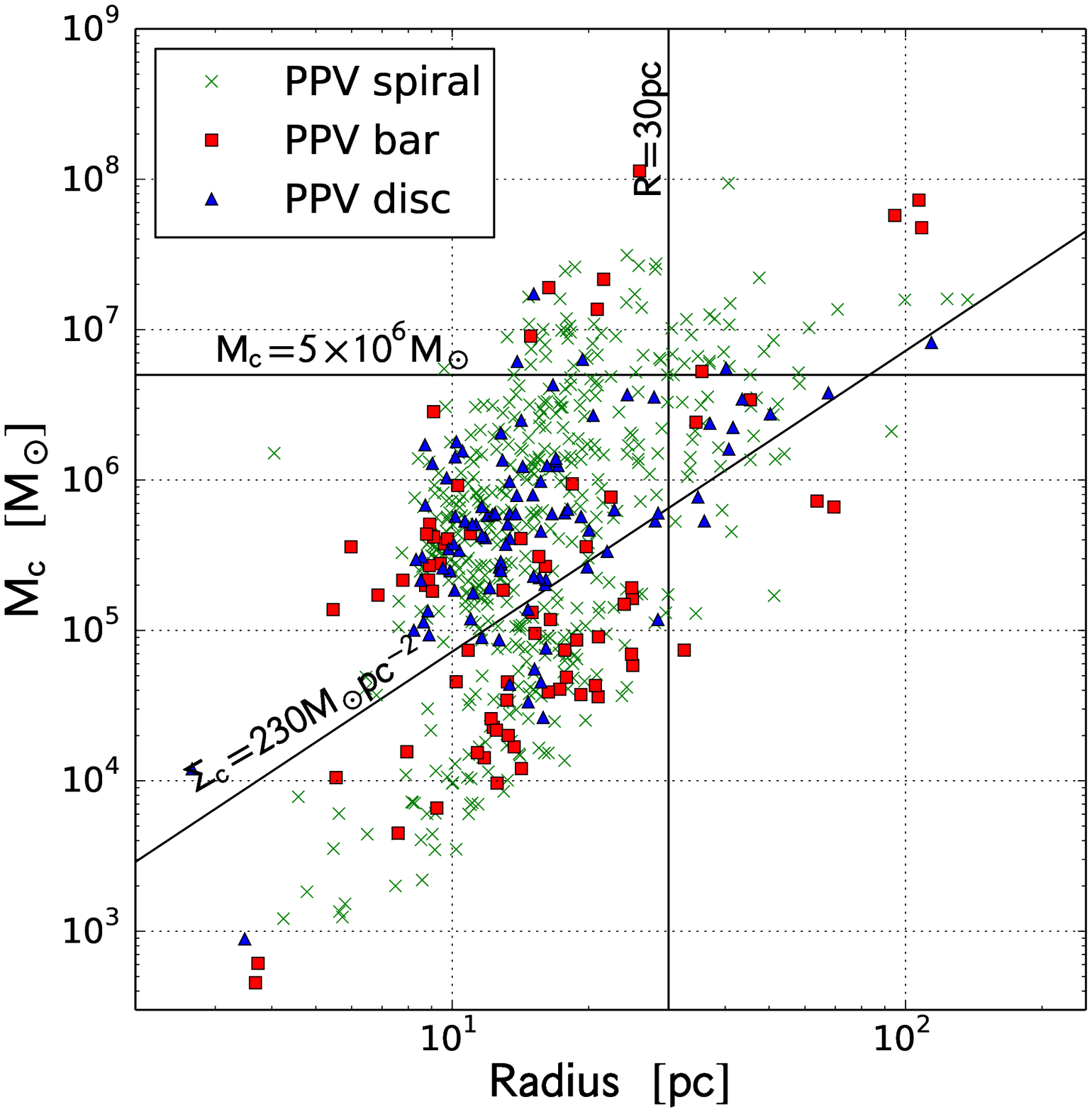}	
\begin{center}
 (b)
\end{center}
    \end{minipage}
    \begin{minipage}{0.43\textwidth}
        \centering
		\includegraphics[width=0.9\textwidth]{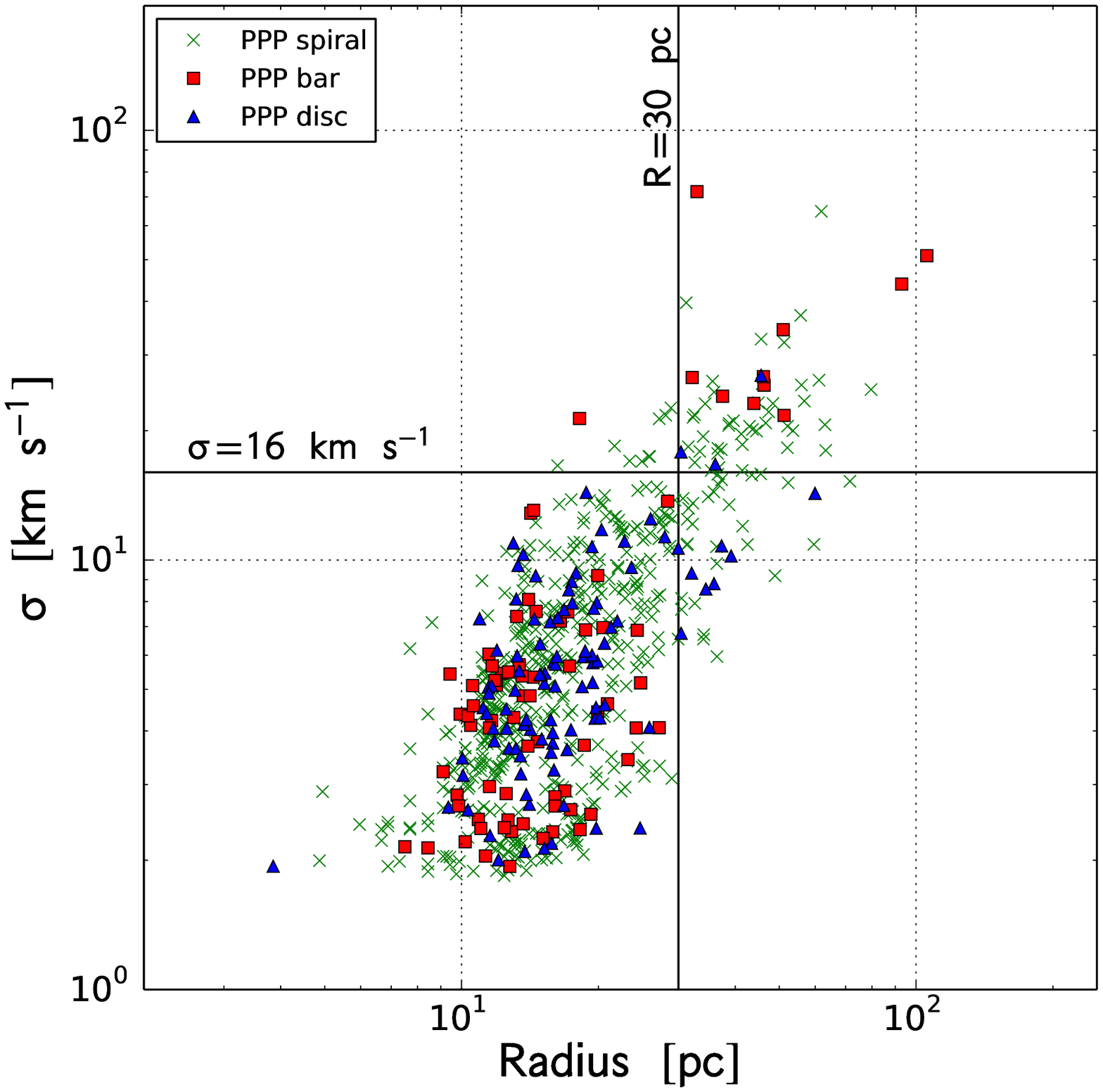}	
\begin{center}
 (c)
\end{center}
    \end{minipage}
    \hspace{35pt}
    \begin{minipage}{0.43\textwidth}
        \centering
		\includegraphics[width=0.9\textwidth]{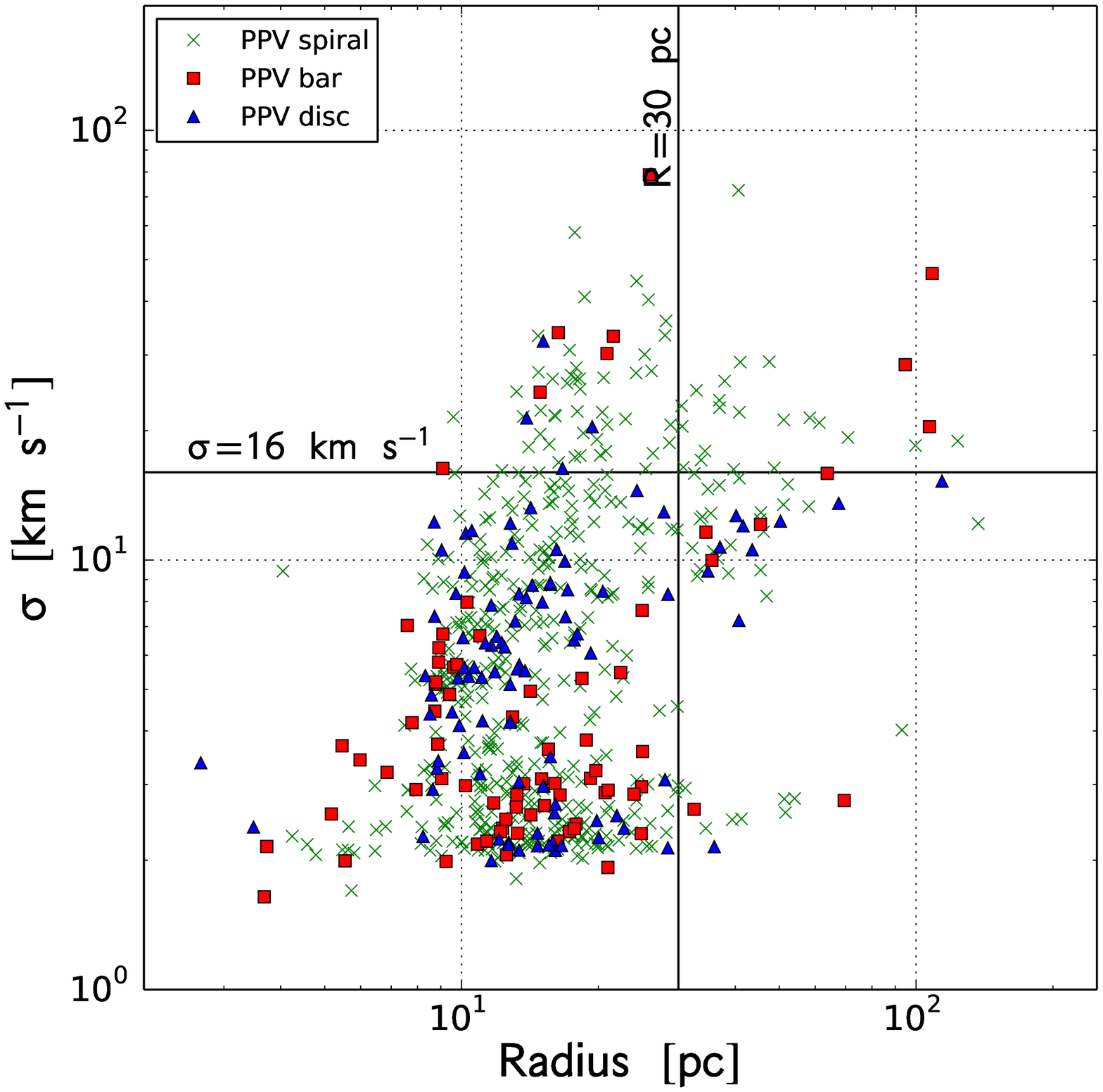}	
\begin{center}
 (d)
\end{center}
    \end{minipage}
	\caption{Scaling relations of the cloud properties. Color symbols denote clouds in different galactic environments: green crosses are spiral clouds, red squares are bar clouds, and blue triangles show disc clouds. Panel (a) and (b) show the relation of cloud mass versus radius for PPP and PPV clouds, respectively. Panel (c) and (d) present the results of velocity dispersion against radius for PPP and PPV clouds, respectively.}
	\label{FIG_PP2_scaling}
\end{figure*}

In the median values of Table~\ref{TAB_phy_prop_gal_str}, the surface density had shown the largest difference between the PPP and PPV populations. The reason for this becomes clear in Figure~\ref{FIG_stucture_prop}(c): while the profile values strongly overlap, the shape has broad peaks which do not perfectly align. These peaks are due to the presence of a bimodal split in the surface density which is more marked in the PPP data than PPV. As described in \citet{Fuj14}, the bimodal surface density cloud populations stem from the production of a transient cloud population that form during tidal interactions. This gives a split around 230\,M$_{\sun}$ pc$^{-2}$, above which sit the majority of the cloud population and below which lie low density, unbound objects formed in the filaments of tidal tails. The two populations are most strongly visible in the bar, where the high number of interactions produces the largest fraction of transient clouds. This is seen more clearly in the scaling relations for GMCs first noted by Larson \citep{Lar81} and plotted in Figure~\ref{FIG_PP2_scaling}(a) to \ref{FIG_PP2_scaling}(d). In the relation between GMC mass and radius in Figure~\ref{FIG_PP2_scaling}(a) and Figure~\ref{FIG_PP2_scaling}(b), two clear trends can be seen, corresponding to the peaks in the surface density. However, the split is far sharper in the (left-hand) PPP data, due to a smaller scatter in the cloud radii values. The previously discussed projection and mass-weighting effects smooth the bimodality in the surface density, leading to off-set peak values in Table~\ref{TAB_phy_prop_gal_str}, a more extended profile in Figure~\ref{FIG_stucture_prop}(c) and a higher level of scatter in the Larson relations.

The velocity dispersion for the clouds is shown in Figure~\ref{FIG_stucture_prop}(d). In both the PPP and PPV data sets, the highest velocity dispersion is found in the bar, due to the high number of cloud-cloud encounters.
The Larson relation between velocity dispersion and cloud radius is shown in Figure~\ref{FIG_PP2_scaling}(c) and Figure~\ref{FIG_PP2_scaling}(d), with the PPP data plotted in the left-hand panel. The dominant trend extends in both cases through about an order of magnitude in $\sigma$ and $R$, ranging from $10 < R < 100$\,pc and $3 < \sigma < 80$\,kms$^{-1}$. As with the mass - radius relation, the scatter in the PPV radius makes the Figure~\ref{FIG_PP2_scaling}(d) considerably more noisy. The two sequences from the transient cloud population are visible in both data sets, with the minor sequence concentrated at low-$\sigma$ ($\sigma$ $<$ 5 km s$^{-1}$), but the PPV-clouds span a wider range in radius from $\sim 10 - 30$\,pc, compared to the $10 - 20$\,pc range of the PPP clouds, in keeping with what is seen in the profile in Figure~\ref{FIG_stucture_prop}(d). At the low velocity dispersion limit for the PPV clouds, an edge is seen at $\sim$ 2 kms$^{-1}$. This corresponds to two velocity channels; the minimal channel width for which CPROPS will accept a continuous structure as a cloud.

The virial mass in Figure~\ref{FIG_stucture_prop}(e) show a close match between the PPP and PPV data sets. However, it is worth noting that this value depends on $M_{\mathrm{c}}$, $R$, and $\sigma$ and therefore combines all the dissimilarity mentioned in this section for these preceeding properties. The associated variable, $\alpha$, in Figure~\ref{FIG_stucture_prop}(f), shows a larger different between the data sets, with the PPV clouds spread across a wider range of values. The median point agrees in all environments, sitting at approximately $\alpha \sim 1$, making the majority of clouds borderline gravitationally bound. The extended range in the PPV data is due to the combined scatter in the cloud properties, especially the radius and velocity dispersion distribution. In the least crowded region of the outer disc, the difference between PPP and PPV is dominated by the PPV mass-weighted radius that acts to decrease $\alpha$, while in the spiral and bar region, the PPP clouds are additionally more often split and combined due to projection effects.

\subsection{Individual cloud comparisons}
\label{SEC_match_cloud}

Moving on from comparing the distribution in cloud properties between the two different techniques, we now directly compare individual clouds that have been identified in both PPP and PPV space. This explores not just if the overall population statistics are equivalent in both methods, but if the properties of individual clouds are also conserved. To do this, we first pair clouds found in both data sets and then directly compare their properties.

\subsubsection{Cloud matching method}

Clouds are matched between the PPP and PPV data sets in an intuitive manner based on their position within the galaxy disc and their radii. However, this process is not entirely straight forward, since there are many permutations which find clouds approximately in the same location. Which of these are considered a match is shown visually in Figure~\ref{FIG_cloud_match_show}, where the grey and black circles each represent a cloud found in the two data sets. Note that since the PPP data set uses a three-dimensional position in ($x$, $y$, $z$) while the PPV data has only the two-dimensional ($x$, $y$) co-ordinates, we have to consider the projected position and radius of the PPP clouds in the $x - y$ plane only. 

The cases depicted by the top three panels are what are accepted as matches. Figure~\ref{FIG_cloud_match_show}(a) is the most straight forward case, where two clouds have a common centre-of-mass. Panels (b) and (c) stretch this criterion to also accept the match if the centre-to-centre distance between the two clouds is smaller than either both (panel (b)) or one (panel (c)) of their radii. The lower row of three diagrams, Figure~\ref{FIG_cloud_match_show}(d) - (f), show cases which do not qualify as a match. Panel (d) is where there is a partial overlap between the two clouds, but their centre-to-centre distance exceeds both their radii. In panel (e), the clouds are relatively close, but have no overlap and in (f) there is a degeneracy where two possible clouds could be matched to a single object in the other data set. 

This is intentionally a conservative set of criteria for matching the clouds. The purpose is to compare the properties when the same objects have been identified in both data sets. We therefore selected a cloud subset with minimal ambiguity. 

\subsubsection{Results of the cloud matching}

\begin{figure}	
\includegraphics[width=0.5\textwidth]{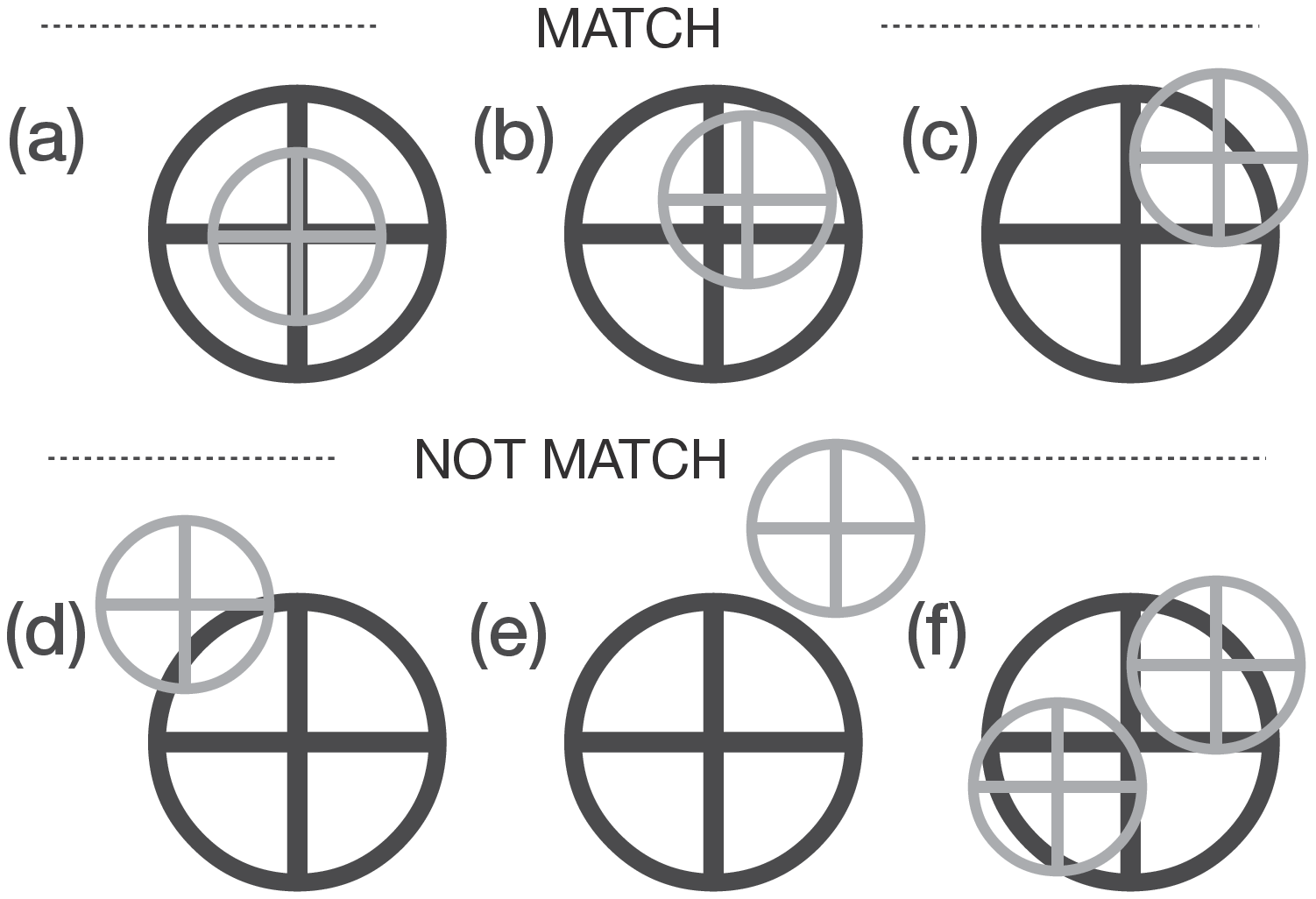} 
\caption{Scheme of the match clouds. Clouds of PPV and PPP are match if their center-to-center distance are smaller than either one or both cloud radii. Grey and black circles denote clouds of two data set. Center of cloud is marked with a cross. Panel (a) -- (c) show the situations of match clouds while panel (d) -- (e) show non-match clouds. (a) Clouds from PPP and PPV overlap at the center. (b) Distance of two clouds is smaller than both radii. (c) Distance of two clouds is smaller than one of radii. (d) Distance of two clouds is larger than both radii. (e) There is no counterpart found in the radius. (f) Multiple clouds share one counterpart (or one cloud found multiple counterparts in its radius).}
\label{FIG_cloud_match_show}
\end{figure}

Despite the stringency of our method, 70\% of the clouds were successfully matched to a single counterpart. Clouds that fell into cases depicted by Figure~\ref{FIG_cloud_match_show}(d) - (f) consisted of 30\% of the cases.

This 70\% match holds for each of the three considered environments. Within the cases that failed to find a matching cloud, the bar region showed more instances of Figure~\ref{FIG_cloud_match_show}(f) with around 20\% finding multiple matches due to the projection effects in the more crowded environment, compared to 17\% in the spiral and disc.

The high match rate is aided by the thin galactic disc which reduces the projection effect. The gas scale height of our simulated galaxy is about 80 pc -- 115 pc, which is similar to the initial value of 100\,pc due to the lack stellar feedback to inject energy \citep{Fuj14}. Nonetheless, recent observations of edge-on (barred-) spiral galaxies show that the scaleheights of molecular gas traced by $^{12}$CO (1--0) are indeed $\leqslant$ 200\,pc and mainly $\leqslant$ 150 pc \citep{Yim14}, so this effect benefits true observations as well.

In a non flat disc environment, \cite{War12} compared simulated and synthetically observed cores inside a GMC and reported a slightly higher probability of 81\% that the PPV and PPP clouds were drawn from the same distribution. This improvement --despite the potentially more challenging geometry-- is due to the high density of the cores, which makes them more compact with a single peak and low sub-structure. By contrast, GMCs are typically non-spherical, with multiple peaks and clad in irregular low-density envelopes. The varied morphology and mass distribution of the GMCs increases the scatter in the identified cloud boundary and properties between the methods, lowering the match rate.

\begin{figure*}
    \begin{minipage}{0.28\textwidth}
        \centering
		\includegraphics[width=0.9\textwidth]{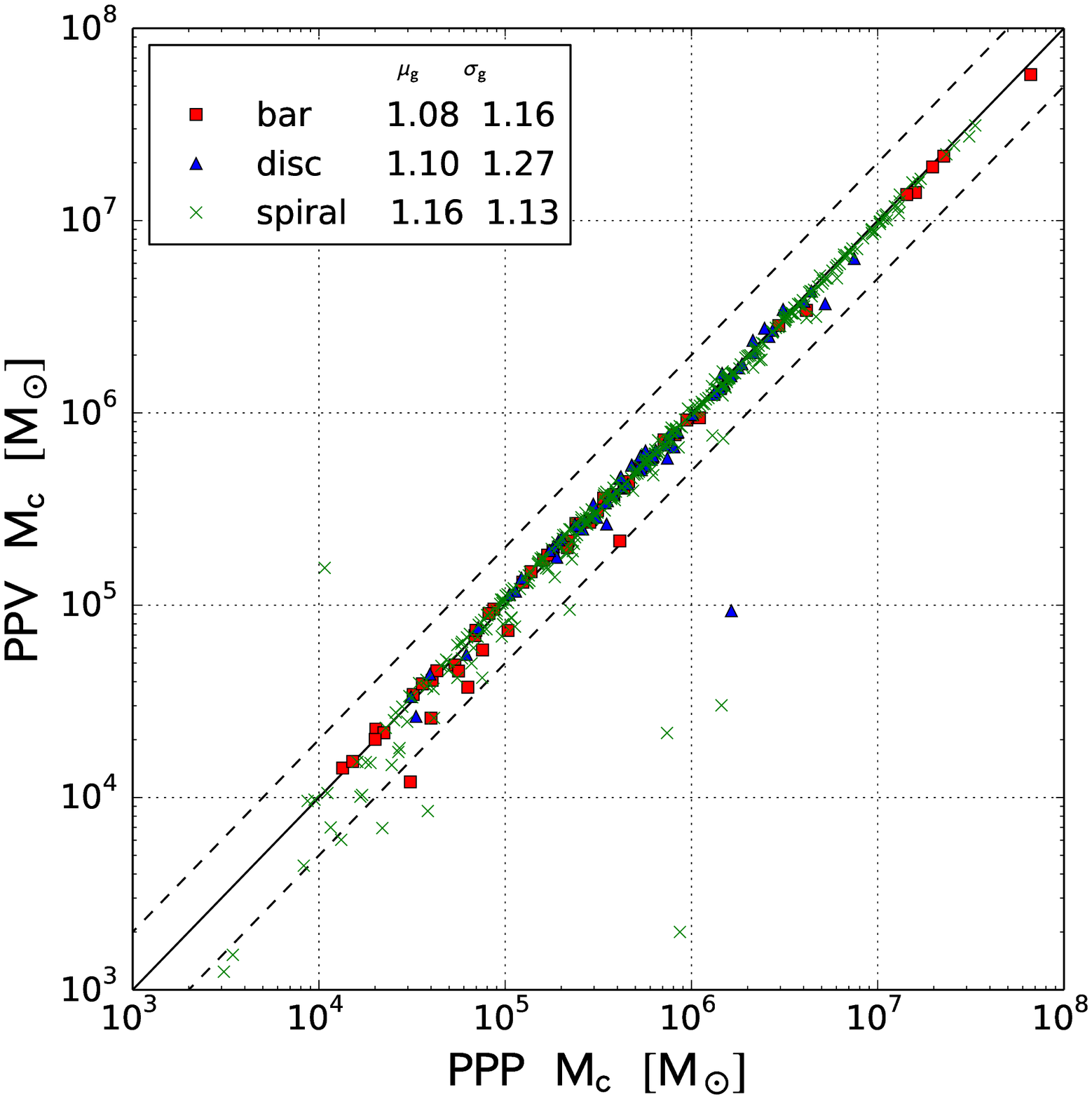}
\begin{center}
 (a)
\end{center}
    \end{minipage}
    \begin{minipage}{0.28\textwidth}
        \centering
		\includegraphics[width=0.9\textwidth]{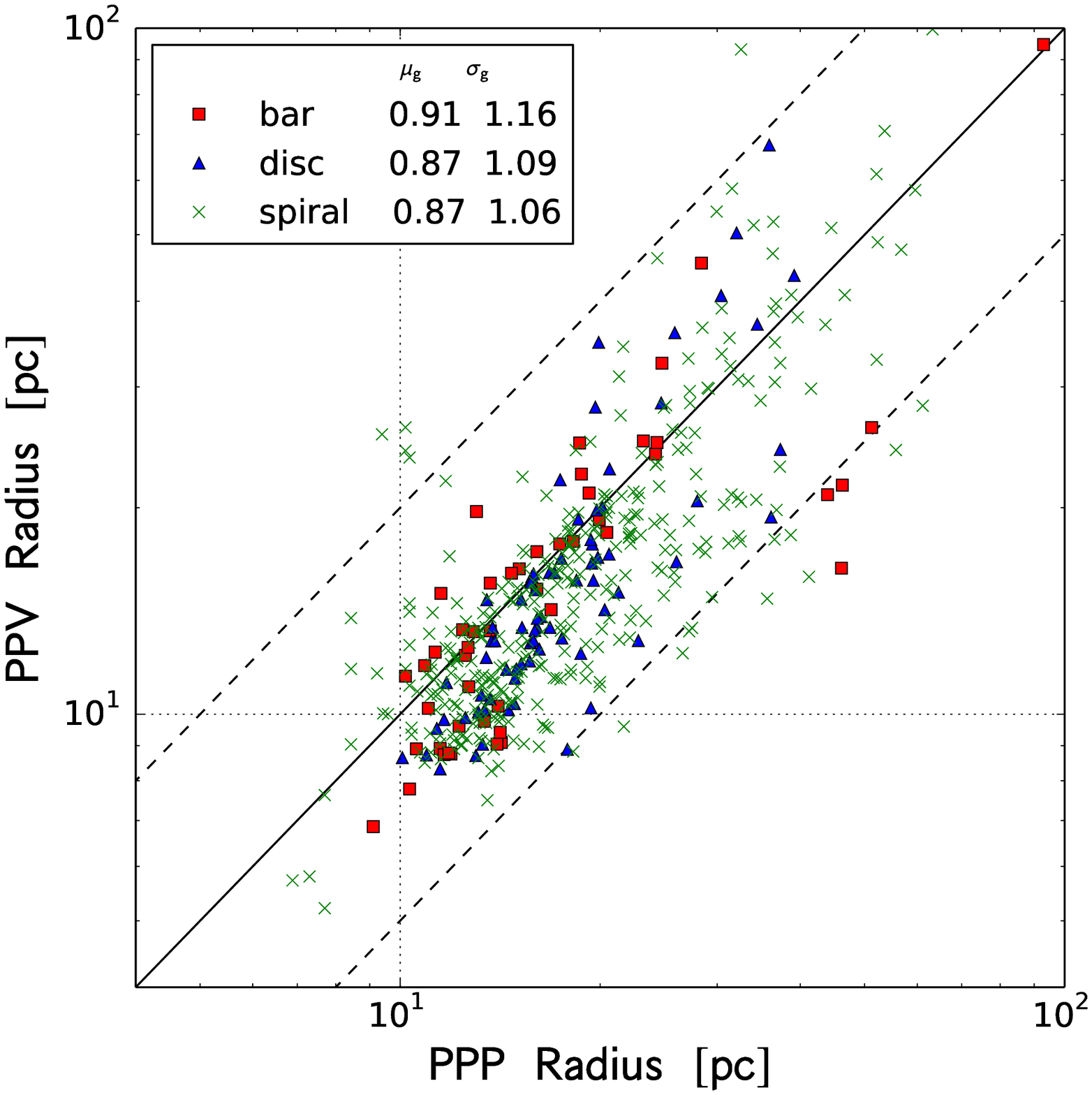}	
\begin{center}
 (b)
\end{center}
    \end{minipage}
    \begin{minipage}{0.28\textwidth}
        \centering
		\includegraphics[width=0.9\textwidth]{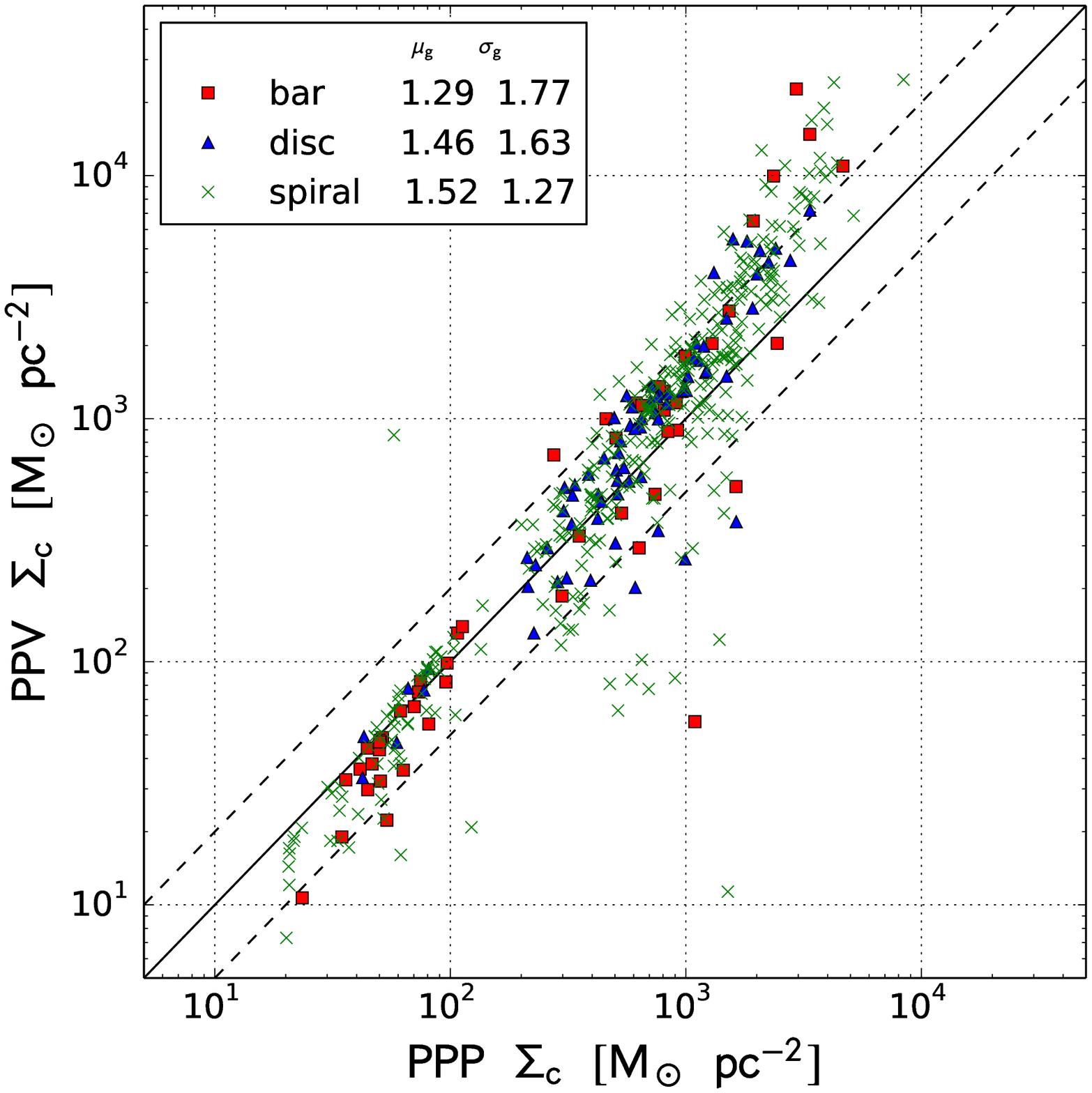}	
\begin{center}
 (c)
\end{center}
    \end{minipage}
            \begin{minipage}{0.28\textwidth}
        \centering
		\includegraphics[width=0.9\textwidth]{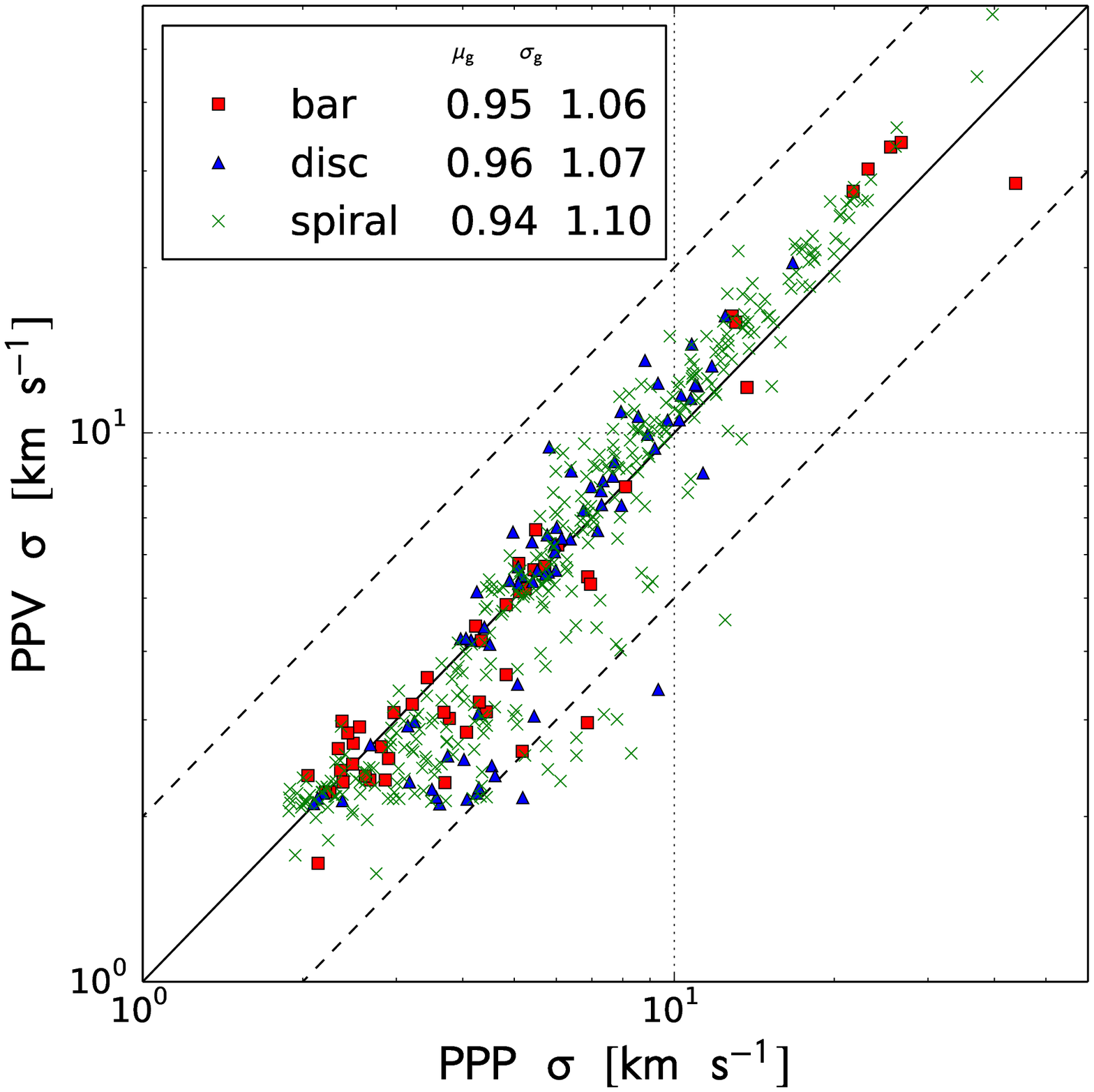}	
\begin{center}
 (d)
\end{center}
    \end{minipage}
    \begin{minipage}{0.28\textwidth}
        \centering
		\includegraphics[width=0.9\textwidth]{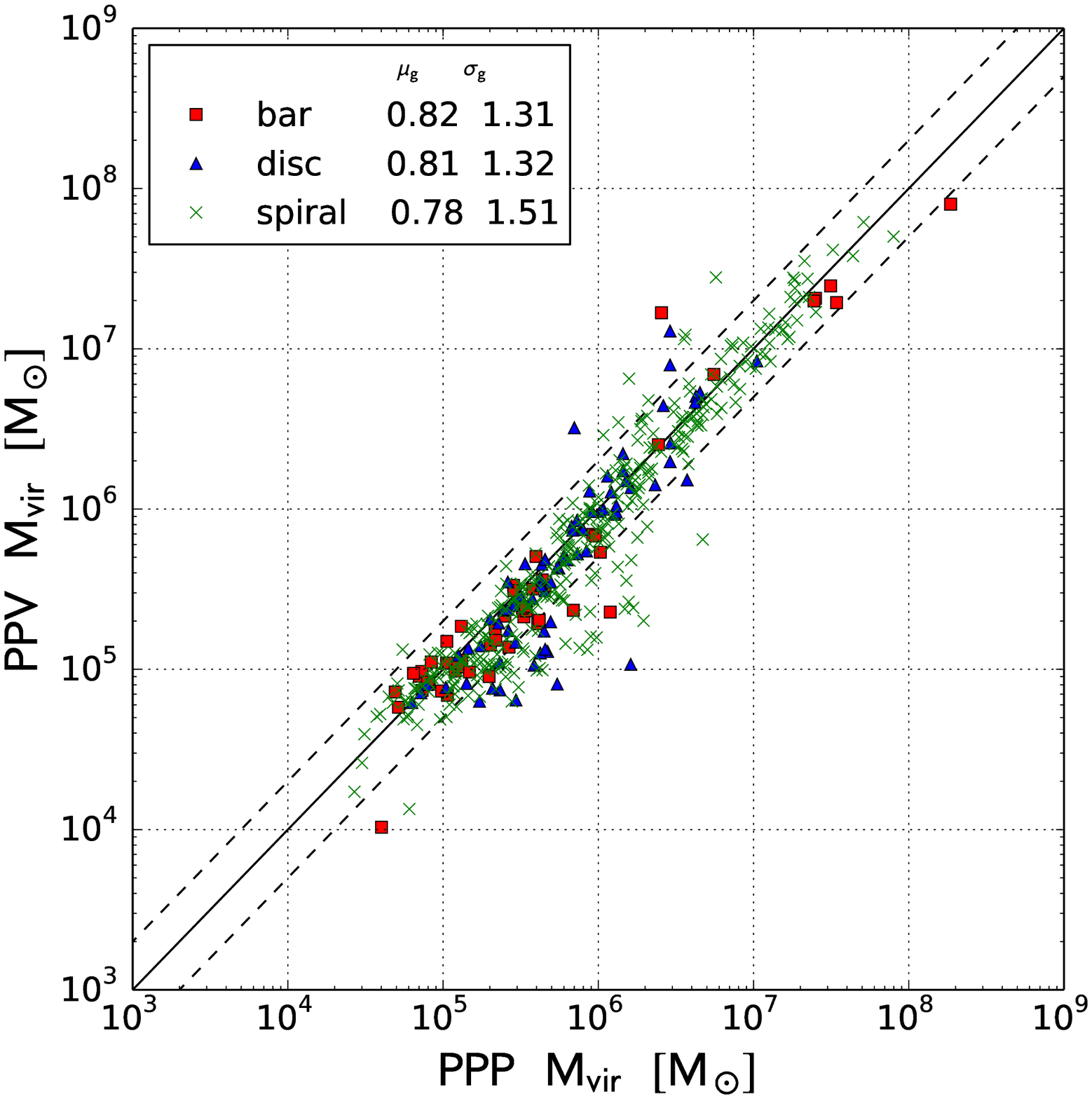}	
\begin{center}
 (e)
\end{center}
    \end{minipage}
    \begin{minipage}{0.28\textwidth}
        \centering
		\includegraphics[width=0.9\textwidth]{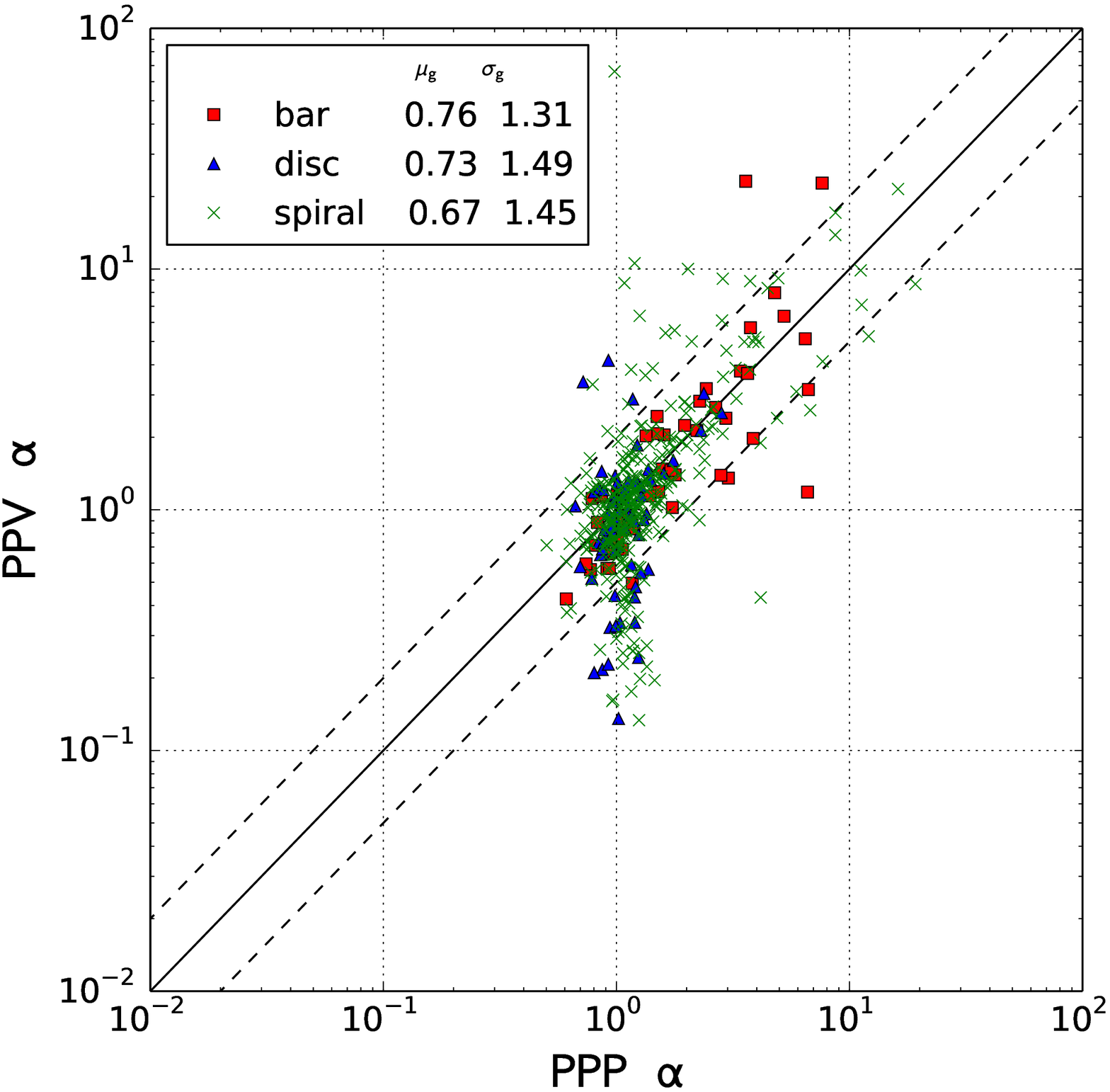}	
\begin{center}
 (f)
\end{center}
    \end{minipage}
	\caption{Comparison of match cloud properties. Color markers are the same as in Figure \ref{FIG_PP2_scaling}. Panel (a) -- (f) show the comparison between cloud mass, radius, surface density of mass, velocity dispersion, virial mass and virial parameter, respectively. Solid line indicate 1 : 1. Dashed lines indicate a factor of two above and below the solid line. Geometric mean ($\mu_{\mathrm{g}}$) and geometric standard deviation ($\sigma_{\mathrm{g}}$) of the ratio of PPV to PPP properties are provided for galactic environments in each panel.}
	\label{FIG_Match_prop}
\end{figure*}

How well the properties of the matched clouds compare is shown in Figure~\ref{FIG_Match_prop}. The PPV value is plotted against the PPP value for the same six properties in Figure~\ref{FIG_Match_prop} for the 70\% clouds that are matched between the data sets. The clouds found in the bar region are marked with red squares, those in the spiral are green crosses and the outer disc clouds are shown as blue triangles. Across the plots are diagonal long-dashed lines that denote a factor of two above and below the solid perfectly matched 1:1 ratio line. In the legend of each plot, the geometric mean ($\mu_{\mathrm{g}}$) and geometric standard deviation ($\sigma_{\mathrm{g}}$) of the ratio between the PPV and PPP are provided for each galactic environment. $\mu_{\mathrm{g}}$ measures the overall match between the PPP and PPV values, while $\sigma_{\mathrm{g}}$ describes the scatter about $\mu_{\mathrm{g}}$.

In all properties, the average ratio between the PPP and PPV values are around 1.0. This strongly implies that the simulation and observational techniques for defining clouds do identify the same objects and estimate similar properties. The tightest correlation between the PPP and PPV values are seen in the cloud mass, with average ratios at $\mu_{\mathrm{g}}\approx 1.0$ (Figure \ref{FIG_Match_prop}(a)). Values of $\mu_{\mathrm{g}}$ show that the PPP clouds are generally more massive than than PPV clouds by $\sim$ 10 \% due to the image of projection. This is considerably lower than the uncertainty of the adopted CO-to-H$_{2}$ conversion factor in real observations \citep[see][]{Bol13} and thus is negligible.  The small number of outliers with a PPV mass smaller than the PPP value are due to the PPP technique including the low density cloud envelope that is missed in the PPV data. In the reverse situation, the single cloud that has a significantly higher PPV mass than PPP is due to two separate PPP clouds being combined. Our 1:1 mass relation shows a tighter correlation than that for the cloud cores in \citet{War12}, even though the cores were found to have a higher probability of locating a match in both data sets. This difference is due to the larger projection effect for the small cores, compared to the GMCs. Unsurprisingly, after the distributions in Figure~\ref{FIG_Match_prop}(a), the scatter is much larger in the radius and velocity dispersion. As we have previously seen, the majority of clouds have a smaller radii by $\sim$ 10\% when selected in the PPV data set than in the PPP due to the mass-weighting of the PPV radius (Figure~\ref{FIG_Match_prop}(b)). A number of the worst cases for this --which lie a factor of two below the ratio of 1.0-- lie in the bar region, which contains the highest fraction of large, merger remnant clouds that tend to have high density centres, giving this region the highest range in values. The effect of the mass-weighted radius would be reduced in real observation since GMCs form stars when the density is sufficiently high and consume the dense gas. On the other hand, the flat floor of velocity dispersion seen in Figure \ref{FIG_Match_prop}(d) due to the unresolved velocity dispersion of small PPV clouds will be improved as well if the resolution of velocity is chosen to be less than our 1 km s$^{-1}$. Such a fine resolution is easily achieved for ALMA.

The two trends in surface density are clearly seen in Figure~\ref{FIG_Match_prop}(c), creating a void of clouds at 200\,M$_{\sun}$\,pc$^{-2}$ in keeping with the profile in Figure~\ref{FIG_Match_prop}(c). The match between the data sets is weakest for the most dense clouds, since these have the most compact centres and are therefore the most sensitive to the mass-weighted radius calculation. This causes the trend in Figure~\ref{FIG_Match_prop}(c) to bend upwards as we move towards the right of the plot; a feature emphasised in the surface density since it uses the square of the radius. The minor trend in the bottom left for the transient cloud population is match well between PPP and PPV.

While the surface density has the poorest match at high values, the velocity dispersion, virial mass and virial parameter suffer more at low values. This is due to the sensitivity of PPV to the line-of-sight direction. While PPP averages over all three spatial dimensions to get the velocity dispersion value, PPV can only use data perpendicular to the disc plane. Clouds that are flattened in this direction therefore hit the PPV resolution limit of two velocity elements, giving a dispersion of 2\,km\,s$^{-1}$. This creates the flat line at the low velocity dispersion end of Figure~\ref{FIG_Match_prop}(d) and affects both the virial mass and virial parameter. At higher values, the match between the PPP and PPV data sets improves, although the virial parameter continues to show the most scatter. Notably, this makes it difficult to tell if a cloud is gravitationally bound, since a spread of a factor of two can turn a bound cloud into an unbound object. 

The  projection effect induced uncertainty in the virial parameter is also suggested by~\cite{Bea13}, who found a factor of two uncertainty in  the virial parameter using their simulated and the synthetic observed clumps. The clumps are considerably smaller than our clouds, having masses and radii of $<$ 10$^{4}$ M$_{\sun}$ and $\ll$ 10 pc.  Our results show that even when the cloud properties are averaged on a larger scale at $>$ 10 pc, this uncertainty still exists. Therefore interpretation based on the observed virial parameter must consider this unignorable effect.

\subsection{Cloud classification based on cloud properties}
\label{sec_cloud_classif_prop}
\subsubsection{Properties of Three Types Clouds}

One of the most exciting results found by \cite{Fuj14} was the identification of three different cloud types in their PPP data. These types consisted of the most common `{\emph Type A}' clouds, with properties that corresponded to the average values measured in observations, the `{\emph Type B}' massive cloud associations that formed during repeated mergers, and the transient `{\emph Type C}' that were born in tidal tails and filaments. The ratio of these three types in a given region depended on the rate of interactions, with the high collision and close encounter rate in the bar resulting in a higher number of {\emph Type B} and {\emph C} clouds. Whether this can be potentially detected in observational results is of key importance, since it would allow a concrete handle on how important cloud interactions are in governing cloud properties and ultimately, star formation.

\begin{figure*}
    \begin{minipage}{0.28\textwidth}
        \centering
		\includegraphics[width=1\textwidth]{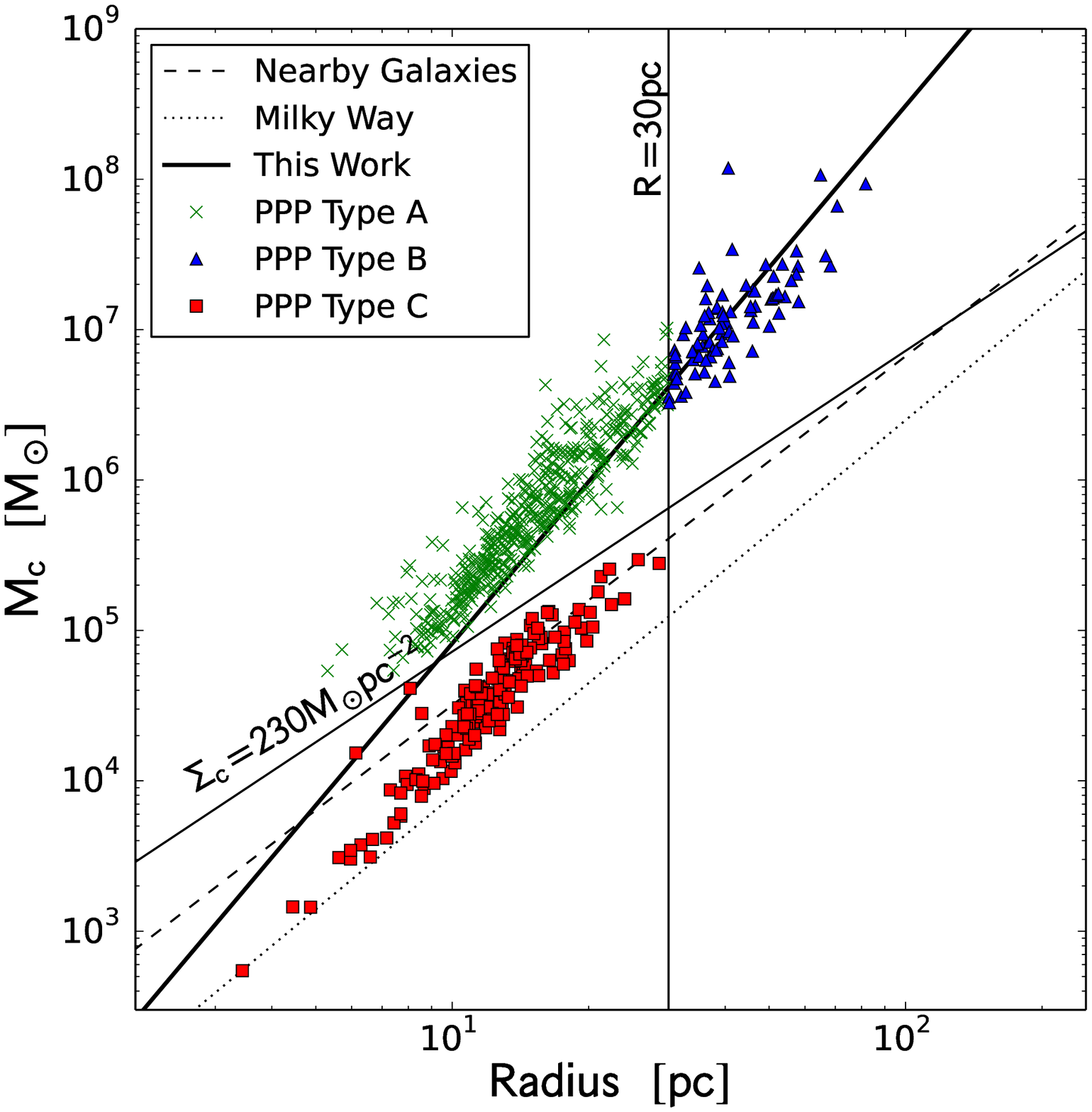} 
\begin{center}
 (a)
\end{center}
    \end{minipage}
    \begin{minipage}{0.28\textwidth}
        \centering
		\includegraphics[width=1\textwidth]{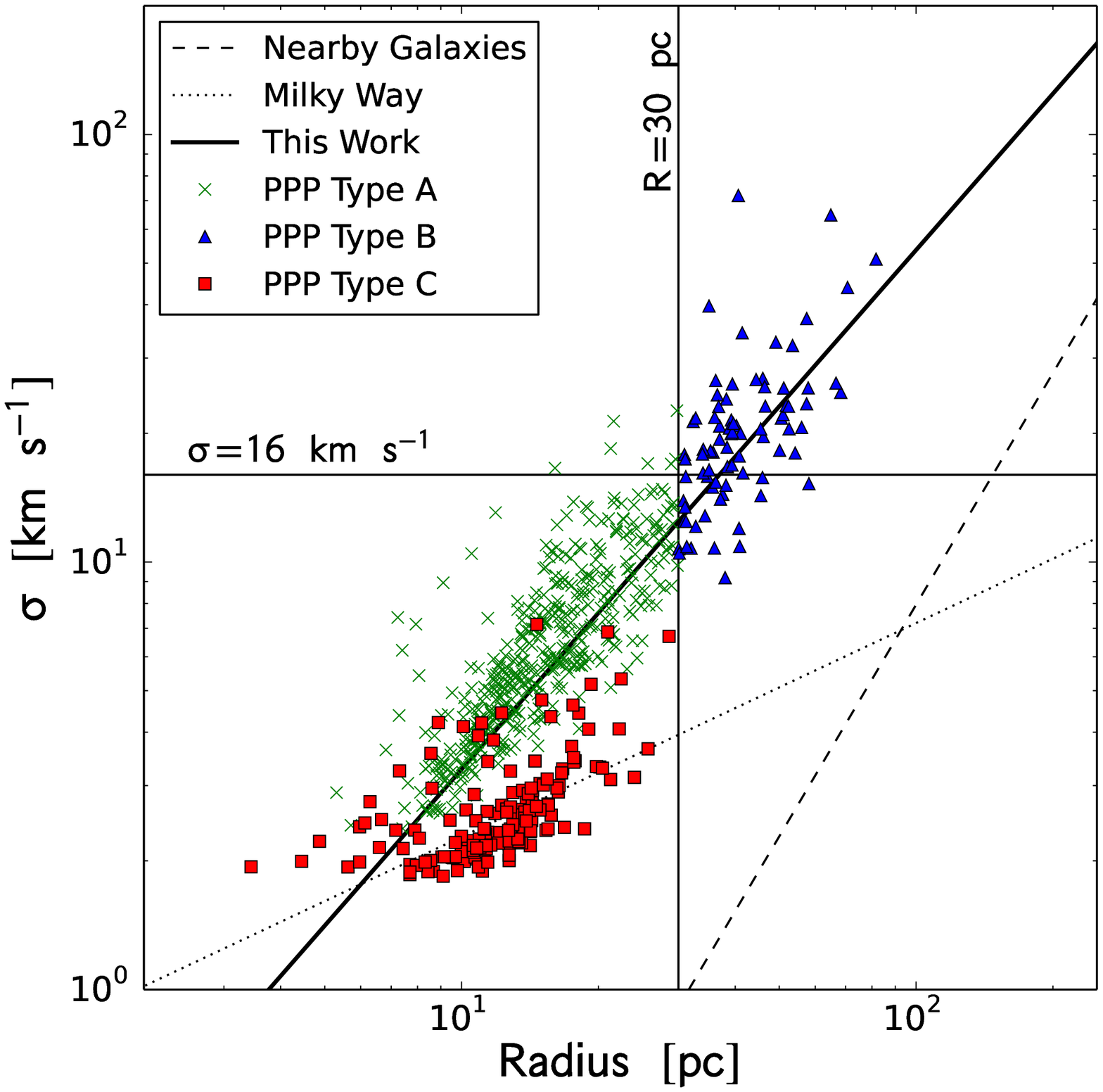} 
\begin{center}
 (b)
\end{center}
    \end{minipage}
    \begin{minipage}{0.28\textwidth}
        \centering
		\includegraphics[width=1\textwidth]{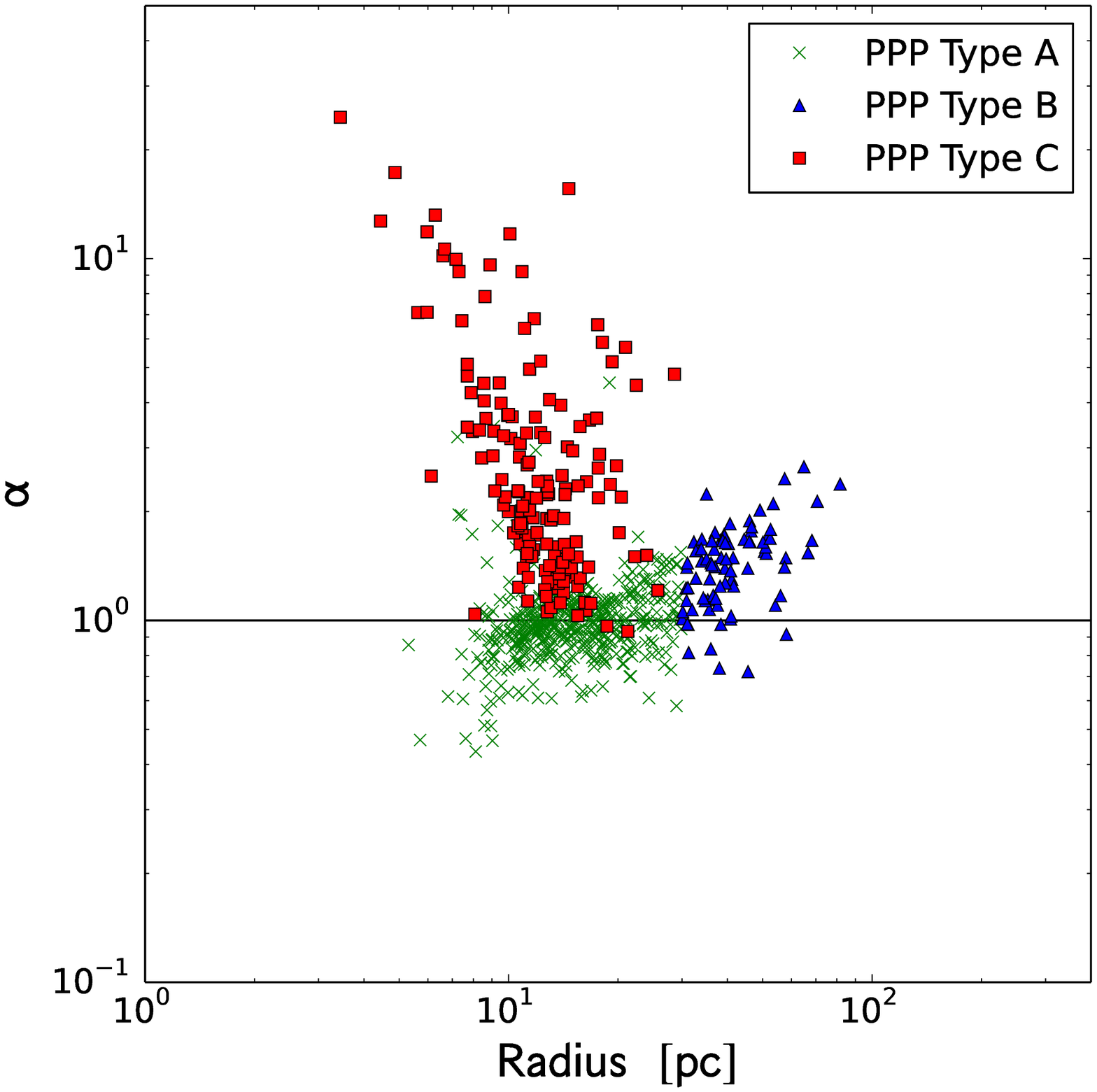} 
\begin{center}
 (c)
\end{center}
    \end{minipage}
            \begin{minipage}{0.28\textwidth}
        \centering
		\includegraphics[width=1\textwidth]{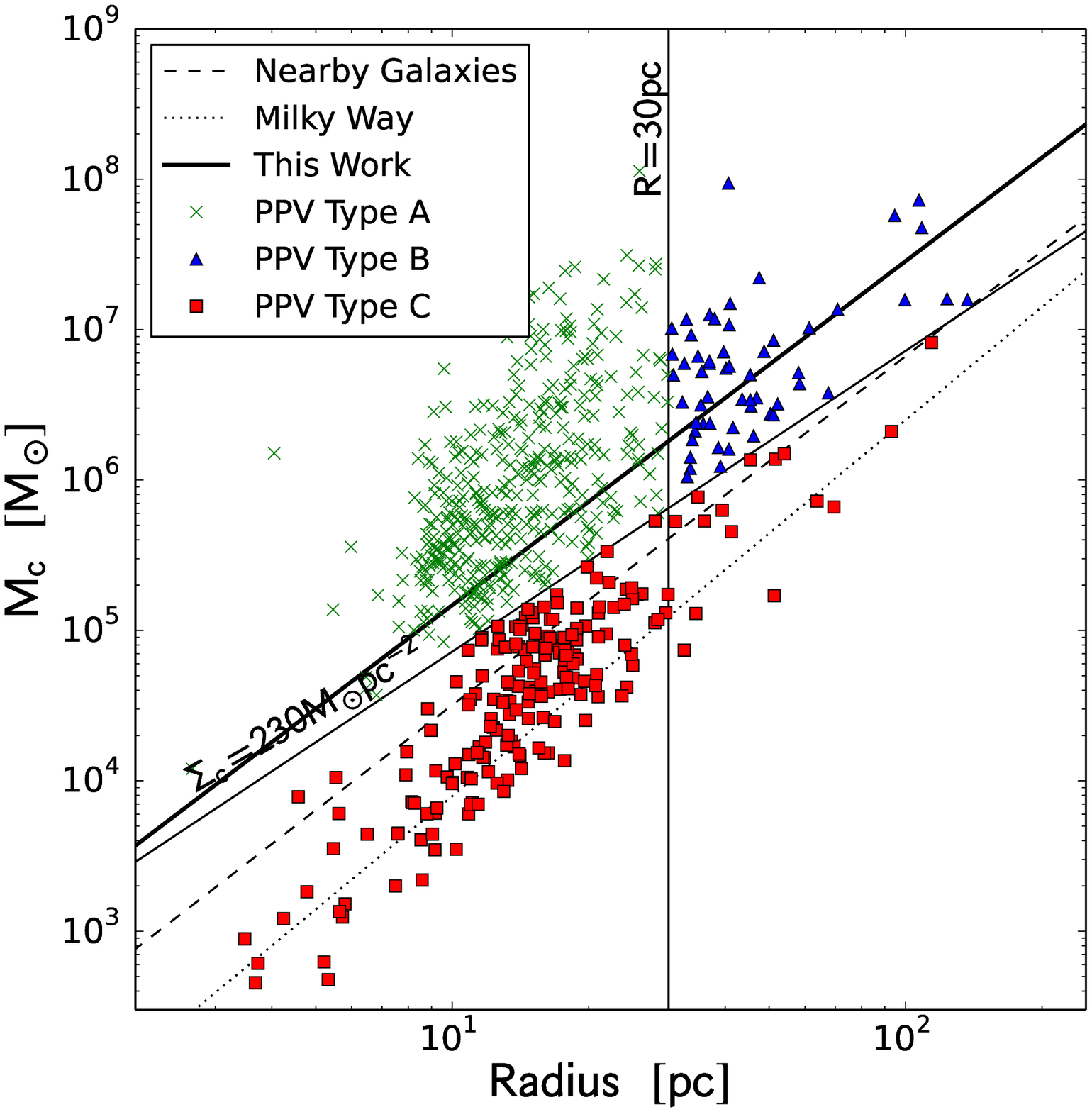} 
\begin{center}
 (d)
\end{center}
    \end{minipage}
    \begin{minipage}{0.28\textwidth}
        \centering
		\includegraphics[width=1\textwidth]{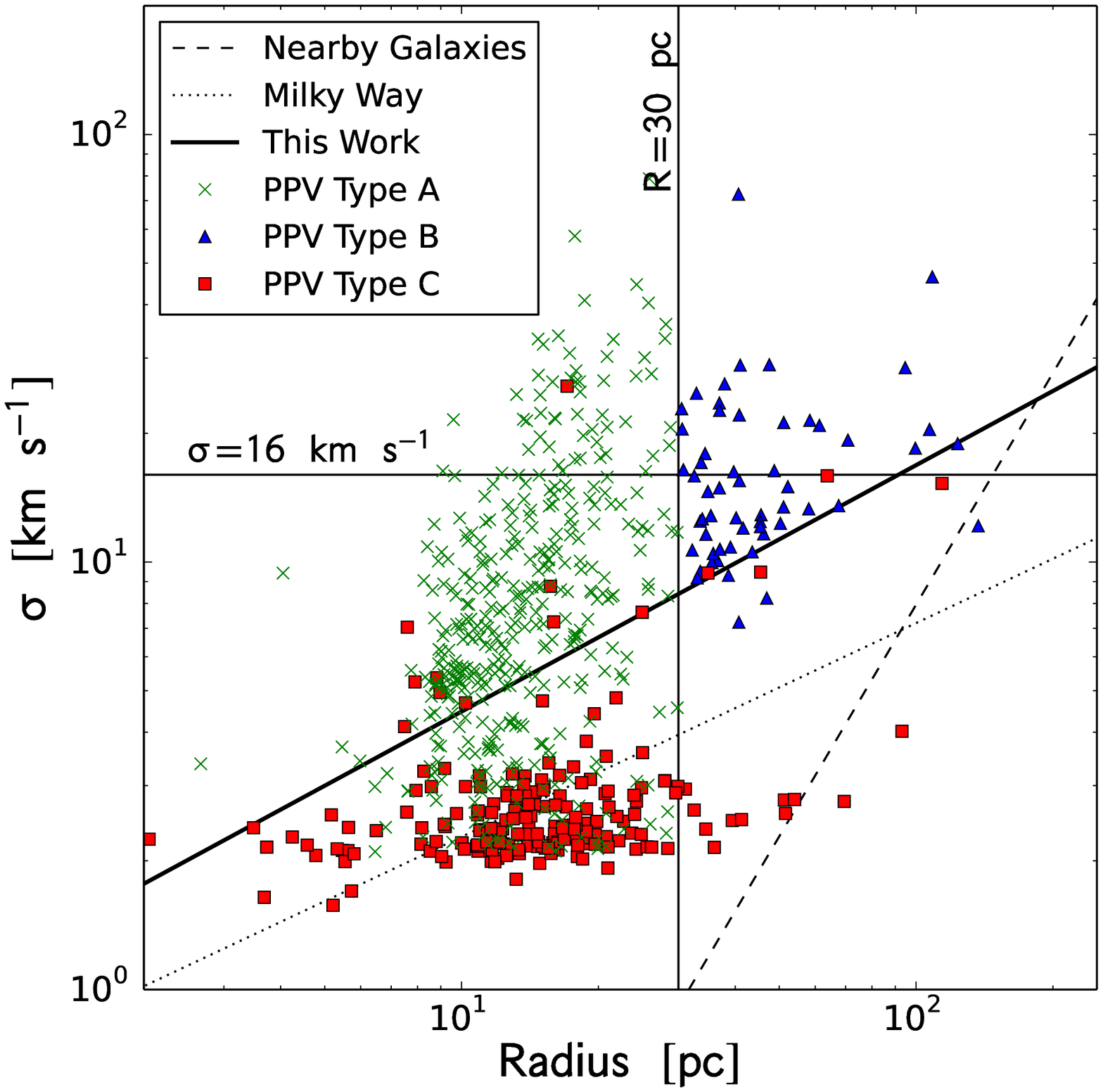} 
\begin{center}
 (e)
\end{center}
    \end{minipage}
    \begin{minipage}{0.28\textwidth}
        \centering
		\includegraphics[width=1\textwidth]{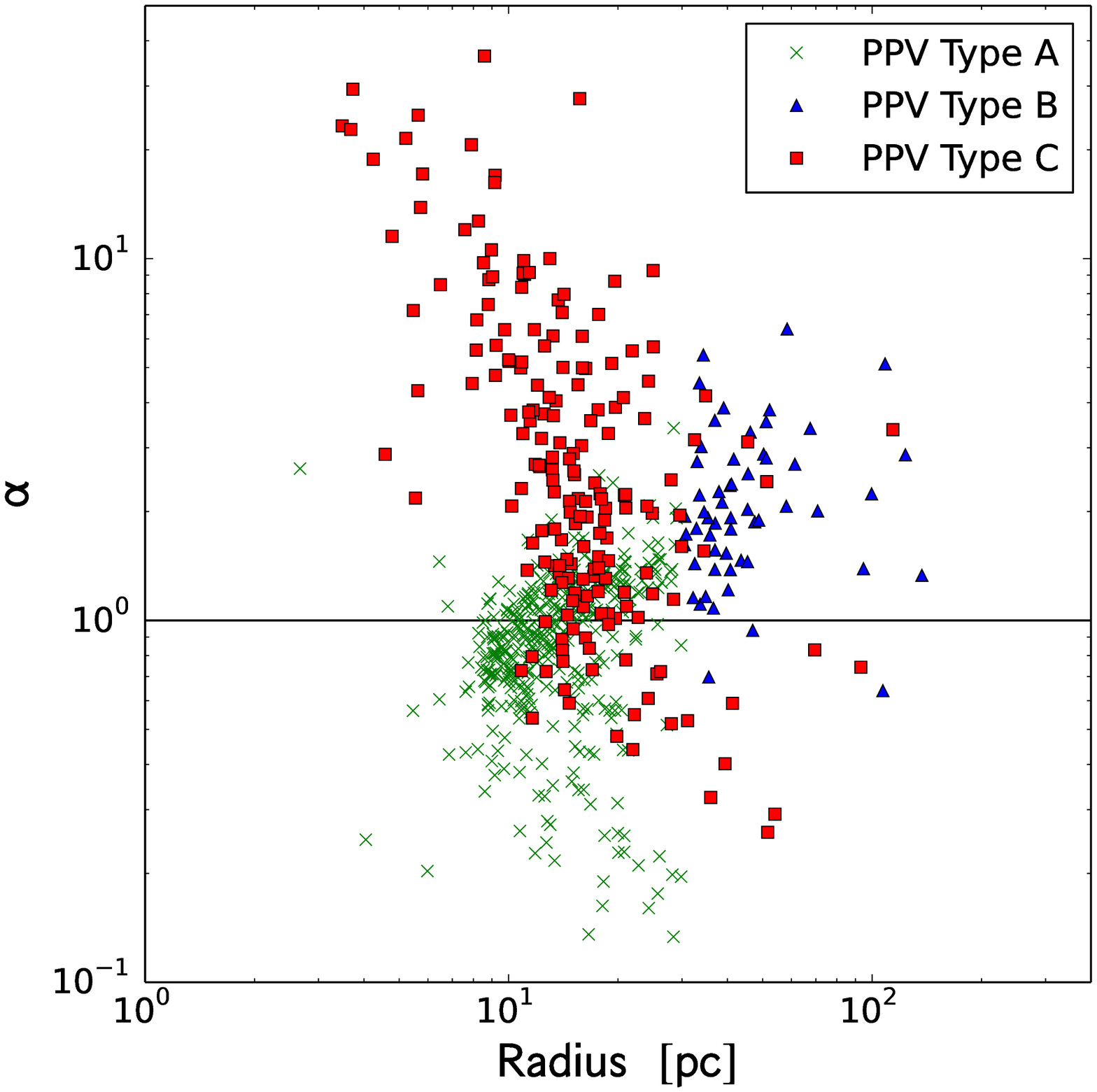} 
\begin{center}
 (f)
\end{center}
    \end{minipage}
	\caption{Scaling relations of the cloud properties. Color markers denote the cloud categories classified by cloud properties. The classification of clouds is introduced by \citet{Fuj14} with PPP clouds. (a)  Classification of cloud types based on PPP clouds on mass -- radius plane.  Cloud with mass surface density greater than 230 M$_{\sun}$ pc$^{2}$ and radius less than 30 pc are \emph{Type A}. Clouds sit on the same sequence but greater than 30 pc are \emph{Type B}. Clouds with  surface density less than 230 M$_{\sun}$ pc$^{2}$ are \emph{Type C}. Boundaries of cloud types are indicated with thin solid lines. p.944:  The fit to the clouds in our PPP is shown as thick solid black line. Observed scaling relations of nearby galaxies (NGC 4736, NGC 4826, and NGC 6946) from \citet{Don13} and the Milky Way from  \citet{Sol87} are shown as dashed and dotted lines, respectively.  (b) The three-type PPP clouds on mass -- radius plane.  (c) Relation of virial parameter versus radius of PPP clouds. Panel (d), (e), and (f) are the same as panel (a), (b), and (c), respectively, but for PPV clouds. The thick solid black lines in panel (d) and (e) represent the fits to the PPV clouds. }
	\label{FIG_PP2_prop_property}
\end{figure*}

The existence of these three populations is shown in the PPP bimodal surface density distributions in Figure~\ref{FIG_stucture_prop}(c) and more clearly in the Larson relations plotted in Figure~\ref{FIG_PP2_prop_property}(a). These plots show a boundary between the Type A and C clouds at around 230\,M$_{\sun}$ pc$^{-2}$, with a further split seen most clearly in the mass-radius relation of the bar clouds in between the Type A and B populations. The populations are identifiable in the PPV data but less strongly than for the PPP clouds. The bimodal split is harder to see in the surface density profiles for PPV clouds, although it is clearly there (albeit with more scatter) in the mass-radius relation in Figure~\ref{FIG_PP2_scaling}(b), with the same division between the two sequences at 230\,M$_{\sun}$ pc$^{-2}$. Moreover, the three populations occur in all galactic environments as see in the one-to-one scatter plot of the cloud surface density in Figure \ref{FIG_Match_prop}(c). 

\begin{figure*}
    \hspace{-4mm}
    \begin{minipage}{0.45\textwidth}
        \centering
		\includegraphics[width=1\textwidth, angle=-90]{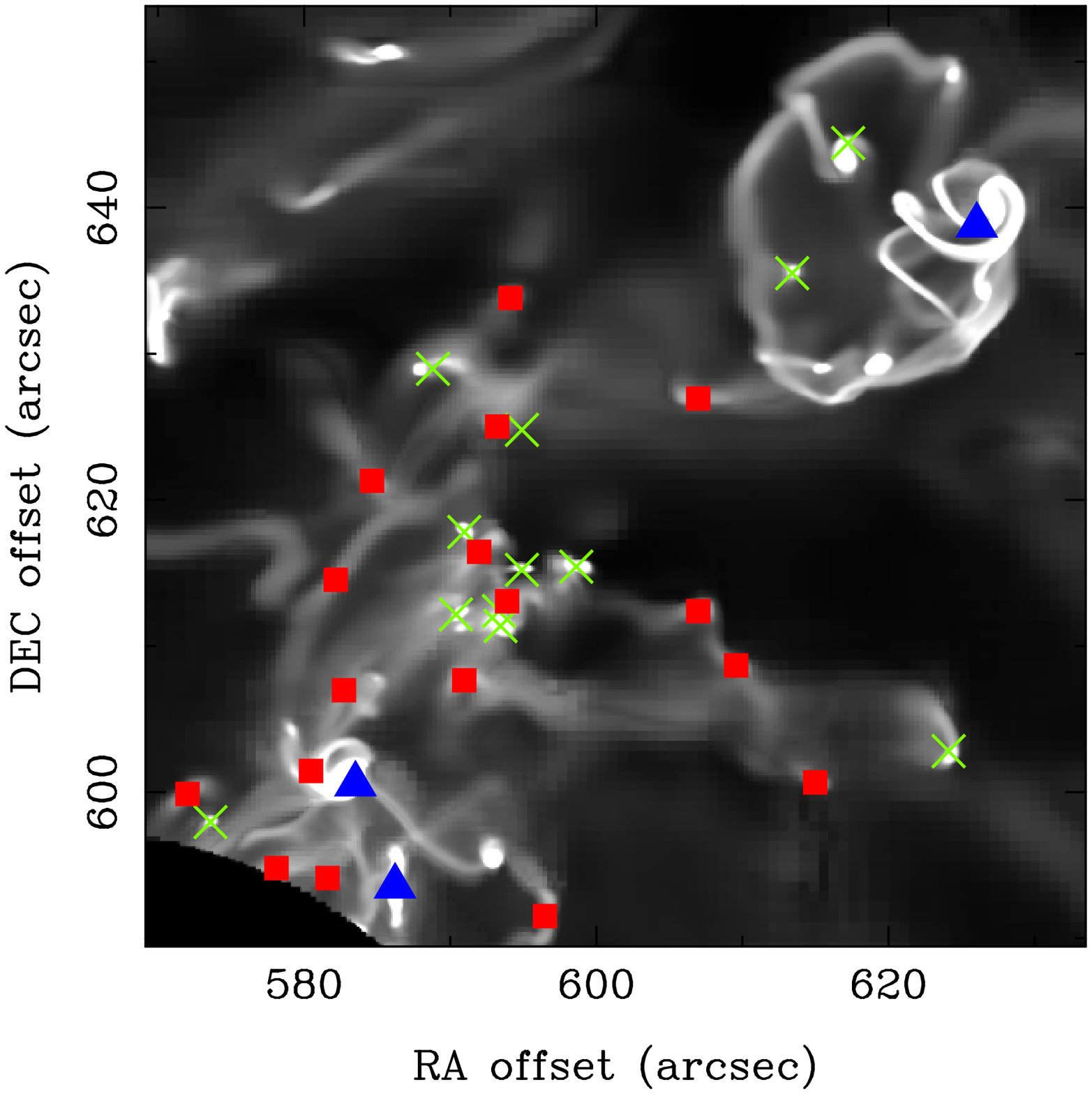} 
\begin{center}
 (a)
\end{center}
    \end{minipage}
    \begin{minipage}{0.45\textwidth}
        \centering
		\includegraphics[width=1\textwidth, angle=-90]{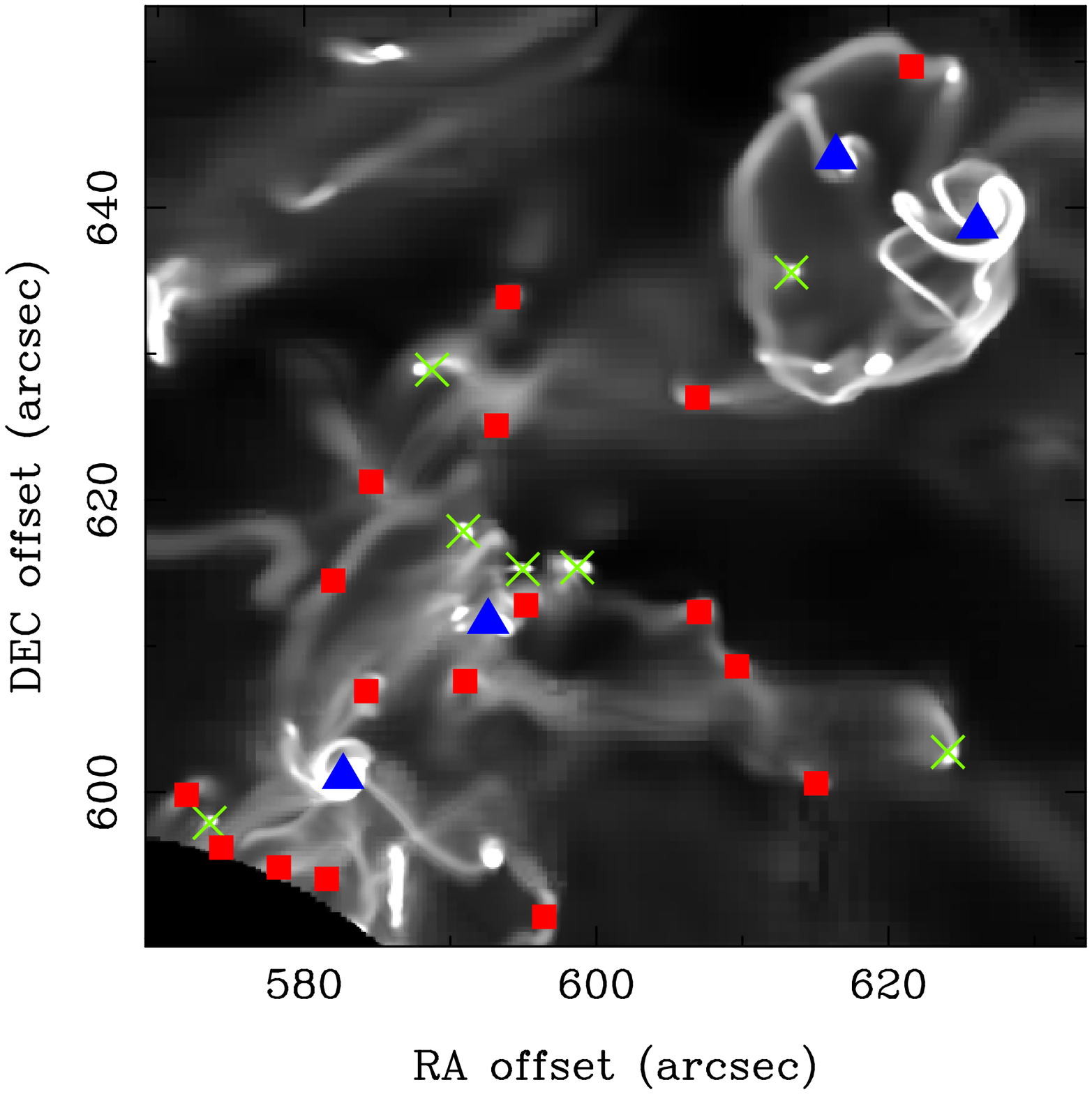} 
\begin{center}
 (b)
\end{center}
    \end{minipage}
      \caption{Examples of three-type clouds in galactic environments. \emph{Type A}, \emph{B}, and \emph{C} clouds are shown with green crosses, blue triangles, and red squares, respectively. (a) PPP-clouds at half of bar region. Scale of the figure is 1.2 kpc $\times$ 1.2 kpc.  (b) PPV-clouds at the same bar region.}
      	\label{FIG_three_types_clouds_pos_examples}
\end{figure*}

Visual examples of these clouds can be seen in Figure~\ref{FIG_three_types_clouds_pos_examples}. The figure shows the surface density in a 1\,kpc $\times$ 1\,kpc section of the bar region. Green crosses mark the \emph{Type A} clouds, while blue triangles show the large \emph{Type B} associations and red squares are the \emph{Type C} clouds. Focusing on the PPP left-hand panel, we can see the \emph{Type B} clouds are tidally distorted massive clouds while the \emph{Type C} clouds are siting primarily along the filaments of tidal tails that stretch between the \emph{Type B} associations. \emph{Type A} clouds are more discrete objects with a clear centre and less extended surrounding structure. Largely, the same cloud types are detected in the PPV data in the right-hand panel of Figure~\ref{FIG_three_types_clouds_pos_examples}. However, there are a few interesting exceptions which will be discussed in \S\ref{sec_change_prop_types}.

To investigate how well the properties of each of these three cloud types are represented in the PPV data, we re-group the clouds in each environment via their position on the mass-radius relation, using the same definition for the three types as \cite{Fuj14}. Here, \emph{Type A} clouds have a mass surface density greater than 230\,M$_{\sun}$ pc$^{-2}$ and radius less than 30 pc, clouds along the same sequence but with a radius above 30\,pc form the \emph{Type B} and small clouds with a mass surface density less than 230\,M$_{\sun}$ pc$^{-2}$ are the \emph{Type C} clouds. 

The results of this categorization are shown in the mass-radius relations plotted in Figures~\ref{FIG_PP2_prop_property}(a) and \ref{FIG_PP2_prop_property}(d). Even while the scatter in the PPV radius measurement blurs the distinction between the upper and lower sequence of clouds, the quantity of clouds in each region appear similar. Most clouds lie in the \emph{Type A} region, with a small number of massive \emph{Type B} and a slightly smaller parallel trend of \emph{Type C}. We note that due to the blur between the two sequences, real observation will need high resolution and sensitivity to confidentially resolve the lower sequence by detecting the clouds with mass $<$ 10$^{4.5}$ M$_{\sun}$  and radius $<$ 10 pc. One interesting difference is  the existence of \emph{large Type C} clouds with radii greater than 30 pc in PPV. These will be explored in \S\ref{sec_change_prop_types}.

The three cloud types also occupy different regions of the Larson scaling relation between velocity dispersion and radius in Figure \ref{FIG_PP2_prop_property}(a) and \ref{FIG_PP2_prop_property}(d). The lower mass \emph{Type C} clouds are small, low mass objects and therefore also have lower velocity dispersions than the other two cloud types. This velocity dispersion is slightly lowered in PPV due to the minimum value and line-of-sight dependence imposed by CPROPS as described in the previous section. \emph{Type A} clouds have a larger scatter and reduced radius range from the mass-weighting in PPV, and extend to higher velocity dispersions due to the merging of PPP
clouds in the projected data set. The massive \emph{Type B} clouds typically have the highest velocity dispersions, although this value follows a weaker trend in PPV. This is due to the tails of material that typically surrounding the \emph{Type B} clouds from regular tidal interactions. Such tails consist of lower density material and are therefore frequently ignored in the PPV data or only partially selected. The tails do not contribute significantly to the mass of the cloud, but give a significant boost to the velocity dispersion. The velocity dispersion is further increased by PPP clouds merging in the PPV data set, and the more complex cloud morphology leading to greater scatter in the line-of-sight velocity estimates compared to the three-dimensional value.

The difference in the \emph{Type C} transient clouds can perhaps be most clearly seen in the relation between the alpha virial parameter and radius, plotted in  \ref{FIG_PP2_prop_property}(c) and \ref{FIG_PP2_prop_property}(f). Forming in filament tails, these clouds are low density and typically unbound, with alpha virial values between $1 - 10$. The other two types of clouds are largely borderline bound with $\alpha \sim 1.0$, although \emph{Type B} clouds are less bound due to their large size and high velocity dispersion. As seen in Figure~\ref{FIG_stucture_prop}(f), the derivation of alpha from three other parameters increases the scatter in the PPV data. Despite this, the basic position of the three cloud types remains distinct and matches the PPP data. The main difference is the extension of \emph{Type A} and \emph{Type C} clouds to lower values of alpha. This is due to the smaller clouds hitting the lower limit of the PPV measurement of the velocity dispersion, as shown in Figure~\ref{FIG_PP2_prop_property}(c).

\subsubsection{Fraction of Three Types Clouds in Galactic Environments}

While not identical, the split of cloud types between the three different galactic regions is very similar in both PPP and PPV. Table~\ref{TAB_frac_types_in_structures} records the percentages of each cloud type in each region for the two identification methods, which mostly differ by $ 30$\% pertween PPP and PPV. The overall features of the divisions noted by \cite{Fuj14} are preserved in the PPV, with the most common type of cloud being the \emph{Type A} in all regions, with the highest fraction of \emph{Type A} clouds found in the quiescent disc region ($\sim 80$\%) and lowest in the interaction-packed bar. The bar and spiral regions have similar fractions of the massive \emph{Type B} clouds but the bar region has the largest percentage of the interaction-spawned \emph{Type C}, at 38\% compared to 23\% and just 11\% in the spiral and disc, respectively.

\begin{table}
\caption{Percentage of each type of clouds in the three environments. The clouds are identified with the island methods.}
\label{TAB_frac_types_in_structures}
\begin{tabular}{lllll}
\hline
                                 &     & bar  & spiral & disc \\
 \hline                               
\multirow{2}{*}{\textit{type A}} & PPP & 49  & 64   & 83  \\
                                 & PPV & 43  & 61    & 76  \\
\multirow{2}{*}{\textit{type B}} & PPP & 13  & 13    & 6   \\
                                 & PPV & 9   & 9     & 7   \\
\multirow{2}{*}{\textit{type C}} & PPP & 38  & 23   & 11  \\
                                 & PPV & 48  & 30    & 17  \\
\hline                         
\end{tabular}
\end{table}

\subsubsection{Change of Clouds Types Between PPV and PPP}
\label{sec_change_prop_types}

The images of bar environment in Figure~\ref{FIG_three_types_clouds_pos_examples} showed that while most clouds are identified with the same type in PPP and PPV, this is not universally true. The number of clouds that do switch type with selection technique is listed in Table~\ref{TAB_ABC_change}, where the clouds were matched together using the method described in \S\ref{SEC_match_cloud}. The vast majority (between $80 - 86$\%) of clouds retain their catagory in both data sets. The type change between the most massive clouds \emph{Type B} and the smallest clouds, \emph{Type C}, is not seen in any galactic environment, while changes between the other types of clouds account for $3-8$\% of the population in each region.

In the mass-radius relation that was used to define the cloud types in Figures~\ref{FIG_PP2_prop_property}(a) and \ref{FIG_PP2_prop_property}(d), the most signficiant difference between the PPP and PPV populations was the existence of \emph{large Type C} clouds in the PPV data set; fifteen clouds with a surface area below 230\,M$_{\sun}$ pc$^{-2}$ (defining them as \emph{Type C}) but with radii above 30\,pc. Since \cite{Fuj14} found \emph{Type C} to be small, transient clouds, such extended objects are very unlikely. 

Further exploration reveals that eight of the fifteen \emph{large Type C} clouds are dense tidal filaments surrounding a massive \emph{Type B} host cloud. In the PPP data set, these filaments are associated with the larger \emph{Type B}, but in the PPV, they have been identified as a separate object. This agrees with Figure~\ref{FIG_PP2_prop_property}(a) which found PPV \emph{Type B} clouds to frequently have lower velocity dispersions than their PPP counteerparts: the tidal tails were not included within the \emph{Type B} boundary. The elongated structure of these filaments results in a low surface density. They are also not associated with any PPP cloud, since their material is part of the host \emph{Type B} in the PPP data set. 

\begin{figure*}
    \hspace{-4mm}
    \begin{minipage}{0.32\textwidth}
        \centering
		\includegraphics[width=1\textwidth, angle=-90]{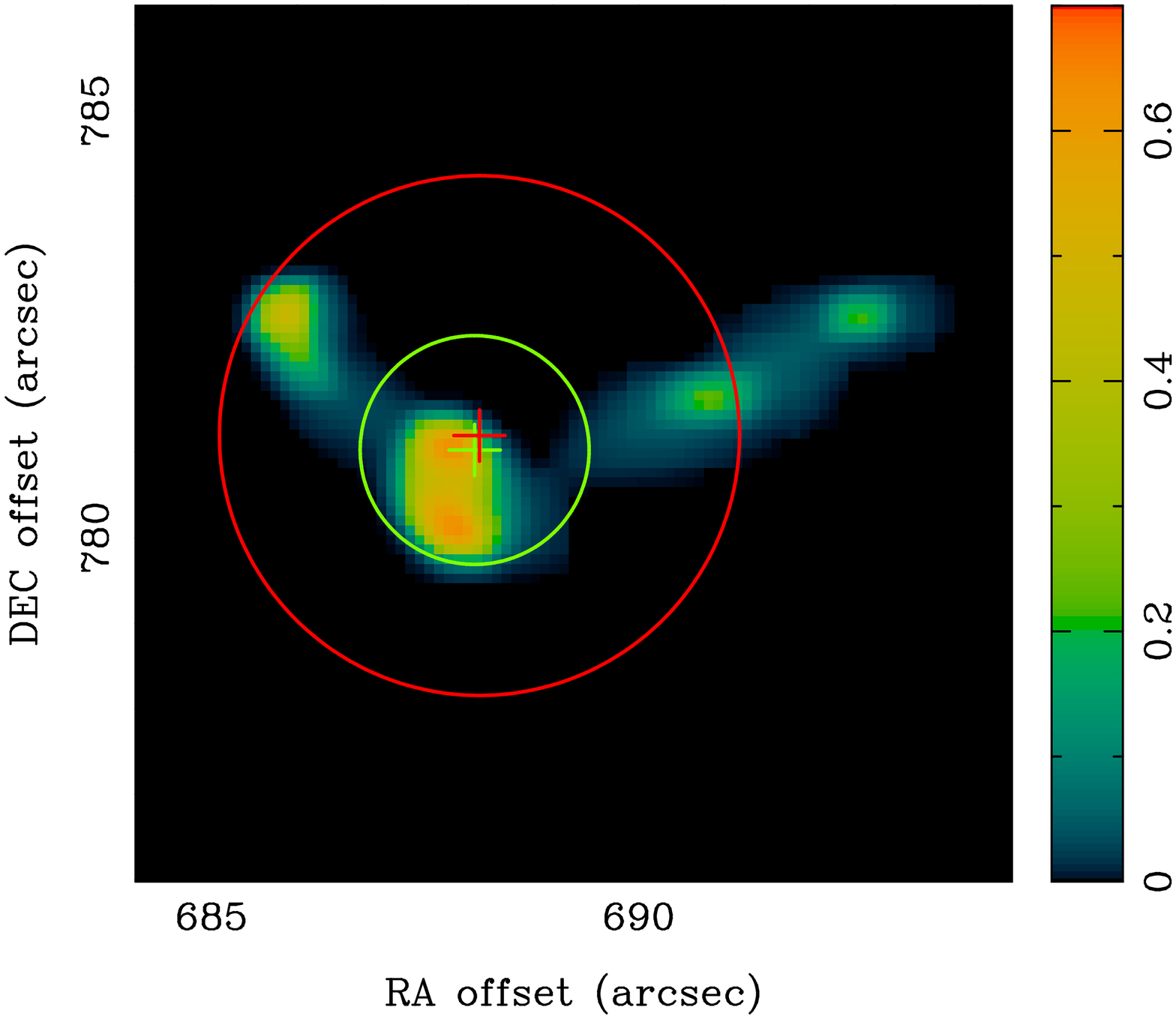} 
\begin{center}
 (a)
\end{center}
    \end{minipage}
    \begin{minipage}{0.32\textwidth}
        \centering
		\includegraphics[width=1\textwidth, angle=-90]{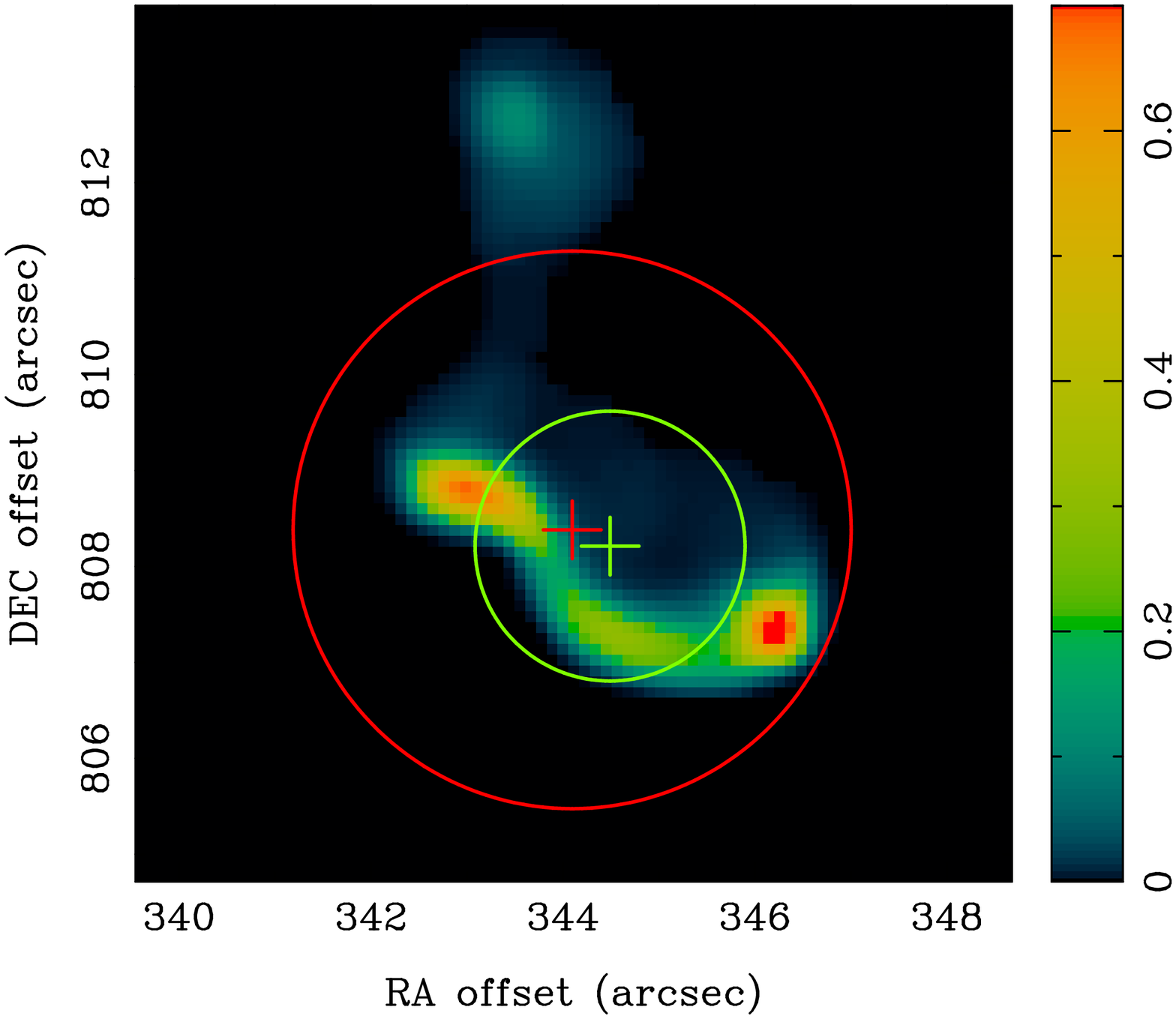} 
\begin{center}
 (b)
\end{center}
    \end{minipage}
        \begin{minipage}{0.32\textwidth}
        \centering
		\includegraphics[width=1\textwidth, angle=-90]{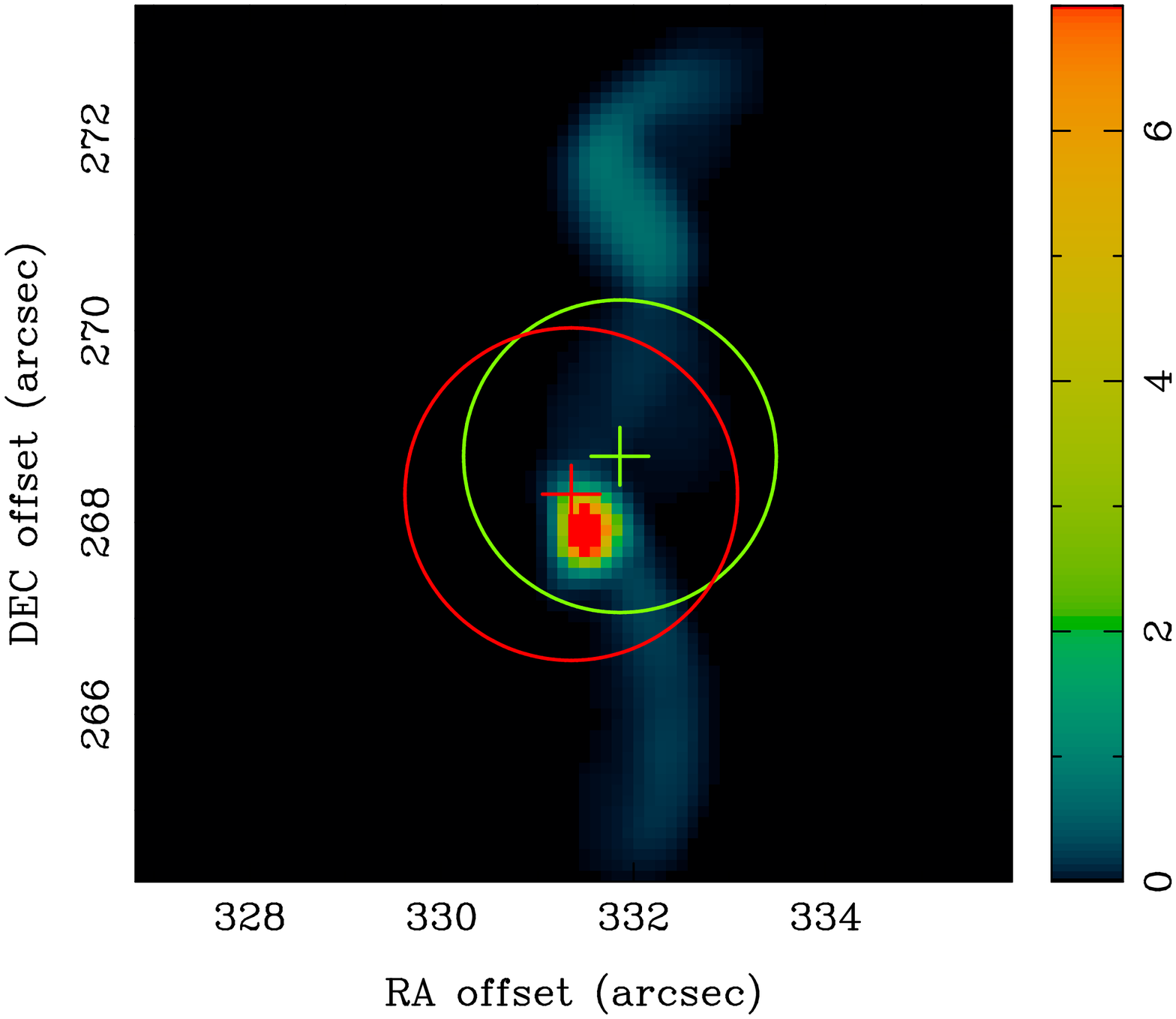} 
\begin{center}
 (c)
\end{center}
    \end{minipage}
  	\caption{Example of the \emph{large type C} clouds of PPV in \S\ref{sec_change_prop_types}. The projection images of the clouds are shown in color scale. Unit of the color bar is g cm$^{-2}$. Red and green circles represent the average size of the cloud identified by PPV and PPP, respectively. Centre of masses are marked with a cross. Panel (a) and (b)  show the clouds which are classified as \emph{Type A}  in PPP but turn the \emph{Type C} in PPV due to the large radii identified by PPV. Panel (c) shows the example of typical  \emph{Type A} in both PPP and PPV. }
	\label{FIG_large_typeC_examples}
\end{figure*}

The remaining seven \emph{large Type C} clouds consist of six PPP \emph{Type A} and one PPP \emph{Type C} but with a radius below 30\,pc. These are not due to splitting, since all seven are isolated with a single counterpart in both data structures. In all cases, the \emph{large Type C} clouds have radii far larger than a typical difference between the PPP and PPV populations, giving a low surface density and dropping them into the \emph{large Type C} regime. These clouds all have common features that include (1) a flattened structure with the $x-y$ area larger that the $x-z$ or $y-z$ area by a factor of 1.5, (2) an elongation with a large ratio between the major and minor axis in the $x-y$ plane and (3) multiple dense peaks with comparable density close to the edge of the cloud. This is shown visually in Figure~\ref{FIG_large_typeC_examples}, where the red circles marks the average radius calculated from the PPV cloud while the green circles shows the average radius in the PPP data for three different cloud cases. The first two panels  show a PPP \emph{Type A} cloud identified as a PPV \emph{large Type C}, with long density profiles and multiple density peaks. The third panel, Figure~\ref{FIG_large_typeC_examples}(c), shows a typical \emph{Type A} cloud found in both data sets. The first property of these clouds reduces the PPV radii compared to PPP, since PPP considers the average area over all planes, while PPV sees only the $x-y$ plane. However, this effect alone is not sufficient to alter the cloud type, since switching the PPP radii definition to use only the $x-y$ plane does not create this cloud class. Rather, the second and third properties identify extreme cases for the CPROPS cloud morphology and profile assumptions. As described in section~\ref{sec_cloud_identification}, CPROPS measures the mass-weighted RMS radii of the cloud and then extends this to an \emph{effective} radius by assuming mass-centered spherical density profile. This assumption is valid for the majority of clouds, since they typically have a dominant peak close to their morphological centre as in Figure~\ref{FIG_large_typeC_examples}(c). However, if the cloud has dense peaks near to its edge, these boosts in the density result in the initial mass-weighted RMS radius measurement being close to the true boundary. The conversion to the effective radius then results in a radius value almost twice as large as the true radius. 
 
This effect is most noticable when it creates a unique PPV class of \emph{large Type C} objects, which Table~\ref{TAB_ABC_change} indicates happens in $4 - 6$\% of cases. However, The table also reveals this effect can change \emph{Type A} clouds into \emph{Type B} if the extended PPV radius is above 30\,pc, but the mass remains sufficiently high to keep the surface density above 230\,M$_{\sun}$ pc$^{-2}$.

Clouds can change type by PPV calculating a smaller radius due to the mass-weighting of the radius measurement. This can cause a PPP \emph{Type B} to change to a PPV \emph{Type A}. This is more frequent in the bar and spiral regions where the gas density is higher, creating more concentrated cloud profiles. 

If a PPP \emph{Type C} cloud has a low density tail that is only partially detected by PPV, then the cloud may become a PPV \emph{Type A} as its radius is truncated and resultant surface density increases. This generates the population of \emph{Type C} to \emph{Type A} clouds that are seen predominantly in the lower density environment of the disc and quiescent regions of the spiral, where \emph{Type C} clouds may be sufficiently undisturbed to hold together an extended structure.

\begin{table}
\caption{Comparison of the change of the cloud types in percentage using the  match clouds between PPP and PPV.  The first column denotes the cloud type change from PPP to PPV.}
\label{TAB_ABC_change}
\begin{tabular}{cccc}
\hline
     Cloud type             & bar        & spiral     & disc       \\
     \hline
No change         & 86          & 80           & 81            \\
A $\rightarrow$ B & 2  & 3   & 6   \\
A $\rightarrow$ C & 4   & 6   & 5   \\
B $\rightarrow$ A & 8   & 8   & 4   \\
B $\rightarrow$ C & 0   & 0   & 0   \\
C $\rightarrow$ A & 0   & 3   & 4   \\
C $\rightarrow$ B & 0   & 0   & 0   \\
\hline
\end{tabular}
\end{table}

\section{Other Cloud Identification Methods}
\label{sec_diff_methods}

In addition to the island method for identifying clouds, the decomposition method that constructs clouds surrounding peaks in the density and intensity fields was introduced in section~\S\ref{sec_cloud_identification}. In this section, we compare the PPP and PPV cloud populations found using this second scheme with the island method choice. 

The number of clouds found by the decomposition method in the PPP and PPV data sets are comparable with one another. For the PPP data, the decomposition method finds a total of 2832 clouds, compared to 3081 clouds in the PPV data. Within each environment, the numbers remain comparable, with PPP having 262 clouds in the bar, 1521 clouds in the spiral and 224 clouds in the disc region and PPV finding 365, 1590 and 213 for the same environments respectively. 

\begin{figure}	
\includegraphics[width=0.45\textwidth]{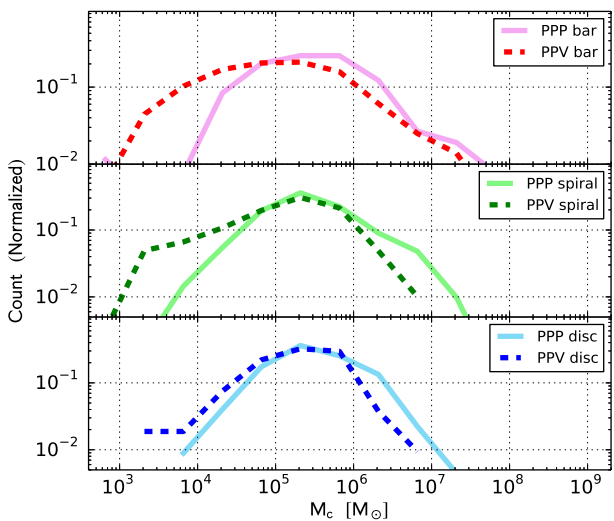} 
\caption{Normalized distribution of  cloud mass of  clouds identified with  decomposition methods. Bar, spiral and disc clouds are displayed with red, green, and blue lines, respectively. PPV clouds are shown with dashed lines while PPP clouds are presented with solid lines.}
\label{FIG_PP2_Mcl_structure_dcomp}
\end{figure}

The decomposition method tends to segment the larger clouds found by the island method into small clouds with more uniform properties. This can be instantly seen by the significantly larger number of clouds found by the decomposition method compared to the island method. It is also reflected in the cloud properties as shown by the mass distribution for the decomposition method in Figure~\ref{FIG_PP2_Mcl_structure_dcomp}. Here, the cloud populations for the bar region are shown in the top panel in red, the spiral region is the middle panal in green and the outer disc is in blue in the bottom panel. As with Figure~\ref{FIG_stucture_prop}, the PPP data cloud population distribution is shown with a solid line, while the PPV clouds are shown with a dashed line. The differences between the three environments are reduced compared to the island method, with many of the massive clouds with mass greater than 10$^{7}$\,M$_{\sun}$ segmented to make smaller clouds in the bar and spiral region. The fraction of small clouds remains similar to the island method, although the PPV data set finds repeatedly more small clouds than the PPP population due to the way islands are divided between multiple peaks. As described in section~\S\ref{sec_cloud_identification}, the PPV decomposition method separates peaks on the same contour island by drawing a second contour that just encases both peaks and then separates these with a third contour above this boundary. Emission that is below this separating third contour is considered a `watershed' and is discarded. In PPP, cells within an island contour containing multiple peaks are simply split between the two separated clouds. This loss of emission in PPV causes a higher number of small PPV clouds to be formed compared to PPP.

This watershed for PPV leads to smaller median cloud properties within the PPV decomposition data set, but both PPP and PPV have median properties that are slightly lower than for the island method by 1.1 -- 2 times, due to cloud segmentation. Between environments, the newly uniformed clouds show little difference. The median cloud mass, radius and velocity dispersion are  $\sim 10^{5}$\,M$_{\sun}$, $\sim 11.1$\,pc and $\sim 3.6$\,km s$^{-1}$ respectively in PPV and $\sim 2.2\times 10^{5}$\,M$_{\sun}$, $\sim 15.4$\,pc and $\sim 5.0$\,km s$^{-1}$ in PPP for all three environments. The virial parameter does show a difference with environment, although not with identification technique. For both the PPP and PPV, the median virial parameter is $\sim 2$ in the bar, $\sim 1.5$ in the spiral and $\sim 1$ in the disc. As with the island method, the strong interactions in the bar region increase the virial parameter as the clouds become less bound.

\begin{figure*}
    \hspace{-4mm}
    \begin{minipage}{0.32\textwidth}
        \centering
		\includegraphics[width=1\textwidth]{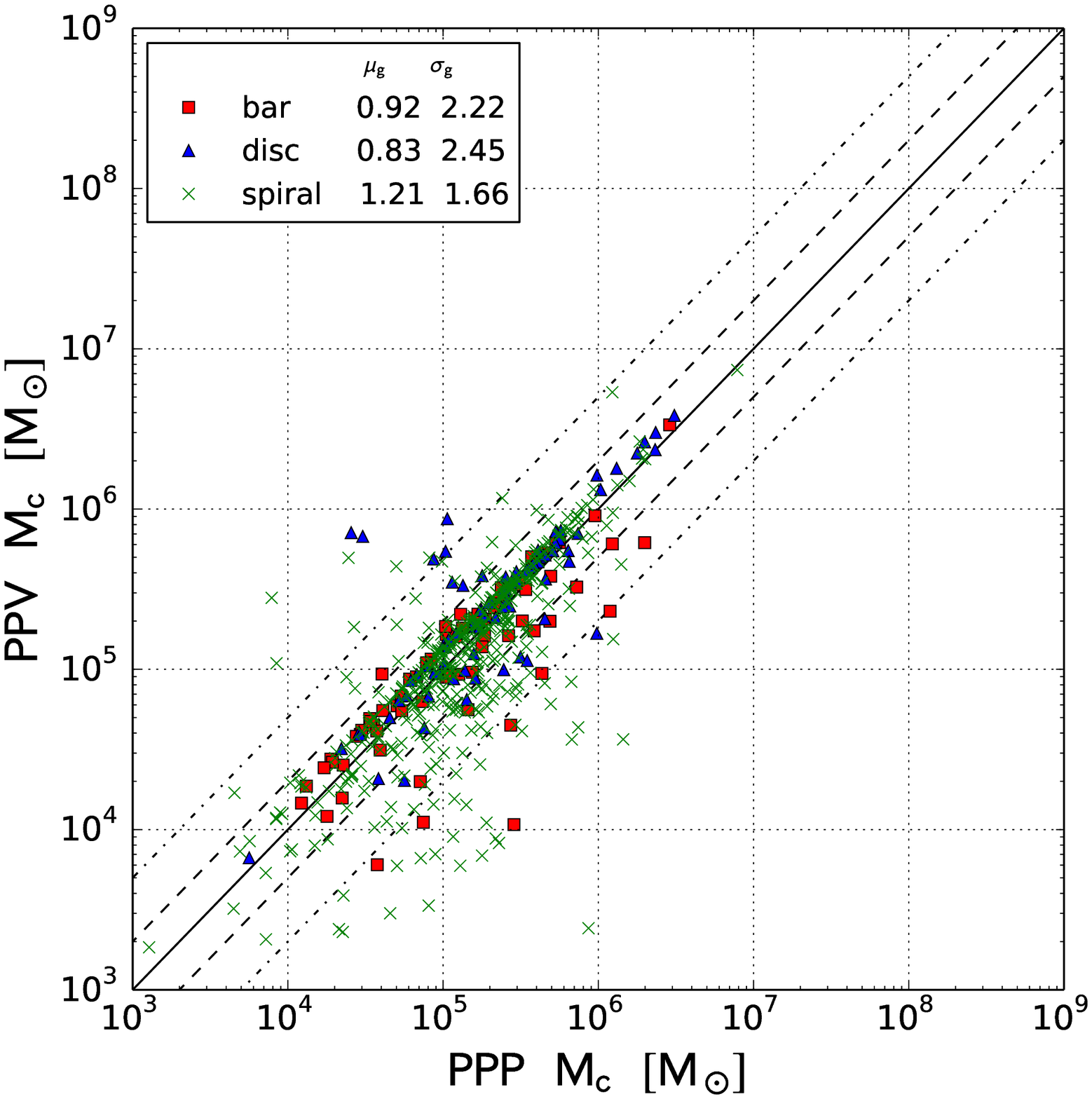} 
\begin{center}
 (a)
\end{center}
    \end{minipage}
    \begin{minipage}{0.32\textwidth}
        \centering
		\includegraphics[width=1\textwidth]{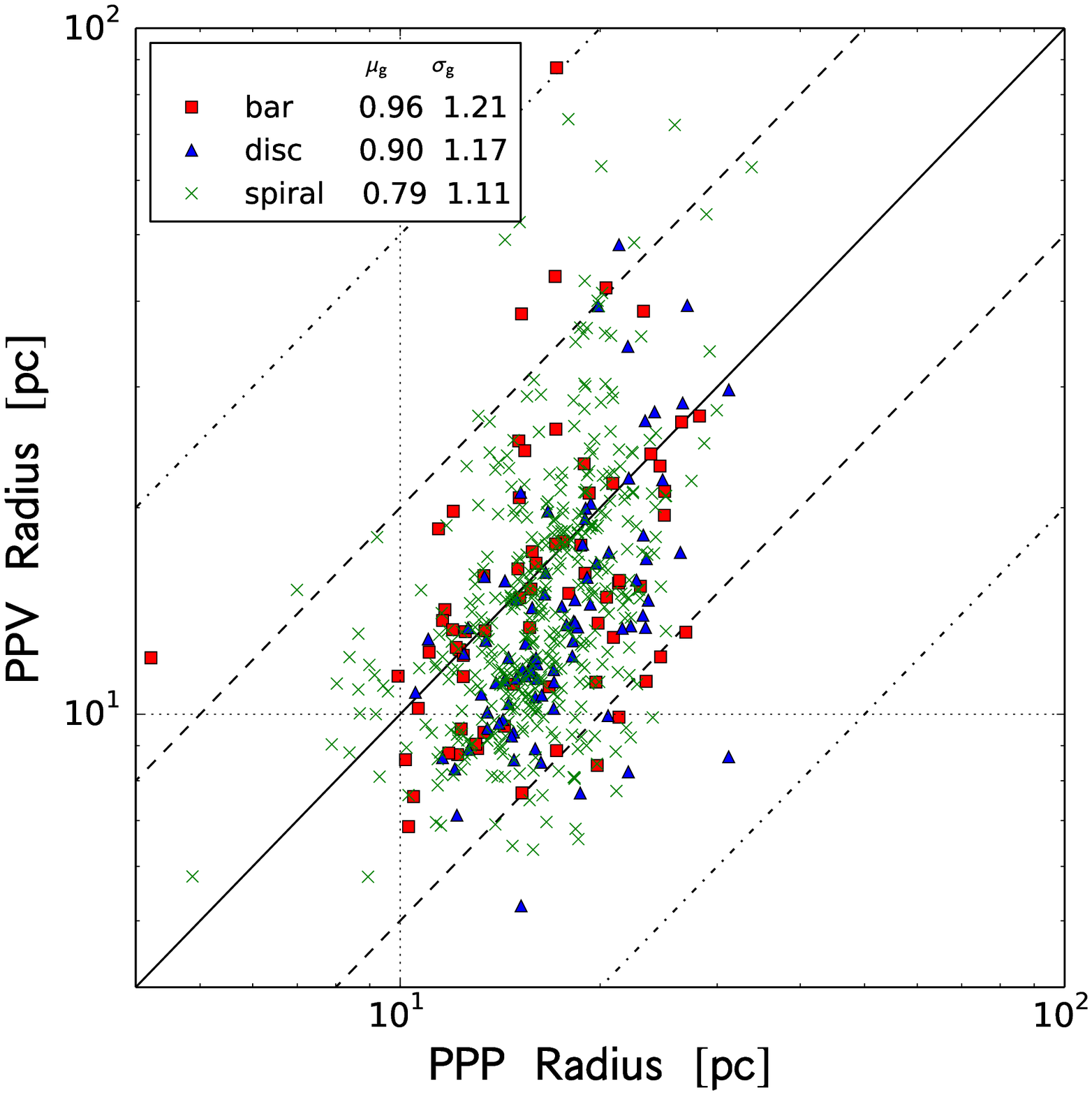} 
\begin{center}
 (b)
\end{center}
    \end{minipage}
        \begin{minipage}{0.32\textwidth}
        \centering
		\includegraphics[width=1\textwidth]{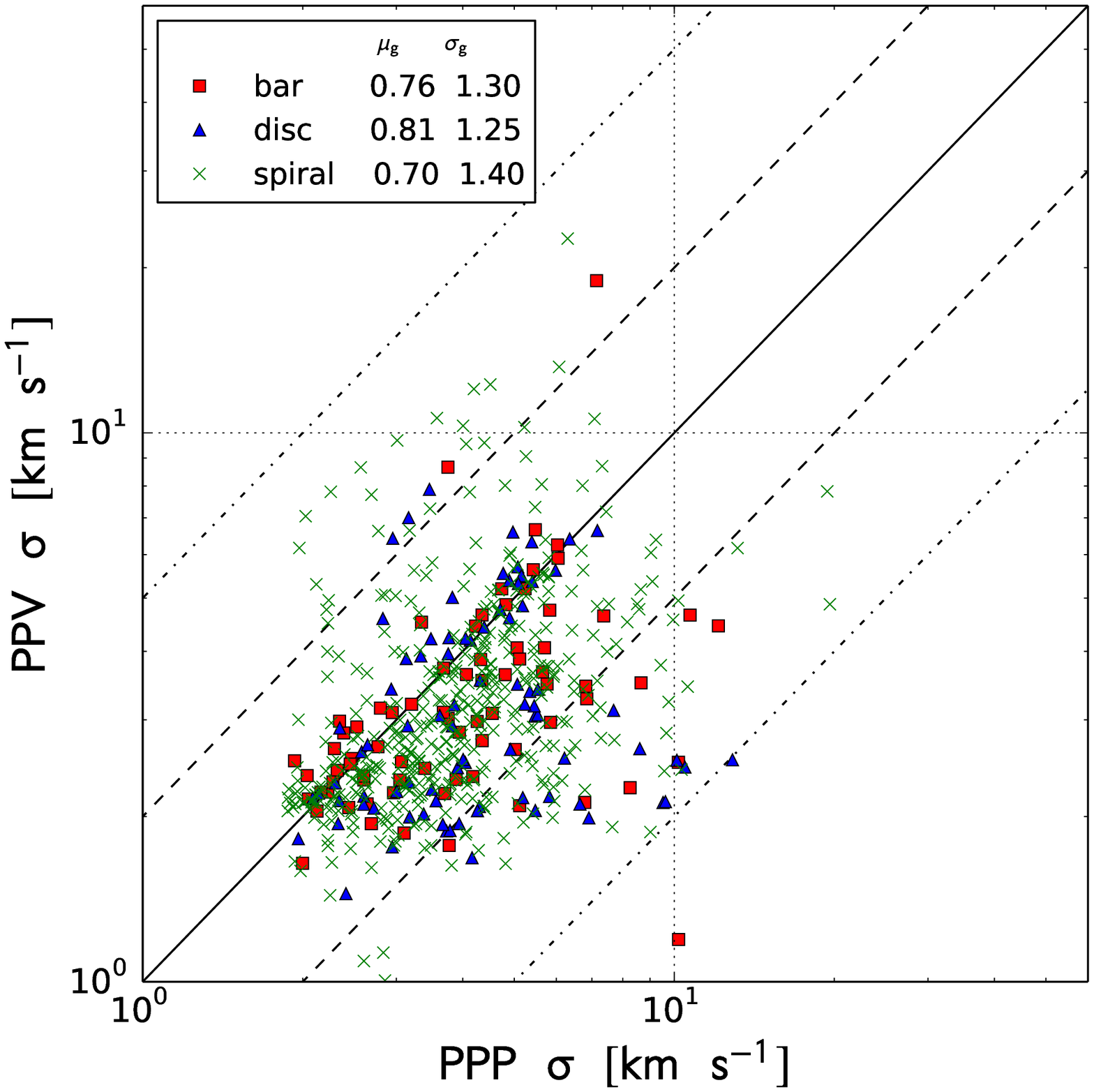} 
\begin{center}
 (c)
\end{center}
    \end{minipage}
	\caption{Comparison of match cloud properties in bar (red squares), spiral (green crosses), and disc (blue triangles) regions. Clouds are identified with  decomposition method.  Panel (a) -- (c) show the comparison of cloud mass, radius, and velocity dispersion, respectively.  Dashed and dashed-dotted lines indicate a factor of two and five above and below the solid 1 : 1 line, respectively. Geometric mean ($\mu_{\mathrm{g}}$) and geometric standard deviation ($\sigma_{\mathrm{g}}$) of PPV-to-PPP ratio of each environments are shown in each panel. }
	\label{FIG_PP2_Match_peak_decomp}
\end{figure*}

Due to the larger number of smaller clouds, the fraction of clouds that are successfully matched one-to-one between PPP and PPV significantly decreases between the island method and decomposition method. However, for the clouds that are matched, their properties continue to agree reasonably well. As with the island method, the scheme for matching clouds between the two data sets is the one described in section~\S\ref{SEC_match_cloud}. It results in a match of only around 40\%, compared to the 70\% match between PPP and PPV clouds in the island method. The comparison between properties of the matched clouds is shown in Figure~\ref{FIG_PP2_Match_peak_decomp} for the mass, radius and velocity dispersion values. The ratio of the PPV to PPP value, $\mu_{\mathrm{g}}$, of the mass radius and velocity dispersion ranged between 0.7 -- 1.2, a slightly wider range than for the island method for the same properties. The scatter is significantly higher than that for the island method, with $\sigma_{\mathrm{g}}$ ranging between 1.1 and 2.5. The solid, dashed and dashed-dotted lines on Figure~\ref{FIG_PP2_Match_peak_decomp} mark out the deviation between the 1:1 agreement by a factor of one, two and five respectively. The differences lie mostly within a factor of two, but there are clouds whose scatter extends up to five. Around 80\% of these high scatter clouds which lie above a factor of two from the 1:1 relation are those that originated from the giant \emph{Type B} clouds in the island method, but were split in the decomposition method. 

Due to the division of the larger clouds, the fraction of large \emph{Type B} clouds shrinks to almost zero in all three environments. The exact fractions are shown in Table~\ref{TAB_decomp_frac_types_in_structures}. In the quiescent disc region, both PPP and PPV have the highest fraction of \emph{Type A} clouds, with roughly two-thirds of the disc's cloud population coming under this type. This cloud type also forms a substantial fraction of the population in the bar and spiral regions, but slightly lower at around 50\%. This distribution between the environments is consistent with the island method results, which also showed the highest number of \emph{Type A} clouds in the disc. Also like the island method, the highest fraction of \emph{Type C} clouds are in the interactive bar region and the lowest fractions are in the passive disc region, where the fractions change from roughly a half to a third. The largest difference between the island and decompositin methods are for the massive \emph{Type B} clouds. This is because 99\% of the \emph{Type B} clouds are split into multiple clouds, with as high as $\sim 85$\% split into more than two clouds, effectively removing the population in the decomposition method. The clouds from this splitting typically result in a \emph{Type A} cloud for the main peak and multiple \emph{Type C} clouds for the extended low density envelopes. This cloud splitting is responsible for the large scatter of cloud properties when clouds are matched between identification methods in Figure~\ref{FIG_PP2_Match_peak_decomp}.

\begin{table}
\caption{Percentage of cloud types in the three galactic environments. Clouds are identified  with the decomposition methods.}
\label{TAB_decomp_frac_types_in_structures}
\begin{tabular}{lllll}
\hline
                                 &     & bar  & spiral & disc \\
 \hline                               
\multirow{2}{*}{\textit{type A}} & PPP  & 54    & 57     & 65   \\
                                 & PPV & 42    & 53      & 67    \\
\multirow{2}{*}{\textit{type B}} & PPP & 0    & 0      & 0     \\
                                 & PPV  & 1     & $<$ 1       & 0     \\
\multirow{2}{*}{\textit{type C}} & PPP & 46    & 43     & 35    \\
                                 & PPV  & 57    & 46      & 33    \\
\hline                         
\end{tabular}
\end{table}

\begin{figure*}
    \hspace{-4mm}
    \begin{minipage}{0.32\textwidth}
        \centering
		\includegraphics[width=1\textwidth]{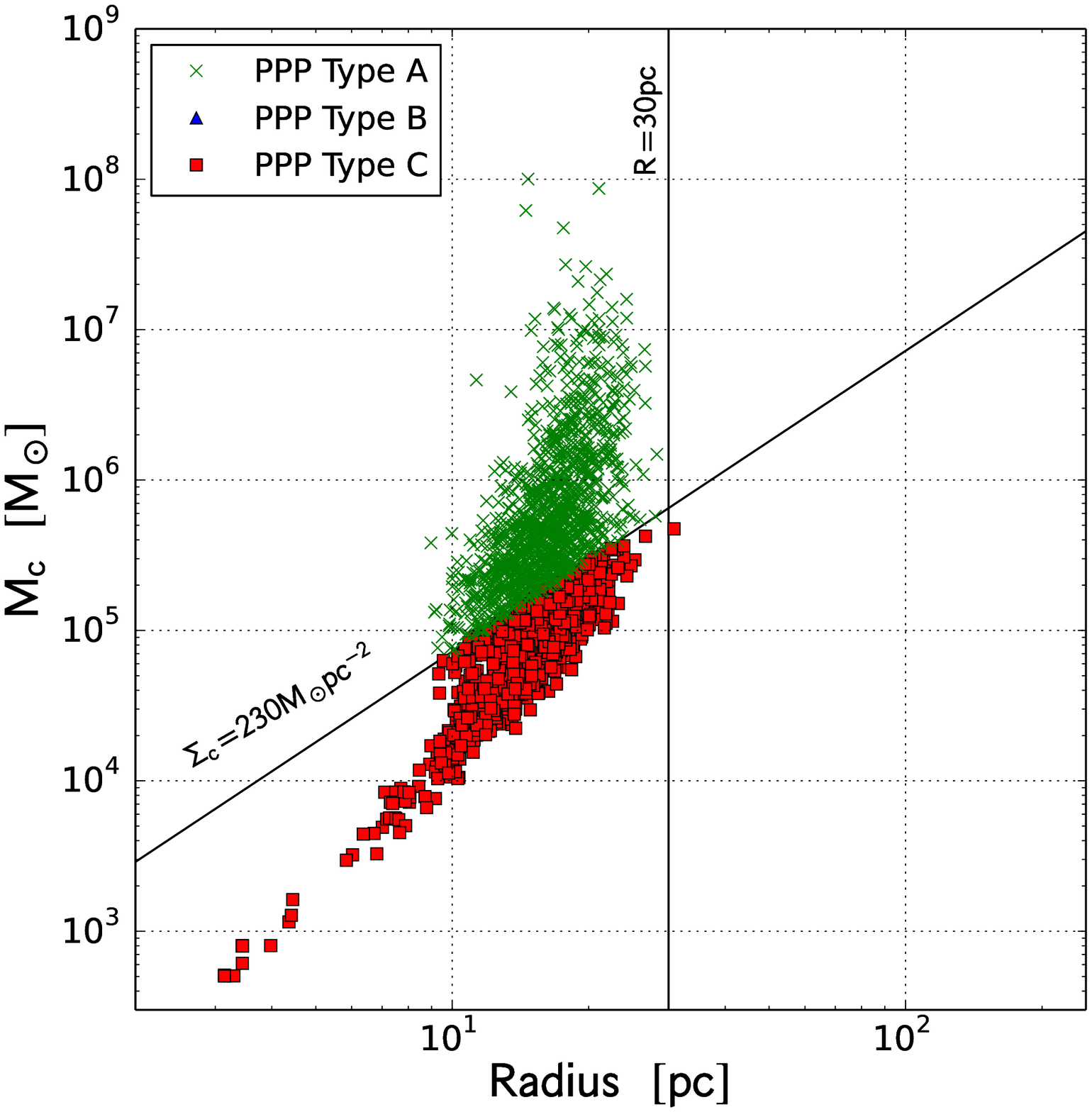} 
\begin{center}
 (a)
\end{center}
    \end{minipage}
    \begin{minipage}{0.32\textwidth}
        \centering
		\includegraphics[width=1\textwidth]{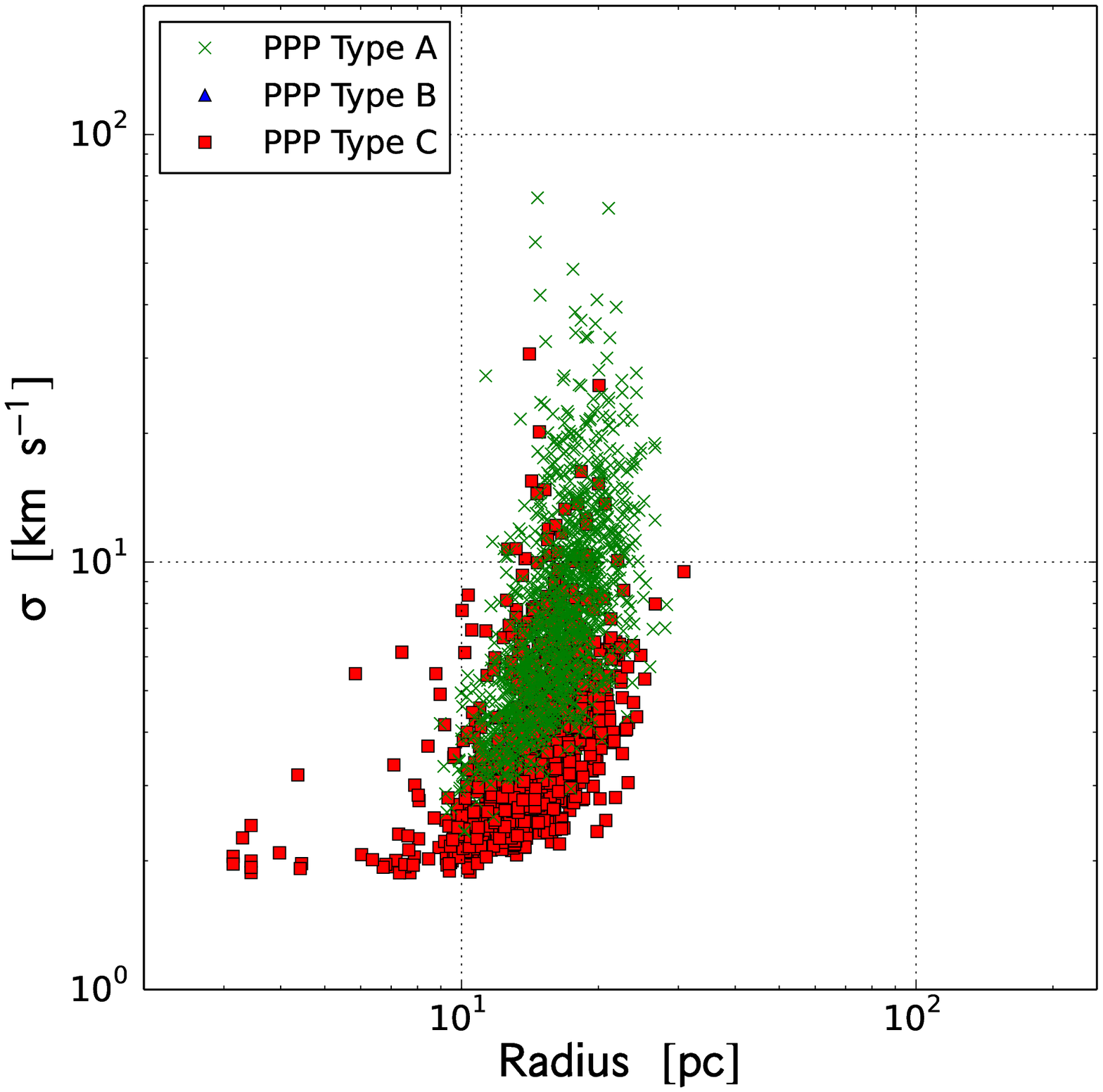} 
\begin{center}
 (b)
\end{center}
    \end{minipage}
        \begin{minipage}{0.32\textwidth}
        \centering
		\includegraphics[width=1\textwidth]{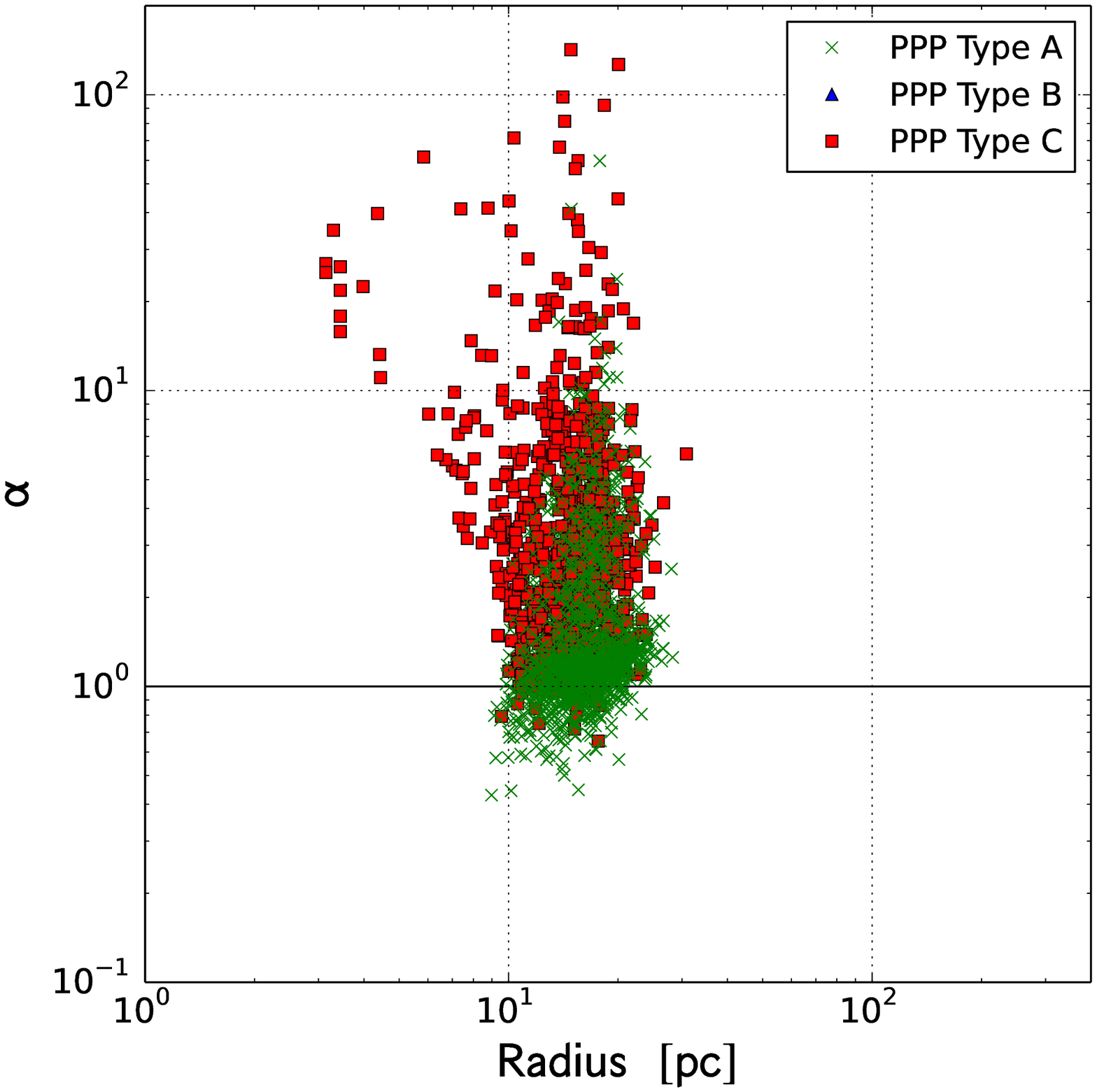} 
\begin{center}
 (c)
\end{center}
    \end{minipage}
            \begin{minipage}{0.32\textwidth}
        \centering
		\includegraphics[width=1\textwidth]{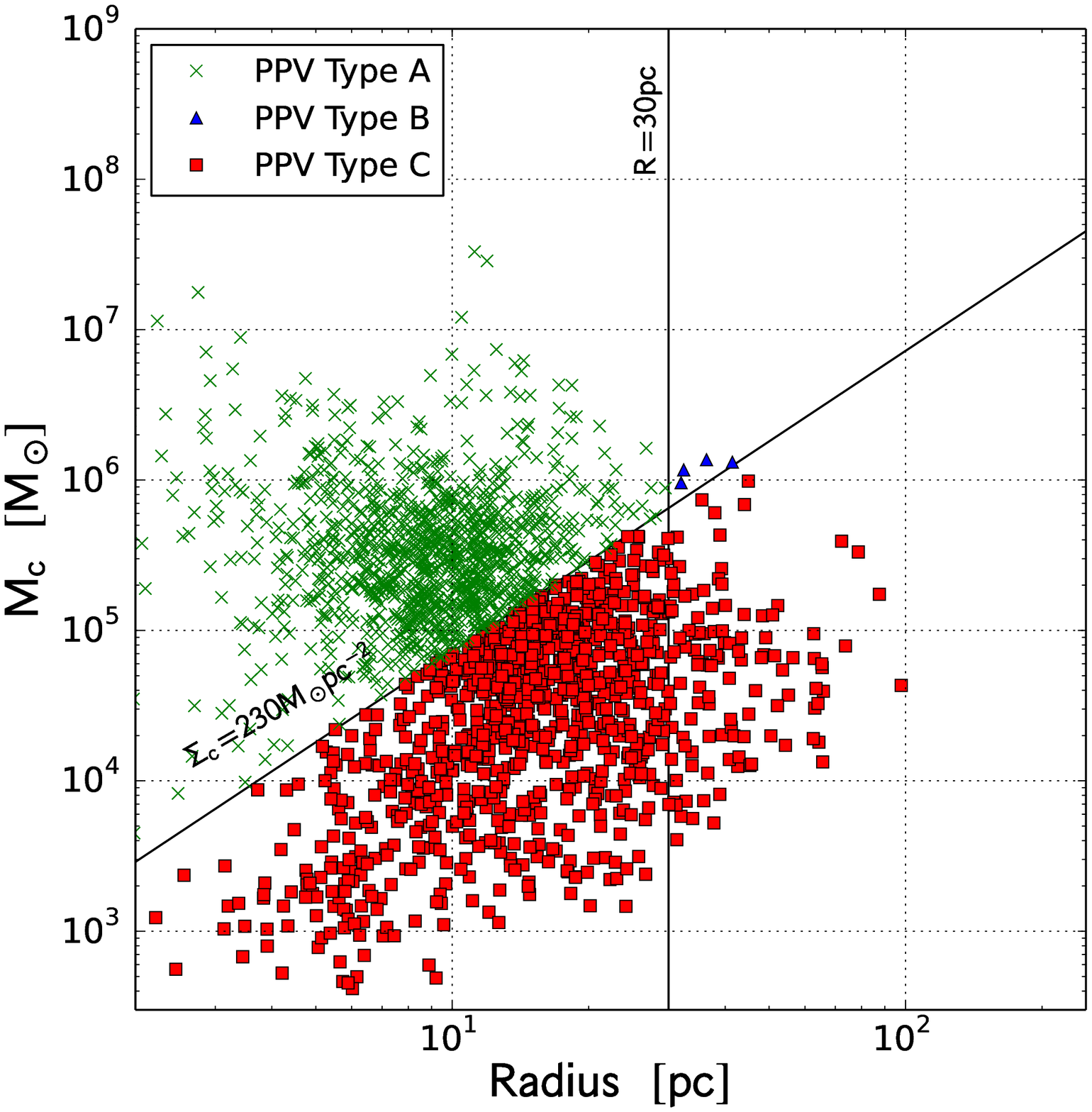} 
\begin{center}
 (d)
\end{center}
    \end{minipage}
            \begin{minipage}{0.32\textwidth}
        \centering
		\includegraphics[width=1\textwidth]{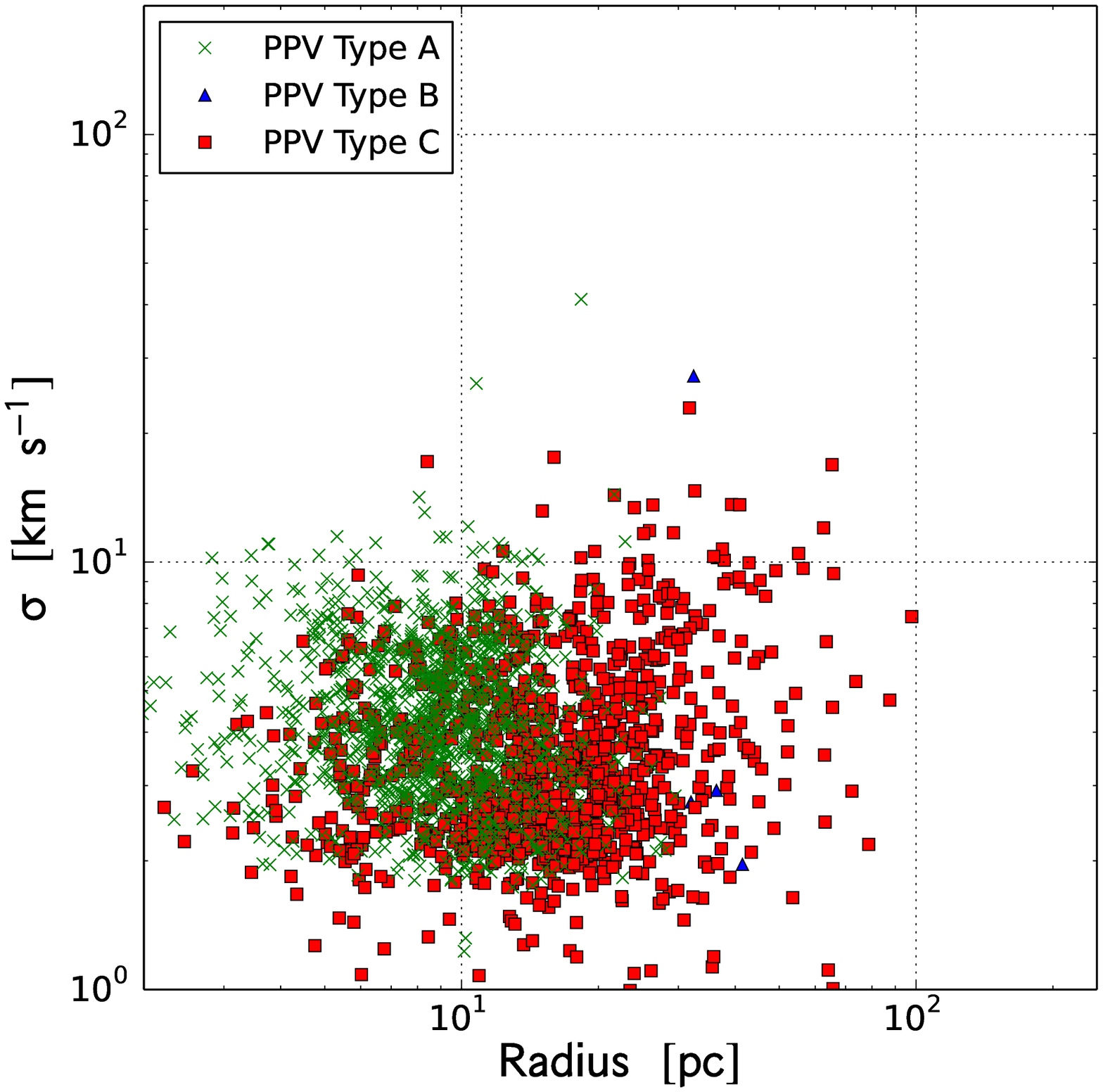} 
\begin{center}
 (e)
\end{center}
    \end{minipage}
            \begin{minipage}{0.32\textwidth}
        \centering
		\includegraphics[width=1\textwidth]{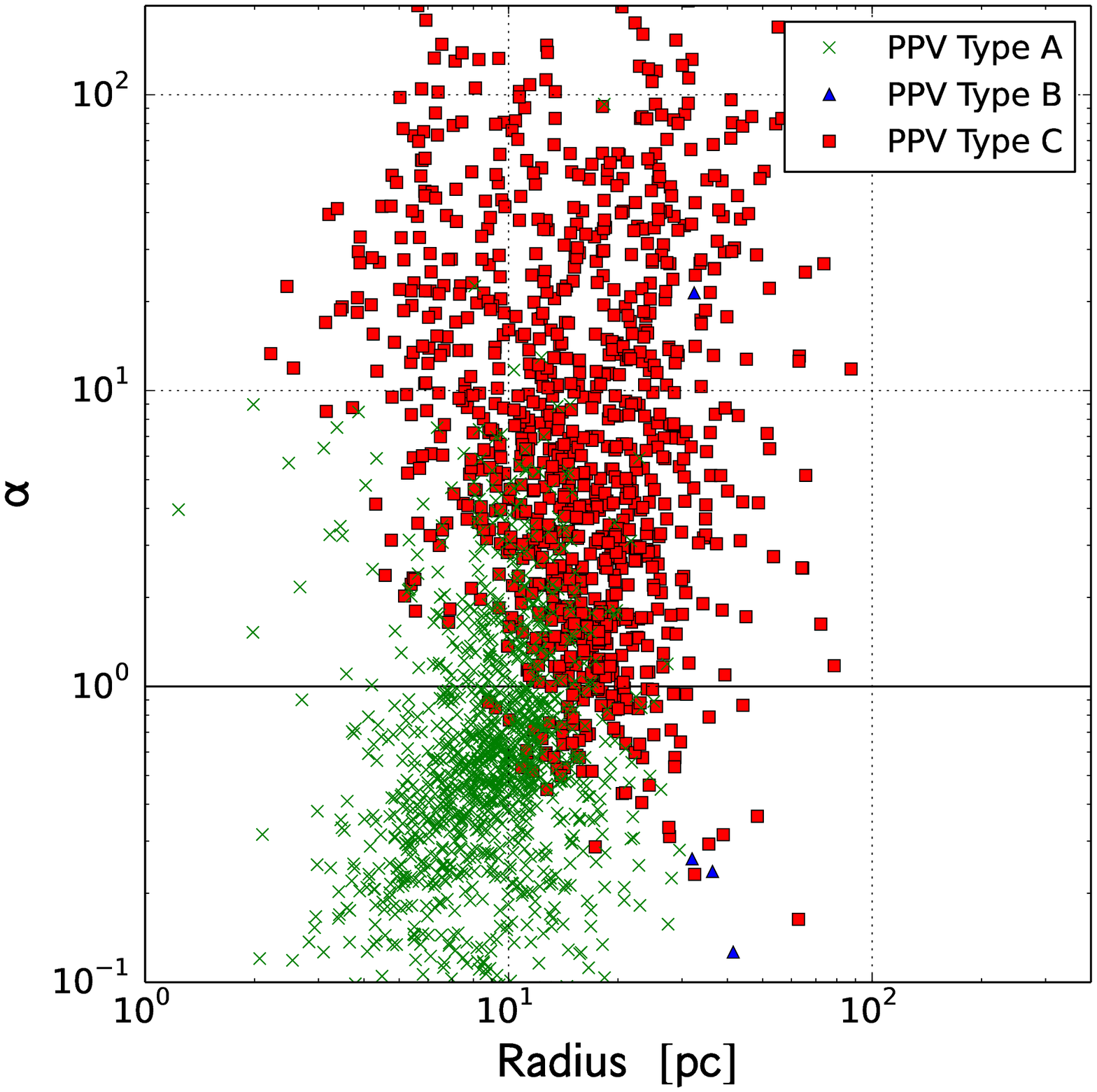} 
\begin{center}
 (f)
\end{center}
    \end{minipage}
	\caption{Scaling relations of the cloud properties using the decomposition methods. Color markers denote the cloud categories classified by cloud properties. The classification of clouds is introduced by \citet{Fuj14} with PPP clouds. (a)  Classification of cloud types based on PPP clouds on mass -- radius plane.  Cloud with mass surface density greater than 230 M$_{\sun}$ pc$^{2}$ and radius less than 30 pc are \emph{Type A}. Clouds sit on the same sequence but greater than 30 pc are \emph{Type B}. Clouds with  surface density less than 230 M$_{\sun}$ pc$^{2}$ are \emph{Type C}. Boundaries of cloud types are indicated with solid lines.  (b) The three-type PPP clouds on mass -- radius plane.  (c) Relation of virial parameter versus radius of PPP clouds. Panel (d), (e), and (f) are the same as panel (a), (b), and (c), respectively, but for PPV clouds.}
	\label{FIG_PP2_prop_property_decomp}
\end{figure*}

The scaling relations for the decomposition method are shown in Figure~\ref{FIG_PP2_prop_property_decomp}. Top panels (a), (b) and (c) show the PPP cloud data set for the two Larson relations and that of the virial parameter with radius. The bottom three panels (d), (e) and (f) show the same relationships for the PPV data. The symbols and colours for each cloud type match those in Figure~ \ref{FIG_PP2_prop_property}. In general, the trends in the PPP clouds are similar to that with the island cloud identification method, but the range of values is reduced due to the splitting of larger structures. The boundary between \emph{Type A} and \emph{Type C} clouds in the mass -- radius relationship in panel (a) is no longer seen as a gap, although two different trends appear to be visible. It vanishes entirely in the velocity -- radius relation in panel (b). The \emph{Type C} clouds remain the most unbound object in the simulation, with virial parameters $> 10$. The spread in the virial parameter value no longer depends greatly on radius, due to the more uniform cloud properties producing a much smaller spread in radius values. For clouds with virial parameter less than 10, the \emph{Type C} and \emph{Type A} clouds overlap more strongly than in the island method, again due to the smaller range in cloud properties.

In the PPV clouds using the decomposition method, the three relations are lost, with the radius, mass and velocity dispersion no longer well correlated. In the mass-radius relation in panel (d), the \emph{Type C} clouds do show the same general trend at the PPP clouds and island method, but with a much wider scatter. By contrast, the \emph{Type A} clouds suggest a reverse trend, where the mass anti-correlates with radius. This huge scatter is due to the watershed effect. In the cases were multiple peaks are within a massive island, all the peaks can share a high density envelope. This causes a large amount of material to be discarded, often up to $\sim 50$\% of the emission\footnote{The total mass of clouds in the decomposition method is 8.2 $\times$ 10$^{8}$ M$_{\sun}$ in PPV, and 1.8 $\times$ 10$^{9}$ M$_{\sun}$ in PPP. For comparison, the total mass of clouds in the island method is 1.6 $\times$ 10$^{9}$ M$_{\sun}$ in PPV, and 1.8 $\times$ 10$^{9}$ M$_{\sun}$ in PPP.}. The result is small contours centered around a high density region, producing the trend in the \emph{Type A} clouds. This effect is most notable in the splitting of the massive \emph{Type B} clouds, which often contain concetrated cores of material that become small objects. Extra scatter is also created from the mass-weighting of CPROPS on the radius, as discussed in previous sections. For the smaller \emph{Type C} clouds, the watershed effect simply increases radius scatter as clouds are divided. Note that a significant number of \emph{Type C} clouds are split from the remains of the \emph{Type B} island method clouds, producing objects both larger and smaller than the original \emph{Type C} clouds. 

The splitting of islands into multiple peaks also reduces the velocity dispersion of the clouds for PPV, as shown in panel (e). As only the partial structure ends up attacked to a peak, the velocity dispersion drops compared to the PPP case, where the extended envelope is not neglected. With the scatter in the radius, the \emph{Type A} and \emph{Type C} clouds strongly overlap and there is no visible trend with velocity dispersion. 

Where the virial parameter is plotted against radius in panel (f), the \emph{Type A} and \emph{Type C} clouds do occupy different region. \emph{Type A} clouds have a virial parameter of $\alpha < 10$ (mostly $\alpha < 1$), while the \emph{C} remain the most unbound objects with $\alpha > 1$. This difference is actually more marked in the PPV data than the PPP, due to the reduced radius and velocity dispersion of the \emph{Type A} clouds which lowers the virial parameter. \emph{Type C} clouds occupy approximately the same plot region in both PPV and PPP, but with a wider scatter in the radius as seen before. 

 Exactly which method --island or decomposition-- is more physical is up for debate. There is no reason why a cloud should contain a single peak and indeed, star formation is unlikely to be centred in only one location of a turbulent cloud. The use of the methods may depend on the quality of the observational data and the purpose of the study.   For the extragalactic observations with typical resolution of $\geq$ 20 pc, most studies adopt the decomposition method to select objects. However, the resolution at these distances means it is hard to discern internal cloud structure, producing clouds that are similar in size to the island method in this work. On the other hand, inspite of the higher resolutions ($<$ 20 pc), most of Galactic studies continue to adopt the decomposition method. This is due to the large dynamic range in the distance to the clouds, varying the resolution within a cloud's boundary. Both these choices originate from the lack of a clear definition for a GMC, leaving is ambiguous as to whether it is resolved. Our study emphasises the importance of this cloud definition and for high resolution data, the island method is the stronger choice for selecting similar objects in both simulation and observation data types; a valuable asset in understanding the evolution of star-forming clouds.

\section{Summary}
\label{sec_summary}

We are rapidly reaching the point where simulations and observations will achieve comparable resolution of molecular cloud populations in a wide variety of galaxies. This opens the door  to truly constrain the mechanisms for cloud formation --and thereby star formation-- in different galaxy environments. The importance of this advance makes it imperative to examine carefully if the two techniques of observation and simulation are discussing the same star-forming molecular clouds. After all, observational data is done in two spatial co-ordinates and one velocity co-ordinate, while simulation data typically uses all three spatial dimensions, but ignores the velocity components. A direct match is not possible.

In this paper, we compared the physical properties of giant molecular clouds formed in a simulation of a barred spiral galaxy using simulation and observational identification methods. The two methods selected clouds from the data in position-position-position (PPP) space typical for simulation and position-position-velocity (PPV) space used in observations. The PPV data cube was assumed to be the product of $^{12}$CO (1--0) observations with optimal resolution and sensitivity. The properties of the clouds found in both methods were compared and the clouds themselves matched to assess whether the two methods identified the same objects and if the properties were method dependent. This process was repeated twice; once where the clouds were identified as continuous structures within a contour (``island method'') and once where each peak in the density or emission was assigned to a seperate cloud (``decomposition method'').

The main results for the general cloud properties when using the island method are as follows:

\begin{enumerate}

  \item The total number of clouds identified was similar in the PPV and PPP methods (971 versus 1029). This was also true when comparing cloud numbers in each galactic environment of bar, spiral and disc. 

  \item The typical (median) cloud properties, such as their mass, radius and velocity dispersion, differ by $\leqslant 20$\% between PPV and PPP.

  \item The bimodal mass surface density distribution of PPP clouds (peaks at $\sim$ 100 and $\sim$ 1000 M$_{\sun}$ pc$^{-2}$)  which had been found by \cite{Fuj14} is reproduced in the PPV clouds, with the boundary between the two populations being roughly consistent.

\end{enumerate}

\noindent We matched clouds one-to-one to compare any changes in an individual cloud's properties when a different method is used. This process demonstrated the following for the island method:

\begin{enumerate}

  \item About 70\% of clouds have single counterpart in both data sets. This match rate was consisted in all three galactic environments. 

  \item The variation in properties between matched clouds typically lies within a scatter of a factor of two. The largest scatter is seen for the derived properties, that depend on multiple cloud variables, such as the surface density and the virial parameter. The differences in these properties suggest care should be taken when interpreting their physical meaning. 

  \item Smaller clouds ($\leqslant$ 10$^{5}$ M$_{\sun}$) have a larger scatter in their velocity dispersion because the velocity spacing of our PPV (1 km s$^{-1}$) data cube is too large to resolve the velocity dispersion of these small objects. This provides a guide of the velocity (instrumental spectral) resolution needed for real observations.
\end{enumerate}

In the analysis of this simulation presented in \citet{Fuj14} using the PPP cloud identification method, clouds were found to fall into three different populations: the \emph{Type A} clouds which have cloud properties agreeing with typical observations and account for the largest fraction of clouds in all galactic environments. The \emph{Type B} massive associations that form through mergers of small clouds and are therefore seen in high-interaction environments like the bar more commonly than the disc. Finally, the \emph{Type C} clouds that are transient, unbound objects, forming in filaments and tidal tails, making them most common where the gravitationally dominant \emph{Type B} clouds are prevalent. When classifying PPV clouds with the same definitions, the main results are:

\begin{enumerate}
  \item  The fraction of each cloud type differs by $<$ 10\% between PPV and PPP in all environments.

  \item Among the $\sim 70$\% of clouds that have a direct match in both data sets, as high as $\sim 80$\% are categorised as same type in all environments. The remaining 10\% that differ change type due to the different mass and/or radius between two data sets.
\end{enumerate}

When we switched from identifying clouds as continuous islands to the peak-based decomposition method, the number of clouds found in both PPP and PPV increases significantly by $\sim$ 3 times. The resulting cloud properties for the overall population remained similar, but the one-to-one cloud match became significantly more difficult in the crowded environment and the match rate dropped to 40\%. The division of clouds such that they contain only a single peak also made the cloud properties more uniform, largely obscuring the three cloud types seen in the island method. This was particularly true in the PPV data set, which had a large amount of scatter in cloud properties. We conclude that the cloud identification method therefore plays a critical role in determining the cloud properties and therefore understanding the influence of galactic environment. 

\begin{figure}	
\includegraphics[width=0.45\textwidth]{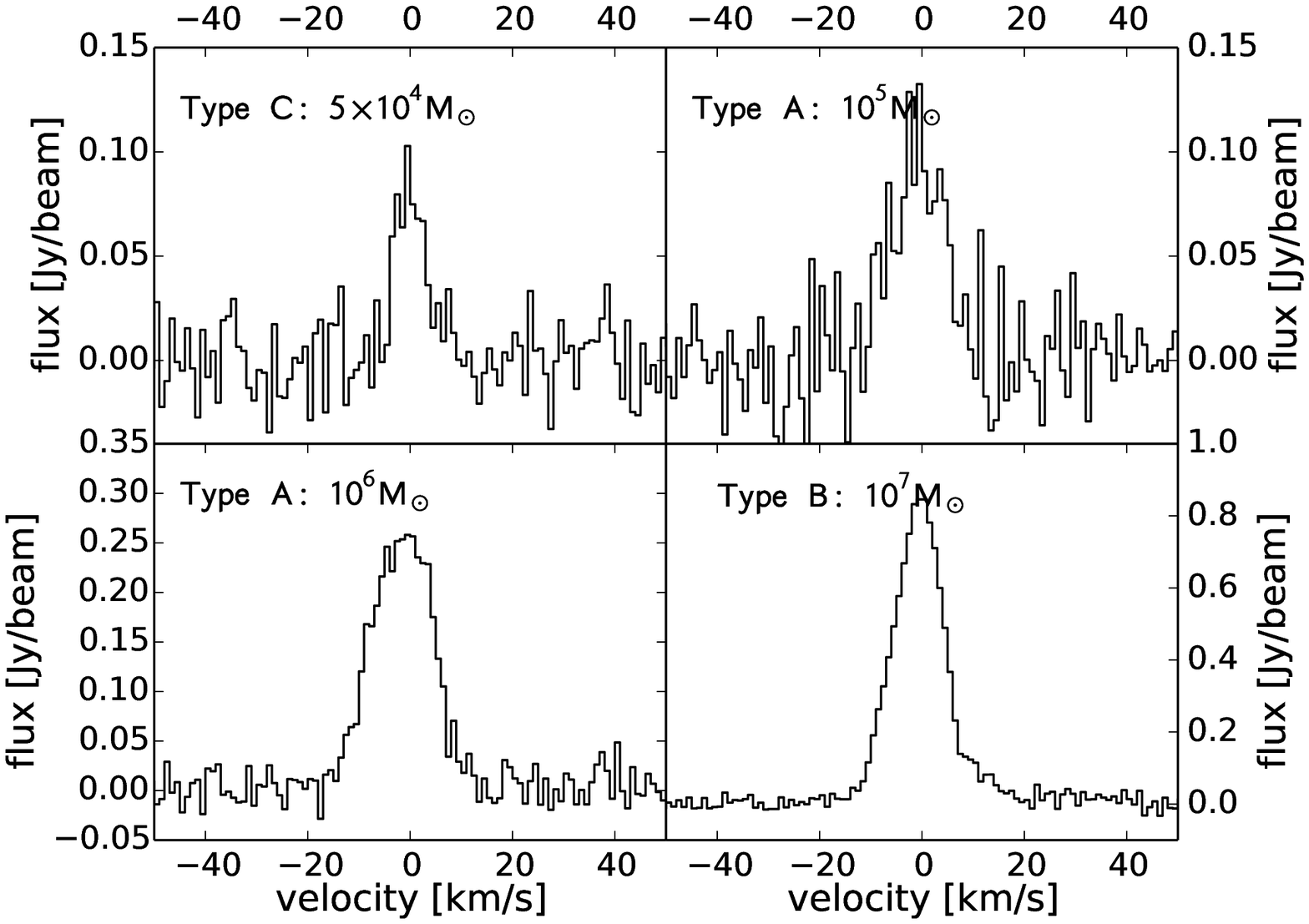} 
\caption{CASA simulated spectrum of the three types of GMCs found in the M83 simulation as observed with ALMA cycle 3 capabilities. The input model is the noise-free PPV cube. All of the clouds presented in this figure are classified as the same category in PPP and PPV. The cloud type and mass (in PPP) are presented in each panel. The simulation is performed with CASA task \texttt{simobserve} and imaged with \texttt{simanalyze}. Only the 12-m array is used in this simulation. The antenna configuration file of ``alma.cycle3.3.cfg'' is used.    Total observing time for the 12-m array is $\sim$ 2 hours ($\sim$ 5 hours is required if the ACA total power observations are included. This is certainly necessary in real observations). The mapping area is 130$\arcsec$ $\times$ 150$\arcsec$ (2.6 $\times$ 3.0 kpc).  Spatial, velocity resolutions and sensitivity are 0.7$\arcsec$ (14 pc), 1 km s$^{-1}$ and 0.012 Jy, respectively.} 
\label{FIG_M83_casa}
\end{figure}

Finally, we emphasize that the resolution of our data means we are only considering the ``best possible'' situation. Our PPV data cube has minimal noise and much higher resolution than typical observations. Based on a simple estimation for ALMA observations using the simulation function of CASA, it is unlikely we can achieve such a fine resolution and deep observation for the entirety of M83 within a reasonable time for Cycle 3 (2015 -- 2016) observations, although a partial galaxy may be possible. As an example, Figure \ref{FIG_M83_casa} shows the simulated spectrum for the three types of cloud derived from a $\sim$ 2-hour ALMA Cycle 3 observations of M83  (including calibrators)  using CASA \citep[the Common Astronomy Software Applications package,][]{McM07}, assuming a 2.6 $\times$ 3.0 kpc observed area in the bar and spiral regions. We have also ignored the inclination of the galaxy, chemical processes and radiation transfer in the ISM. Such effects would alter the observed column density of molecular clouds and also raise the difficulty in identifying the clouds themselves. 

For our main results using the island method, we conclude that both PPP and PPV can potentially identify the same objects with close properties. Therefore, the techniques themselves are able to be compared well. The question then becomes one of fighting down noise and achieving the necessary resolution to understand cloud and star formation in galaxies.

The next step in this work is to evaluate the  projection effect as a function of different  inclinations of galactic discs and physical resolutions. This will provide a set of reference GMCs to help observers to declare the intrinsic bias in the their observed GMC properties.

\label{lastpage}

\begin{thebibliography}{99}
\bibitem[Andr{\'e} et 
al.(2010)]{And10} Andr{\'e}, P., Men'shchikov, A., Bontemps, S., et al.\ 2010, A\&A, 518, LL102 

\bibitem[Alves et 
al.(2007)]{Alv07} Alves, J., Lombardi, M., \& Lada, C.~J.\ 2007, A\&A, 462, L17 


\bibitem[Battisti 
\& Heyer(2014)]{Bat14} Battisti, A.~J., \& Heyer, M.~H.\ 2014, ApJ, 780, 173 

\bibitem[Beaumont et al.(2013)]{Bea13} Beaumont, C.~N., 
Offner, S.~S.~R., Shetty, R., Glover, S.~C.~O., 
\& Goodman, A.~A.\ 2013, ApJ, 777, 173 


\bibitem[Bern{\'e} et al.(2014)]{Ber14} Bern{\'e}, O., Marcelino, N., \& Cernicharo, J.\ 2014, ApJ, 795, 13 

\bibitem[Beuther et al.(2002)]{Beu02} Beuther, H., Schilke, 
P., Menten, K.~M., et al.\ 2002, ApJ, 566, 945 


\bibitem[Bolatto et 
al.(2013)]{Bol13} Bolatto, A.~D., Wolfire, M., \& Leroy, A.~K.\ 2013, ARA\&A, 51, 207 

\bibitem[Brunt et al.(2003)]{Bru03} Brunt, C.~M., Kerton, 
C.~R., \& Pomerleau, C.\ 2003, ApJS, 144, 47 

\bibitem[Bryan et al.(2014)]{Bry14} Bryan, G.~L., Norman, 
M.~L., O'Shea, B.~W., et al.\ 2014, ApJS, 211, 19 

\bibitem[Burton et al.(2013)]{Bur13} Burton, M.~G., Braiding, 
C., Glueck, C., et al.\ 2013, PASA, 30, e044

\bibitem[{{Colombo} {et~al.}(2014){Colombo}, {Hughes}, {Schinnerer}, {Meidt},
  {Leroy}, {Pety}, {Dobbs}, {Garc{\'{\i}}a-Burillo}, {Dumas}, {Thompson},
  {Schuster}, \& {Kramer}}]{Col14}
{Colombo}, D., {Hughes}, A., {Schinnerer}, E., {et~al.} 2014, ApJ, 784, 3 


\bibitem[Dobbs et al.(2011)]{Dob11} Dobbs, C.~L., Burkert, 
A., \& Pringle, J.~E.\ 2011, MNRAS, 413, 2935 

\bibitem[Donovan Meyer et al.(2013)]{Don13} Donovan Meyer, 
J., Koda, J., Momose, R., et al.\ 2013, ApJ, 772, 107 

\bibitem[{{Fujimoto} {et~al.}(2014){Fujimoto}, {Tasker}, {Wakayama}, \&
  {Habe}}]{Fuj14}
{Fujimoto}, Y., {Tasker}, E.~J., {Wakayama}, M., \& {Habe}, A. 2014, MNRAS,
  439, 936

\bibitem[Goldsmith et al.(2008)]{Gol08} Goldsmith, P.~F., 
Heyer, M., Narayanan, G., et al.\ 2008, ApJ, 680, 428 

\bibitem[Goodman et al.(2009)]{Goo09} Goodman, A.~A., Pineda, 
J.~E., \& Schnee, S.~L.\ 2009, ApJ, 692, 91  

\bibitem[Heyer et al.(2009)]{Hey09} Heyer, M., Krawczyk, C., 
Duval, J., \& Jackson, J.~M.\ 2009, ApJ, 699, 1092 

\bibitem[Hughes et al.(2013)]{Hug13} Hughes, A., Meidt, 
S.~E., Colombo, D., et al.\ 2013,  ApJ, 779, 46 

\bibitem[Jarrett et al.(2003)]{Jar03} Jarrett, T.~H., 
Chester, T., Cutri, R., Schneider, S.~E., 
\& Huchra, J.~P.\ 2003, AJ, 125, 525 

\bibitem[Larson(1981)]{Lar81} Larson, R.~B.\ 1981, MNRAS, 
194, 809 


\bibitem[Lada 
\& Lada(2003)]{Lad03} Lada, C.~J., \& Lada, E.~A.\ 2003, ARA\&A, 41, 57 

\bibitem[McMullin et al.(2007)]{McM07} McMullin, J.~P., 
Waters, B., Schiebel, D., Young, W., 
\& Golap, K.\ 2007, Astronomical Data Analysis Software and Systems XVI, 376, 127 

\bibitem[Ostriker et al.(2001)]{Ost01} Ostriker, E.~C., 
Stone, J.~M., \& Gammie, C.~F.\ 2001, \apj, 546, 980 

\bibitem[Roman-Duval et al.(2010)]{Rom10} Roman-Duval, J., 
Jackson, J.~M., Heyer, M., Rathborne, J., 
\& Simon, R.\ 2010, ApJ, 723, 492 

\bibitem[Rosolowsky \& Leroy(2006)]{Ros06} Rosolowsky, E., \& Leroy, A.\ 2006, PASP, 118, 590 

\bibitem[Schinnerer et al.(2013)]{Sch13} Schinnerer, E., 
Meidt, S.~E., Pety, J., et al.\ 2013, ApJ, 779, 42 

\bibitem[Solomon et al.(1987)]{Sol87} Solomon, P.~M., Rivolo, 
A.~R., Barrett, J., \& Yahil, A.\ 1987, ApJ, 319, 730 

\bibitem[Tasker 
\& Bryan(2006)]{Tas06} Tasker, E.~J., \& Bryan, G.~L.\ 2006, ApJ, 641, 878 


\bibitem[Tasker \& Tan(2009)]{Tas09} Tasker, E.~J., \& Tan, J.~C.\ 2009, ApJ, 700, 358

\bibitem[Thim et al.(2003)]{Thi03} Thim, F., Tammann, G.~A., 
Saha, A., et al.\ 2003, ApJ, 590, 256 

\bibitem[Ward et al.(2012)]{War12} Ward, R.~L., Wadsley, J., 
Sills, A., \& Petitclerc, N.\ 2012, ApJ, 756, 119 

\bibitem[Wong et al.(2011)]{Won11} Wong, T., Hughes, A., Ott, 
J., et al.\ 2011,  ApJS, 197, 16 


\bibitem[Whitmore et al.(2014)]{Whi14} Whitmore, B.~C., 
Brogan, C., Chandar, R., et al.\ 2014,  ApJ, 795, 156 

\bibitem[Yim et al.(2014)]{Yim14} Yim, K., Wong, T., Xue, R., 
et al.\ 2014, arXiv:1408.5905 

\end{thebibliography}
\end{document}